\documentclass[preprint,NumberedRefs]{JASA}
\usepackage{multirow}
\usepackage{subfigure}
\usepackage{tikz}
\usetikzlibrary{positioning}
\tikzset{>=stealth}

\begin{document}
\title[]{A Chebyshev--Tau Spectral Method for Coupled Modes of Underwater Sound Propagation in Range-Dependent Ocean Environments}

	\author{Houwang Tu}
\email{tuhouwang@nudt.edu.cn}
\affiliation{College of Computer, National University of Defense Technology, Changsha, 410073, China}

	\author{Yongxian Wang*}
\email{yxwang@nudt.edu.cn}
\affiliation{College of Meteorology and Oceanography, National University of Defense Technology, Changsha, 410073, China}

	\author{Chunmei Yang}
\email{ycm@fio.org.cn}
\affiliation{First Institute of Oceanography and Key Laboratory of Marine Science and Numerical Modeling, Ministry of Natural Resources, Qingdao, 266061, China}

	\author{Wei Liu}
\affiliation{College of Meteorology and Oceanography, National University of Defense Technology, Changsha, 410073, China}

	\author{Wenbin Xiao}
\affiliation{College of Meteorology and Oceanography, National University of Defense Technology, Changsha, 410073, China}

	\author{Xiaodong Wang}
\affiliation{College of Computer, National University of Defense Technology, Changsha, 410073, China}

\begin{abstract}
\noindent \textbf{ABSTRACT}
\noindent The stepwise coupled-mode model is a classic approach for solving range-dependent sound propagation problems. Existing coupled-mode programs have disadvantages such as high computational cost, weak adaptability to complex ocean environments and numerical instability. In this paper, a new algorithm is designed that uses an improved range normalization and global matrix approach to address range dependence in ocean environments. Due to its high accuracy in solving differential equations, the spectral method has recently been applied to range-independent normal modes and has achieved remarkable results. This algorithm uses the Chebyshev--Tau spectral method to solve for the eigenmodes in the range-independent segments. The main steps of the algorithm are parallelized, so OpenMP multithreading technology is also applied for further acceleration. Based on this algorithm, an efficient program is developed, and numerical simulations verify that this algorithm is reliable, accurate and capable. Compared with the existing coupled-mode programs, the newly developed program is more stable and efficient at comparable accuracies and can solve waveguides in more complex and realistic ocean environments.
\end{abstract}

	\maketitle

\section{Introduction}
The numerical sound field of a range-dependent waveguide is a research hot spot in computational ocean acoustics. At present, techniques for solving range-dependent acoustic propagation problems include coupled modes, adiabatic modes, rays \cite{Jensen2011,Etter2018}, the parabolic approximation \cite{RAM} and direct solutions to the Helmholtz equation using finite difference \cite{Liuw2021} or finite element methods \cite{Murphy1988a,Murphy1988b,Murphy1989,Murphy1996}. Each method or model has its own advantages and disadvantages. Coupled-mode theory is a classic model to solve sound propagation problems in range-dependent ocean environments. It is often used to provide benchmark solutions to test the reliability of other numerical models because of its high accuracy.

The classic normal mode theory proposed by Pekeris \cite{Pekeris1948} provides solutions suitable only for range-independent acoustic waveguides and is powerless for range-dependent problems. The theory of coupled modes was proposed by Pierce \cite{Pierce1954} and Miller \cite{Miller1954} in 1954; they asserted that energy is exchanged between normal modes in a horizontally changing waveguide. Subsequently, Rutherford and Hawker \cite{Rutherford1981} noted that Pierce and Miller's use of vertical derivative operators to replace normal derivative operators resulted in nonconservation of energy in sloping terrains; consequently, they proposed a first-order modification to coupled-mode theory to maintain first-order conservation of energy on slopes; Fawcett provided a full, analytically exact evaluation of these same terms \cite{Fawcett1992}. In 1983, Evans \cite{Evans1983} proposed the idea of using a stair-step geometry to discretize sloping terra, where each step was considered a flat segment. In combination with boundary conditions, the propagator matrix between the coupling coefficients of each segment can be obtained, and the coupling coefficients of the segments can be obtained by considering radiation conditions. The acoustic field solution of each segment contains both the forward scattering mode, which exponentially decays with increasing range, and the backward scattering mode, which exponentially grows with increasing range. When considering leaky modes, the traditional superposition method suffers from numerical instability. In 1985, Mattheij \cite{Mattheij1985} proposed a decoupling matrix algorithm to solve the two-point boundary value problem. Soon after, Evans \cite{Evans1986} applied this decoupling algorithm to stepwise coupled modes, successfully resolved the numerical instability caused by leaky modes, and developed the numerical program COUPLE. The latest version, COUPLE07 \cite{Couple}, can accurately calculate the fully elliptic two-way solution of the Helmholtz equation, which is considered to be an outstanding representative of coupled modes and has been widely used for many years to provide accurate solutions for numerical experiments.

However, Luo et al. \cite{Luowy2012a,Luowy2012b,Luowy2012c,Luowy2012d} and Yang et al. \cite{Yangcm2012,Yangcm2015a} reported that COUPLE exhibited numerical instability due to unreasonable normalized range solutions. In solving for the range-independent normal modes, COUPLE employs the Galerkin method, which forms a generalized eigenvalue problem of symmetric matrices $\mathbf{A}$ and $\mathbf{B}$ \cite{Couple} in each segment:
\linenomath
\begin{subequations}
		\begin{gather}
			\mathbf{Au}=\lambda\mathbf{Bu}\\
			A_{n,m}=\int_{0}^{H} \frac{k^{2}(z) \phi_{n}(z) \phi_{m}(z)}{\rho(z)} \mathrm{d} z-\int_{0}^{H} \frac{1}{\rho(z)} \frac{\mathrm{d} \phi_{n}(z)}{\mathrm{d} z} \frac{\mathrm{d} \phi_{m}(z)}{\mathrm{d} z} \mathrm{d} z \\
			B_{n,m}=\int_{0}^{H} \frac{\phi_{n}(z) \phi_{m}(z)}{\rho(z)} \mathrm{d} z
		\end{gather}
	\end{subequations}
where $\phi(x)$ are the basis/weight functions in the Galerkin method. Although the matrices $\mathbf{A}$ and $\mathbf{B}$ are both formally symmetrical (symmetry means that such a generalized eigenvalue problem is efficient to solve), the elements in matrices $\mathbf{A}$ and $\mathbf{B}$ must be individually obtained through numerical quadrature, which requires many calculations. In addition, the Galerkin method must construct basis functions that satisfy the boundary conditions in each segment, which imposes considerable computational cost. Furthermore, COUPLE considers only two layers of media, which is a limitation in many cases. For example, for the lower boundary of the acoustic half-space, the bottom sediment of COUPLE needs to be set as an absorbing layer, which precludes flexibility for complicated waveguides. The KRAKEN program based on the finite difference method has good flexibility in solving for range-independent normal modes, but it can calculate only one-way coupled modes, and the stability of the coupling is often unsatisfactory \cite{Kraken2001}.

In recent years, many studies have begun to address acoustic propagation problems by applying more accurate spectral methods \cite{Tuhw2020a,Tuhw2021a,Tuhw2021b,Wangyx2021a,SMPE,Tuhw2021c,Wangyx2021b}. In 1993, Dzieciuch \cite{Dzieciuch1993,aw} first used the Chebyshev--Tau spectral method to solve for the normal modes of the water column. Evans \cite{rimLG} in 2016 devised a Legendre--Galerkin spectral method to solve the problem of acoustic propagation in a two-layer ocean environment that contained bottom sediment. In 2020, Tu et al. \cite{Tuhw2020a,Tuhw2021a} used the Chebyshev--Tau spectral method to more efficiently solve this problem. Numerical experiments have shown that the NM-CT program based on the Chebyshev--Tau spectral method \cite{NM-CT} is faster than the rimLG program \cite{rimLG} based on the Legendre--Galerkin spectral method and more accurate than the classic finite difference method \cite{Kraken2001}. Recently, Sabatini et al. \cite{Sabatini2019} and Tu et al. \cite{Tuhw2021c,MultiLC} used the Chebyshev collocation method and Legendre collocation method, respectively, to solve the problem of acoustic propagation in multilayer media. Existing studies have shown that spectral methods can solve underwater acoustic propagation problems with high accuracy. However, the current programs \cite{aw,rimLG,NM-CT,MultiLC} based on spectral methods can provide solutions for only range-independent acoustic waveguides. The present article combines stepwise coupled modes with the Chebyshev--Tau spectral method to develop a new algorithm that can efficiently provide solutions for range-dependent acoustical waveguides. Compared with the existing program-based coupled modes, the capability and computational efficiency of the algorithm proposed in this paper are greatly improved while maintaining the same accuracy.

\section{Physical Model}
\subsection{Range-independent normal modes}
We consider a two-dimensional point source acoustic field in a cylindrical axisymmetric environment, where the angular frequency of the acoustic source is $\omega$ and the simple harmonic point source is located at $r=0$ with $z=z_\mathrm{s}$. Let the acoustic pressure be $p=p(r,z)$, and omit the time factor $\exp(-\mathrm{i}\omega t)$. The acoustic governing equation (Helmholtz equation) can be written as \cite{Jensen2011}:
\begin{equation}
		\label{eq:2}
		\frac{1}{r}\frac{\partial}{\partial r}\left(
		r\frac{\partial p}{\partial r}
		\right) + 
		\rho (z) \frac{\partial}{\partial z}\left(
		\frac{1}{\rho (z)}\frac{\partial p}{\partial z}
		\right) +
		\frac{\omega ^2}{c^2(z)}p=-\frac{\delta(r)\delta(z-z_\mathrm{s})}{2 \pi r}
	\end{equation}
where $\omega=2\pi f$, $f$ is the frequency of the sound source, and $c(z)$ and $\rho(z)$ are the sound speed and density profiles, respectively.

Using the technique of the separation of variables \cite{Pekeris1948}, the acoustic pressure can be decomposed into:
\begin{equation}
		\label{eq:3}
		p(r,z)=\psi(z)R(r)
	\end{equation}
where $R(r)$ is related only to the range $r$ and satisfies:
\begin{equation}
		\label{eq:4}
		\frac{1}{r}\frac{\mathrm{d}}{\mathrm{d}r}\left(r\frac{\mathrm{d}R(r)}{\mathrm{d}r}\right)+k_{r}^2R(r)=-\frac{\delta(r)\psi(z_\mathrm{s})}{2 \pi r \rho(z_\mathrm{s})}
	\end{equation}
where $k_{r}$ is the horizontal wavenumber. By solving the above formula, we obtain:
\begin{equation}
		\label{eq:5}
		R(r)=\frac{\mathrm{i}}{4\rho (z_\mathrm{s})}\psi(z_\mathrm{s})\mathcal{H}_0^{(1)}(k_{r}r)
	\end{equation}
where $\mathcal{H}_0^{(1)}(\cdot)$ is the Hankel function of the first type and $\psi(z)$ in Eq.~\eqref{eq:3} satisfies the following modal equation:
\begin{equation}
		\label{eq:6}
		\rho(z)\frac{\mathrm{d}}{\mathrm {d}z}\left(
		\frac{1}{\rho(z)}\frac{\mathrm{d}\psi(z)}{\mathrm {d}z}
		\right) + k^2\psi(z) = k_{r}^2 \psi(z),\quad k=(1+\mathrm{i}\eta\alpha)\omega/c(z)
	\end{equation}
where $k$ is the complex wavenumber, $\alpha$ is the attenuation coefficient in dB$/\lambda$ ($\lambda$ is the wavelength), and $\eta=(40\pi \log_{10}{\mathrm{e}})^{-1}$. This is the essential equation to be solved in this paper. When supplemented by boundary conditions, Eq.~\eqref{eq:6} has a set of solutions $(k_{r,m},\psi_m), m=1, 2, \dots$, where $\psi_m$ is also called the eigenmode. The eigenmodes of Eq.~\eqref{eq:6} satisfy orthogonal normalization:
\begin{equation}
		\label{eq:7}
		\int_{0}^{H} \frac{{\psi_m(z)}{\psi_n(z)}}{\rho(z)}\mathrm {d} z=\delta_{mn},
		\quad m,n = 1, 2, \dots
	\end{equation}
where $H$ is the ocean depth and $\delta$ is the Kronecker delta function. Finally, the fundamental solution of the Helmholtz equation can be written as:
\begin{equation}
		\label{eq:8}
		p(r,z)=\frac{\mathrm{i}}{4\rho(z_\mathrm{s})}\sum_{m=1}^{\infty}\psi_m(z_\mathrm{s})\psi_m(z)\mathcal{H}_0^{(1)}(k_{r,m}r)
	\end{equation}
To accurately obtain the sound pressure, it is necessary to synthesize an infinite number of eigenmodes, which is impossible in actual calculations. It is usually more practical to take $M$ physically meaningful eigenmodes to synthesize the sound field. The specific value of $M$ can usually be estimated from the depth of the ocean $H$, the speed of sound $c$, and the frequency of the sound source $f$.

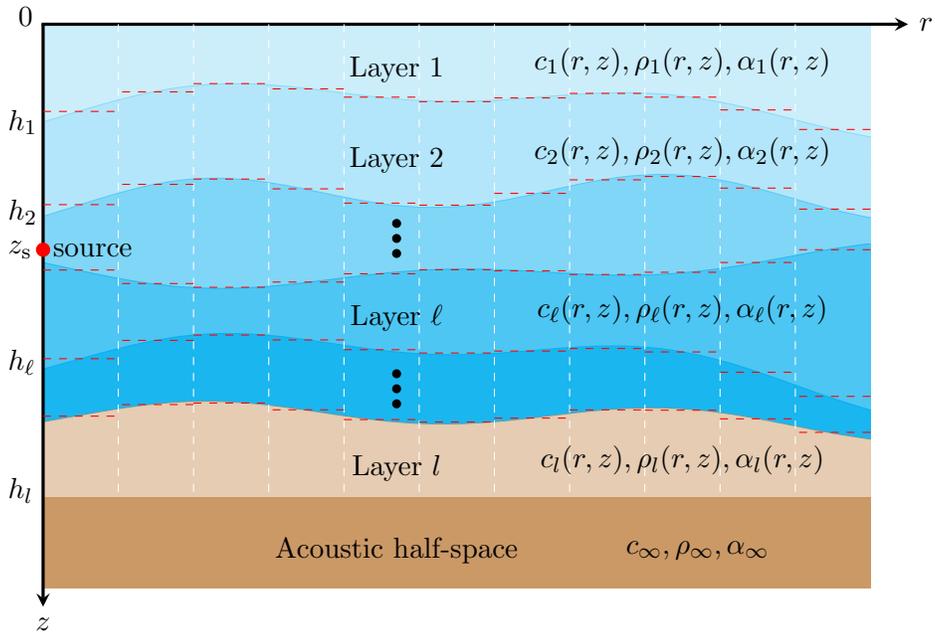
\begin{figure}[htbp]
	\centering
\begin{tikzpicture}[node distance=3cm]
		\filldraw [cyan,opacity=0.2]  plot[domain=-0.5:10.5,smooth] (\x ,{2.6+0.3*(-1.5+0.5*sin(\x r)+3*cos(0.1*\x r)+2*sin(0.2*\x r)+cos(0.3*\x r)+0.1*\x)})  -- (10.5,4.5) -- (-0.5,4.5)--cycle;
		\filldraw [cyan,opacity=0.3]  plot[domain=-0.5:10.5,smooth] (\x ,{2.6+0.3*(-1.5+0.5*sin(\x r)+3*cos(0.1*\x r)+2*sin(0.2*\x r)+cos(0.3*\x r)+0.1*\x)})  -- plot[domain=10.5:-0.5,smooth]  (\x ,{0.5+0.8*(0.1*\x+0.3*sin(\x r)+2*cos(0.1*\x r))}) -- cycle;
		\filldraw [cyan,opacity=0.5]  plot[domain=-0.5:10.5,smooth] (\x ,{0.5+0.8*(0.1*\x+0.3*sin(\x r)+2*cos(0.1*\x r))})  -- plot[domain=10.5:-0.5,smooth] (\x ,{2+0.6*(-0.16*\x+2-0.2*sin(\x r)-3*cos(0.1*\x r)-0.3*cos(0.3*\x r))}) -- cycle;
		\filldraw [cyan,opacity=0.7]  plot[domain=-0.5:10.5,smooth] (\x ,{2+0.6*(-0.16*\x+2-0.2*sin(\x r)-3*cos(0.1*\x r)-0.3*cos(0.3*\x r))})  -- plot[domain=10.5:-0.5,smooth] (\x ,{-1+0.7*(0.25*sin(\x r)+1.5*cos(0.1*\x r)+0.5*sin(0.3*\x r))}) -- cycle;
		\filldraw [cyan,opacity=0.9]  plot[domain=-0.5:10.5,smooth] (\x ,{-1+0.7*(0.25*sin(\x r)+1.5*cos(0.1*\x r)+0.5*sin(0.3*\x r))})  -- plot[domain=10.5:-0.5,smooth] (\x ,{-1.7+0.5*(0.06*\x+0.3*sin(\x r)+2*cos(0.1*\x r))}) -- cycle;
		\filldraw [brown,opacity=0.4]  plot[domain=-0.5:10.5,smooth] (\x ,{-1.7+0.5*(0.06*\x+0.3*sin(\x r)+2*cos(0.1*\x r))})--(10.5,-1.8)--(-0.5,-1.8)--cycle;
		\filldraw [brown,opacity=0.8] (-0.5,-1.8)--(10.5,-1.8)--(10.5,-3)--(-0.5,-3)--cycle;	\draw[very thick,->] (-0.02-0.5,4.5) --(11,4.5) node[right] {$r$};
		\draw[very thick,->] (0-0.5,4.5) --(0-0.5,-3.25) node[below] {$z$};		
		\draw[dashed,white](0.5,4.5)--(0.5,-1.8);
		\draw[dashed,white](0.5*3,4.5)--(0.5*3,-1.8);
		\draw[dashed,white](0.5*5,4.5)--(0.5*5,-1.8);
		\draw[dashed,white](0.5*7,4.5)--(0.5*7,-1.8);
		\draw[dashed,white](0.5*9,4.5)--(0.5*9,-1.8);
		\draw[dashed,white](0.5*11,4.5)--(0.5*11,-1.8);
		\draw[dashed,white](0.5*13,4.5)--(0.5*13,-1.8);
		\draw[dashed,white](0.5*15,4.5)--(0.5*15,-1.8);
		\draw[dashed,white](0.5*17,4.5)--(0.5*17,-1.8);
		\draw[dashed,white](0.5*19,4.5)--(0.5*19,-1.8);
		\draw[dashed,red](0-0.5,3.34)--(0.5,3.34);\draw[dashed,red](1.5,3.6)--(0.5,3.6);\draw[dashed,red](1.5,3.71)--(2.5,3.71);\draw[dashed,red](3.5,3.64)--(2.5,3.64);\draw[dashed,red](3.5,3.53)--(4.5,3.53);\draw[dashed,red](4.5,3.47)--(5.5,3.47);\draw[dashed,red](5.5,3.52)--(6.5,3.52);\draw[dashed,red](6.5,3.58)--(7.5,3.58);\draw[dashed,red](7.5,3.53)--(8.5,3.53);\draw[dashed,red](8.5,3.36)--(9.5,3.36);\draw[dashed,red](10.5,3.1)--(9.5,3.1);			
		\draw[dashed,red](0-0.5,2.1)--(0.5,2.1);\draw[dashed,red](1.5,2.37)--(0.5,2.37);\draw[dashed,red](1.5,2.44)--(2.5,2.44);\draw[dashed,red](3.5,2.31)--(2.5,2.31);\draw[dashed,red](3.5,2.12)--(4.5,2.12);\draw[dashed,red](4.5,2.093)--(5.5,2.093);\draw[dashed,red](5.5,2.25)--(6.5,2.25);\draw[dashed,red](6.5,2.43)--(7.5,2.43);\draw[dashed,red](7.5,2.475)--(8.5,2.475);\draw[dashed,red](8.5,2.32)--(9.5,2.32);\draw[dashed,red](10.5,2.04)--(9.5,2.04);			
		\draw[dashed,red](0-0.5,1.23)--(0.5,1.23);\draw[dashed,red](1.5,1.05)--(0.5,1.05);\draw[dashed,red](1.5,1.00)--(2.5,1.0);\draw[dashed,red](3.5,1.07)--(2.5,1.07);\draw[dashed,red](3.5,1.18)--(4.5,1.18);\draw[dashed,red](4.5,1.234)--(5.5,1.234);\draw[dashed,red](5.5,1.22)--(6.5,1.22);\draw[dashed,red](6.5,1.17)--(7.5,1.17);\draw[dashed,red](7.5,1.2)--(8.5,1.2);\draw[dashed,red](8.5,1.33)--(9.5,1.33);\draw[dashed,red](10.5,1.5)--(9.5,1.5);			
		\draw[dashed,red](0-0.5,0.05)--(0.5,0.05);\draw[dashed,red](1.5,0.294)--(0.5,0.294);\draw[dashed,red](1.5,0.365)--(2.5,0.365);\draw[dashed,red](3.5,0.3)--(2.5,0.3);\draw[dashed,red](3.5,0.17)--(4.5,0.17);\draw[dashed,red](4.5,0.126)--(5.5,0.126);\draw[dashed,red](5.5,0.152)--(6.5,0.152);\draw[dashed,red](6.5,0.186)--(7.5,0.186);\draw[dashed,red](7.5,0.14)--(8.5,0.14);\draw[dashed,red](8.5,-0.13)--(9.5,-0.13);\draw[dashed,red](10.5,-0.451)--(9.5,-0.451);			
		\draw[dashed,red](0-0.5,-0.714)--(0.5,-0.714);\draw[dashed,red](1.5,-0.56)--(0.5,-0.56);\draw[dashed,red](1.5,-0.54)--(2.5,-0.54);\draw[dashed,red](3.5,-0.63)--(2.5,-0.63);\draw[dashed,red](3.5,-0.76)--(4.5,-0.76);\draw[dashed,red](4.5,-0.792)--(5.5,-0.792);\draw[dashed,red](5.5,-0.74)--(6.5,-0.74);\draw[dashed,red](6.5,-0.631)--(7.5,-0.631);\draw[dashed,red](7.5,-0.635)--(8.5,-0.635);\draw[dashed,red](8.5,-0.75)--(9.5,-0.75);\draw[dashed,red](10.5,-0.93)--(9.5,-0.93);
		\filldraw [red] (0-0.5,1.5) circle [radius=2.5pt] node[right,black]{$\text{source}$}; 
		\node at (-0.6-0.5,1.5)[right]{$z_\mathrm{s}$};
		\node at (0-0.5,4.6)[left]{$0$};
		\node at (-0.6-0.5,3.2)[right]{$h_1$};
		\node at (-0.6-0.5,2)[right]{$h_2$};
		\node at (-0.6-0.5,0)[right]{$h_\ell$};
		\node at (-0.6-0.5,-1.7)[right]{$h_l$};
		\node at (4.2,3.9){Layer 1};
		\node at (4.2,2.7){Layer 2};
		\filldraw (4.2,1.85) circle [radius=1.5pt];
		\filldraw (4.2,1.65) circle [radius=1.5pt];
		\filldraw (4.2,1.45) circle [radius=1.5pt];
		\node at (4.2,0.6){Layer $\ell$};
		\filldraw (4.2,-0.15) circle [radius=1.5pt];
		\filldraw (4.2,-0.35) circle [radius=1.5pt];
		\filldraw (4.2,-0.55) circle [radius=1.5pt];
		\node at (4.2,-1.4){Layer $l$};
		\node at (8.2,-2.5){$c_{\infty},\rho_{\infty},\alpha_{\infty}$};
		\node at (4.2,-2.5){Acoustic half-space};				
		\node at (8,4.0){$c_1(r,z),\rho_1(r,z),\alpha_{1}(r,z)$};
		\node at (8,2.8){$c_2(r,z),\rho_2(r,z),\alpha_2(r,z)$};			
		\node at (8,0.7){$c_\ell(r,z),\rho_\ell(r,z),\alpha_\ell(r,z)$};
		\node at (8,-1.3){$c_l(r,z),\rho_l(r,z),\alpha_l(r,z)$};
	\end{tikzpicture}
\caption{Schematic diagram of the ocean environment with multilayer media.}
	\label{Figure1}
\end{figure}

For ocean environments containing multilayer sediments, $\rho(z)$ and $k(z)$ are usually discontinuous at the interfaces $z=\{h_\ell\}_{\ell=1}^{l-1}$. Considering an intermittent environment, the ocean is divided into $l$ discontinuous layers, as shown by the red dotted line in \autoref{Figure1}. The environmental parameters are separately defined in the columns as:
\linenomath
\begin{subequations}
		\label{eq:9}
		\begin{align}
			c(z) &= \begin{cases}
				c_\ell(z),&h_{\ell-1} \leq z \leq h_\ell,\quad \ell=1,\cdots,l\\
				c_\infty,&z\geq H
			\end{cases}\\
			\rho(z) &= \begin{cases}
				\rho_\ell(z),&h_{\ell-1} \leq z \leq h_\ell,\quad \ell=1,\cdots,l\\
				\rho_\infty,&z\geq H
			\end{cases}\\
			\alpha(z) &= \begin{cases}
				\alpha_\ell(z),&h_{\ell-1} \leq z \leq h_\ell,\quad \ell=1,\cdots,l\\
				\alpha_\infty,&z\geq H
			\end{cases}
		\end{align}
	\end{subequations}
where $h_0=0$ and $h_l=H$, respectively.

Boundary conditions should be imposed at the sea surface ($z=0$) and seabed ($z=H$), and interface conditions should be imposed at the discontinuous interfaces ($z=\{h_\ell\}_{\ell=1}^{l-1}$). Taking the pressure release boundary condition as an example, the upper boundary condition is:
\begin{equation}
		\label{eq:10}
		\psi(0)=0
	\end{equation}
The bottom boundary is either perfectly free or rigid:
\begin{subequations}
		\label{eq:11}
		\begin{gather}
			\psi(H)=0\\
			\psi'(H) = 0 
		\end{gather}
	\end{subequations}
In addition, the use of an acoustic half-space is common in underwater acoustic modeling \cite{Jensen2011}:
\begin{equation}
		\label{eq:12}
		\psi(H)+\frac{\rho_\infty}{\rho_b(H)\gamma_\infty} \psi'(H)=0,\quad \gamma_{\infty}=\sqrt{k_r^{2}-k_\infty^2},\quad k_\infty=(1+\mathrm{i}\eta\alpha_\infty)\omega/c_\infty
	\end{equation}
At the interfaces ($z=\{h_\ell\}_{\ell=1}^{l-1}$), both acoustic pressure and normal particle velocity must be continuous. Thus, two constraints on continuity are explicitly imposed by:
\begin{subequations}
		\label{eq:13}
		\begin{gather}
			\label{eq:13a}
			\psi(h_\ell^{-})=\psi(h_\ell^{+}) \\
			\label{eq:13b}
			\frac{1}{\rho(h_\ell^{-})} \frac{\mathrm{d}\psi(h_\ell^{-} )}{\mathrm {d}z}= \frac{1}{\rho(h_\ell^{+})}\frac{\mathrm{d}\psi(h_\ell^{+} )}{\mathrm {d}z}
		\end{gather}
	\end{subequations}

where the superscripts $-$ and $+$ of $h_\ell$ indicate the limits from above and below, respectively.

\subsection{Improved global matrix of coupled modes}
	\label{section2.2}
For range-dependent ocean environments, the classic technique is to divide the terrain into many sufficiently narrow segments \cite{Evans1983}, e.g., to resemble stair steps, as shown by the white dotted line in \autoref{Figure1}. Segments are treated as independent of range; after the eigenmodes and horizontal wavenumbers of each segment are obtained, the segment conditions of $J$ segments are used to couple the subfields of each segment to obtain the acoustic field of the entire waveguide.

Referring to the form of Eq.~\eqref{eq:8}, the acoustic field of the $j$-th segment can generally be represented as:
\linenomath
\begin{equation}
		\label{eq:14}
		p^{j}(r, z)\approx \sum_{m=1}^{M}\left[a_{m}^{j} H 1_{m}^{j}(r)+b_{m}^{j} H2_{m}^{j}(r)\right] \psi_{m}^{j}(z),\quad j = 1,2,\cdots,J
	\end{equation}
where $M$ is the total number of normal modes to synthesize the acoustic field, $\psi_{m}^{j}(z)$ is the $m$-th eigenmode of the $j$-th segment, and $\{a_{m}^{j}\}_{m=1}^M$ and $\{b_{m}^{j}\}_{m=1}^M$ are the coupling coefficients, which denote the amplitudes of the forward and backward propagating modes in the $j$-th segment, respectively. $H1_{m}^{j}(r)$ and $H2_{m}^{j}(r)$ are the ratios of the two types of Hankel functions and are defined as follows:
\begin{subequations}
		\label{eq:15}
		\begin{gather}
			\label{eq:15a}
			H1_{m}^{j}(r)=\frac{\mathcal{H}_{0}^{(1)}\left(k_{r,m}^{j} r\right)}{\mathcal{H}_{0}^{(1)}\left(k_{r,m}^{j} r_{j-1}\right)}\simeq \sqrt{\frac{r_{j-1}}{r}} \mathrm{e}^{\mathrm{i} k_{r,m}^{j}\left(r-r_{j-1}\right)}\\
			\label{eq:15b}
			H2_{m}^{j}(r)=\frac{\mathcal{H}_{0}^{(2)}\left(k_{r,m}^{j} r\right)}{\mathcal{H}_{0}^{(2)}\left(k_{r,m}^{j} r_{j}\right)}\simeq \sqrt{\frac{r_{j}}{r}} \mathrm{e}^{-\mathrm{i} k_{r,m}^{j}\left(r-r_{j}\right)}
		\end{gather}
	\end{subequations}
where $k_{r,m}^j$ is the horizontal wavenumber of the $m$-th mode in the $j$-th segment. For special cases, $j=1$ and $r_{j-1}=r_1$. The definition of $H1_{m}^{j}(r)$ here is identical to that in COUPLE \cite{Couple}, but the definition of $H2_{m}^{j}(r)$ is different. In COUPLE, $H2_{m}^{j}(r)$ is defined as:
\begin{equation}
		\label{eq:16}
		H2_{m}^{j}(r)=\frac{\mathcal{H}_{0}^{(2)}\left(k_{r,m}^{j} r\right)}{\mathcal{H}_{0}^{(2)}\left(k_{r,m}^{j} r_{j-1}\right)}\simeq \sqrt{\frac{r_{j}}{r}} \mathrm{e}^{-\mathrm{i} k_{r,m}^{j}\left(r-r_{j-1}\right)}
	\end{equation}
This improved definition \eqref{eq:15b} was proposed by Luo \cite{Luowy2012a,Luowy2012b,Luowy2012c,Luowy2012d} and Yang \cite{Yangcm2012,Yangcm2015a}. Leaky modes and long-range flat ocean environments may cause the value of $H2_{m}^{j}(r)$ defined in COUPLE to overflow. This is specifically the case for leaky mode $k_{r,m}^j=\mathcal{R}+\mathcal{I}\mathrm{i}$, where $\mathcal{R}$ and $\mathcal{I}$ denote the real and imaginary parts of $k_{r,m}^j$, respectively, and where $\mathcal{R}\geq 0, \mathcal{I}> 0$. In Eq.~\eqref{eq:16}, since $r-r_{j-1}>0$, then Eq.~\eqref{eq:16} contains $\exp[\mathcal{I}(r-r_{j-1})]$. When $\mathcal{I}$ or $r-r_{j-1}$ is large, using Eq.~\eqref{eq:16} may cause numerical overflow. In contrast, in Eq.~\eqref{eq:15b}, the exponential part contains $\exp[\mathcal{I}(r-r_j)]$, and because $r-r_j<0$, regardless of how large $\mathcal{I}$ is, the value of $H2_{m}^{j}(r)$ is limited, and no numerical overflow occurs. In other words, in this improved global matrix of coupled modes, the left boundary is used to normalize the forward acoustic field, and the right boundary is used to normalize the backward acoustic field, which ensures the numerical stability of the simulation. Therefore, a reasonable normalized range solution eliminates the numerical overflow that may occur in previous two-way models and is unconditionally stable.

The method of coupling segments explicitly imposes two segment continuity conditions on the sides of the segments. The first segment condition is that the acoustic pressure must be continuous at the $j$-th side, and the second is that the radial velocity is continuous at the $j$-th side:
\begin{subequations}
		\begin{gather}
			\label{eq:17a}
			p^{j+1}\left(r_{j}, z\right)=p^{j}\left(r_{j}, z\right) \\
			\label{eq:17b}
			\frac{1}{\rho_{j+1}(z)} \frac{\partial p^{j+1}\left(r_{j}, z\right)}{\partial r}=\frac{1}{\rho_{j}(z)} \frac{\partial p^{j}\left(r_{j}, z\right)}{\partial r}
		\end{gather}
	\end{subequations}
For the first segment condition, we have:
\begin{equation}
		\label{eq:18}
		\sum_{m=1}^{M}\left[a_{m}^{j+1}H 1_{m}^{j+1}(r_j)+b_{m}^{j+1}H 2_{m}^{j+1}(r_j) \right] \psi_{m}^{j+1}(z)=\sum_{m=1}^{M}\left[a_{m}^{j} H 1_{m}^{j}\left(r_{j}\right)+b_{m}^{j}H 2_{m}^{j}\left(r_{j}\right)\right] \psi_{m}^{j}(z)
	\end{equation}
where $H 1_{m}^{j+1}(r_j)=H 2_{m}^{j}(r_j)=1$. We apply the following operator to both sides of the above equation:
\begin{equation*}
		\int_0^H(\cdot) \frac{\psi_{n}^{j+1}(z)}{\rho_{j+1}(z)} \mathrm{d} z
	\end{equation*}
Then, we use the orthogonal normalization relationship Eq.~\eqref{eq:7} of the eigenmodes in the $(j+1)$-th segment. Accordingly, Eq.~\eqref{eq:18} is equivalent to:
\begin{subequations}
		\begin{gather}
			\label{eq:19a}
			a_{n}^{j+1}+b_{n}^{j+1}H 2_{n}^{j+1}\left(r_{j}\right) =\sum_{m=1}^{M}\left[a_{m}^{j} H 1_{m}^{j}\left(r_{j}\right)+b_{m}^{j}\right] \tilde{c}_{n m}, \quad n=1, \ldots, M\\
			\label{eq:19b}
			\tilde{c}_{n m}=\int_0^H \frac{\psi_{n}^{j+1}(z) \psi_{m}^{j}(z)}{\rho_{j+1}(z)} \mathrm{d} z
		\end{gather}
	\end{subequations}

The above formula can be easily written in the following matrix-vector form:
\begin{equation}
		\label{eq:20}
		\mathbf{a}^{j+1}+\mathbf{H}_{2}^{j+1}\mathbf{b}^{j+1}=\widetilde{\mathbf{C}}^{j}\left(\mathbf{H}_{1}^{j} \mathbf{a}^{j}+\mathbf{b}^{j}\right)  
	\end{equation}
To similarly address the segment condition \eqref{eq:17b}, we first write the derivative expression of $p$ with respect to $r$, which can be derived from Eq.~\eqref{eq:14}:
\begin{equation}
		\label{eq:21}
		\frac{1}{\rho_{j}} \frac{\partial p^{j}(r, z)}{\partial r} \simeq \frac{1}{\rho_{j}} \sum_{m=1}^{M} k_{r,m}^{j}\left[a_{m}^{j} H 1_{m}^{j}(r)-b_{m}^{j} H 2_{m}^{j}(r)\right] \psi_{m}^{j}(z)
	\end{equation}
Then, the second segment condition is equivalent to:
\begin{equation}
		\label{eq:22}
		\frac{1}{\rho_{j+1}}\sum_{m=1}^{M}k_{r,m}^{j+1}\left[a_{m}^{j+1} -b_{m}^{j+1}H 2_{m}^{j+1}\left(r_{j}\right)\right]\psi_{m}^{j+1}(z)=\frac{1}{\rho_{j}}\sum_{m=1}^{M}k_{r,m}^{j}\left[a_{m}^{j} H 1_{m}^{j}\left(r_{j}\right)-b_{m}^{j}\right] \psi_{m}^{j}(z)
	\end{equation}
Similarly, we apply the following operator to the above equation:
\begin{equation*}
		\int_0^H(\cdot) \psi_{n}^{j+1}(z) \mathrm{d} z
	\end{equation*}
Next, we utilize the orthogonal normalization relationship Eq.~\eqref{eq:7} of the eigenmodes in the $(j+1)$-th segment to obtain:
\begin{subequations}
		\begin{gather}
			\label{eq:23a}
			a_{n}^{j+1}-b_{n}^{j+1}H 2_{n}^{j+1} =\sum_{m=1}^{M}\left[a_{m}^{j} H 1_{m}^{j}\left(r_{j}\right)-b_{m}^{j}\left(r_{j}\right)\right] \hat{c}_{n m}, \quad n=1, \ldots, M\\
			\label{eq:23b}
			\hat{c}_{n m}=\frac{k_{r,m}^{j}}{k_{r,n}^{j+1}} \int \frac{\psi_{n}^{j+1}(z) \psi_{m}^{j}(z)}{\rho_{j}(z)} \mathrm{d} z
		\end{gather}
	\end{subequations}

The above formula can be easily written in the following matrix-vector form:
\begin{equation}
		\label{eq:24}
		\mathbf{a}^{j+1}-\mathbf{H}_{2}^{j+1}\mathbf{b}^{j+1}=\widehat{\mathbf{C}}^{j}\left(\mathbf{H}_{1}^{j} \mathbf{a}^{j}-\mathbf{b}^{j}\right)
	\end{equation}
Eqs.~\eqref{eq:20} and \eqref{eq:24} can be combined into the following form:
\begin{subequations}
		\label{eq:25}
		\begin{align}
			\label{eq:25a}
			\left[\begin{array}{l}
				\mathbf{a}^{j+1} \\
				\mathbf{b}^{j+1}
			\end{array}\right]&=\left[\begin{array}{ll}
				\mathbf{R}_{1}^{j} & \mathbf{R}_{2}^{j} \\
				\mathbf{R}_{3}^{j} & \mathbf{R}_{4}^{j}
			\end{array}\right]\left[\begin{array}{c}
				\mathbf{a}^{j} \\
				\mathbf{b}^{j}
			\end{array}\right]\\			
			\mathbf{R}_{1}^{j} &=\frac{1}{2}\left(\widetilde{\mathbf{C}}^{j}+\widehat{\mathbf{C}}^{j}\right) \mathbf{H}_{1}^{j} \\
			\mathbf{R}_{2}^{j} &=\frac{1}{2}\left(\widetilde{\mathbf{C}}^{j}-\widehat{\mathbf{C}}^{j}\right) \\
			\mathbf{R}_{3}^{j} &=\frac{1}{2}\left(\mathbf{H}_{2}^{j+1}\right)^{-1}\left(\widetilde{\mathbf{C}}^{j}-\widehat{\mathbf{C}}^{j}\right) \mathbf{H}_{1}^{j} \\
			\mathbf{R}_{4}^{j} &=\frac{1}{2}\left(\mathbf{H}_{2}^{j+1}\right)^{-1}\left(\widetilde{\mathbf{C}}^{j}+\widehat{\mathbf{C}}^{j}\right)
		\end{align}
	\end{subequations}
Finally, the segment condition and radiation condition should be imposed at the acoustic source $r=0$ and $r\rightarrow \infty$. The segment condition at the acoustic source $r=0$ is:
\begin{equation}
		\label{eq:26}
		a_{m}^{1}=\frac{\mathrm{i}}{4 \rho_{1}\left(z_\mathrm{s}\right)} \psi_{m}^{1}\left(z_\mathrm{s}\right) \mathcal{H}_{0}^{(1)}\left(k_{r,m}^{1} r_{1}\right)+b_{m}^{1} \frac{\mathcal{H}_{0}^{(1)}\left(k_{r,m}^{1} r_{1}\right)}{\mathcal{H}_{0}^{(2)}\left(k_{r,m}^{1} r_{1}\right)}, \quad m=1, \ldots, M
	\end{equation}
This condition can be written in a matrix-vector form:
\begin{subequations}
		\label{eq:27}
		\begin{gather}
			\mathbf{a}^{1}-\mathbf{D}\mathbf{b}^{1} =\mathbf{s}\\
			D_{mm}=\frac{\mathcal{H}_{0}^{(1)}\left(k_{r,m}^{1} r_{1}\right)}{\mathcal{H}_{0}^{(2)}\left(k_{r,m}^{1} r_{1}\right)},\quad s_{m}=\frac{\mathrm{i}}{4 \rho_{1}\left(z_\mathrm{s}\right)} \psi_{m}^{1} \left(z_\mathrm{s}\right) \mathcal{H}_{0}^{(1)}\left(k_{r,m}^{1} r_{1}\right)
		\end{gather}    
	\end{subequations}
For the radiation condition at $r\rightarrow\infty$, $\mathbf{b}^J=\mathbf{0}$ is sufficient.

Combining the continuity conditions at the boundaries of the $J$ segments with the boundary condition at the acoustic source and the radiation condition at infinity, the following system of linear algebraic equations is obtained:
\begin{equation}
		\label{eq:28}
		\left[\begin{array}{cccccccc}
			\mathbf{E} & -\mathbf{D} & \mathbf{0} & & & & & \\
			\mathbf{R}_{1}^{1} & \mathbf{R}_{2}^{1} & -\mathbf{E} & \mathbf{0} & & & & \\
			\mathbf{R}_{3}^{1} & \mathbf{R}_{4}^{1} & \mathbf{0} & -\mathbf{E} & & & & \\
			& \ddots & \ddots & \ddots & \ddots & & & \\
			& & & \mathbf{R}_{1}^{J-2} & \mathbf{R}_{2}^{J-2} & -\mathbf{E} & \mathbf{0} & \\
			& & & \mathbf{R}_{3}^{J-2} & \mathbf{R}_{4}^{J-2} & \mathbf{0} & -\mathbf{E} & \\
			& & & & & \mathbf{R}_{1}^{J-1} & \mathbf{R}_{2}^{J-1} & -\mathbf{E} \\
			& & & & & \mathbf{R}_{3}^{J-1} & \mathbf{R}_{4}^{J-1} & \mathbf{0}
		\end{array}\right]\left[\begin{array}{c}
			\mathbf{a}^{1} \\
			\mathbf{b}^{1} \\
			\mathbf{a}^{2} \\
			\vdots \\
			\mathbf{b}^{J-2} \\
			\mathbf{a}^{J-1} \\
			\mathbf{b}^{J-1} \\
			\mathbf{a}^{J}
		\end{array}\right]=\left[\begin{array}{c}
			\mathbf{s} \\
			\mathbf{0} \\
			\mathbf{0} \\
			\vdots \\
			\mathbf{0} \\
			\mathbf{0} \\
			\mathbf{0} \\
			\mathbf{0}
		\end{array}\right]
	\end{equation}
where $\mathbf{E}$ denotes the identity matrix. This system of linear algebraic equations can be solved to obtain the coupling coefficients $\left(\{\mathbf{a}^j\}_{j=1}^J,\{\mathbf{b}^j\}_{j=1}^J\right)$; then, Eq.~\eqref{eq:14} is used to synthesize the acoustic pressure field.

Since $r_{j-1}=r_1$ when $j=1$ is defined above, in the first segment, $H1_m^j(r)$ is normalized to the right side. When $\exp[\mathcal{I}(r_1-r)]$ is large, calculating $H1_m^j(r)$ may cause numerical instability. To avoid this problem, the superposition principle is used to solve for the first subfield. Substituting $\mathbf{a}^1$ in Eq.~\eqref{eq:27} into Eq.~\eqref{eq:14} reveals:
\begin{equation}
		\label{eq:29}
		p^{1}(r, z)\approx\frac{\mathrm{i}}{4 \rho\left(z_\mathrm{s}\right)} \sum_{m=1}^{M} \psi_{m}^{1}\left(z_\mathrm{s}\right) \psi_{m}^{1}(z) \mathcal{H}_{0}^{(1)}\left(k_{r,m}^{1} r\right) + 2 \sum_{m=1}^{M} b_{m}^{1} \frac{\mathcal{J}_{0}\left(k_{r,m}^{1} r\right)}{\mathcal{H}_{0}^{(2)}\left(k_{r,m}^{1} r_{1}\right)} \psi_{m}^{1}(z)
	\end{equation}
where $\mathcal{J}_{0}(\cdot)$ is the Bessel function, the first term on the right side represents the range-independent acoustic field, and the second term represents the scattered acoustic field caused by range dependency \cite{Luowy2012a}.

\section{Methodology and Algorithm}
\subsection{Chebyshev--Tau spectral method}
The classic spectral method is the Galerkin-type spectral method, which is derived from the Galerkin method of the weighted residual method. A special feature of the Galerkin-type spectral method is that the basis/weight functions are selected as the same set of orthogonal polynomials. Since the classic Galerkin-type spectral method requires the basis function to satisfy the boundary conditions (generally a linear combination of orthogonal polynomials of a certain kind), it is not easy to apply to differential equations with complex boundary conditions. To resolve this problem, Lanczos proposed the Tau method in 1938 \cite{Lanczos1938}. This method also uses the same set of orthogonal polynomials as the basis/weight functions but does not require the basis function to satisfy the boundary conditions and imposes boundary constraints on only the coefficients of the spectral expansion. In other words, the spectral coefficients are forced to satisfy the boundary conditions in the spectral space. The Chebyshev--Tau spectral method is a type of spectral method that uses Chebyshev polynomials as the basis/weight functions. In our previous research \cite{Tuhw2020a,Tuhw2021a}, we concisely introduced the Chebyshev--Tau spectral method and its application to normal modes of range-independent two-layer media (water column and bottom sediment). We developed the related NM-CT program, which is included in the open-source code and available in the Ocean Acoustics Library (OALIB) \cite{NM-CT}. Similarly, for range-independent segments containing multiple layers of media, the Chebyshev--Tau spectral method can still solve for the horizontal wavenumbers and eigenmodes of the modal equation (Eq.~\eqref{eq:6}). In addition, for the acoustic half-space boundary condition, an eigenvalue transformation technique, not just an absorbing layer technique, is adopted.

When the Chebyshev--Tau spectral method is used to solve the modal equation, the modal equation should be scaled to the domain of the Chebyshev polynomials $\{T_i(x)\}$:
\begin{equation}
		\frac{4}{|\Delta h|^2}\rho(x)\frac{\mathrm{d}}{\mathrm{d}x}\left(\frac{1}{\rho(x)}\frac{\mathrm{d}\psi(x)}{\mathrm {d}x}\right) +k^2\psi(x) = k_r^2 \psi(x),
		\quad
		x \in [-1, 1]
	\end{equation}
Moreover, the modal function $\psi(x)$ must be transformed into the spectral space formed by the Chebyshev orthogonal polynomials $\{T_i(x)\}_{i=0}^N$:
\linenomath
\begin{equation}
		\label{eq:31}
		\psi(x) \approx \sum_{i=0}^{N}\hat{\psi}_{i}T_i(x)
	\end{equation}
where $\{\hat{\psi}_i\}_{i=0}^N$ are the spectral coefficients of $\psi(x)$ and $N$ denotes the spectral truncated order. Due to the good properties of Chebyshev polynomial/basis functions, the following relations are easily derived \cite{Boyd2001,Canuto2006}:
\begin{subequations}
		\label{eq:32}
		\begin{gather}
			\label{eq:32a}		
			\hat{\psi}'_i \approx \frac{2}{c_i}
			\sum_{\substack{j=i+1,\\ 
					j+i=\text{odd}
			}}^{N} j \hat{\psi}_j, \quad c_0=2,c_{i>1}=1
			\Longleftrightarrow \bm{\hat{\Psi}}' \approx \mathbf{D}_N \bm{\hat{\Psi}} \\
			\label{eq:32b}
			\widehat{(v\psi)}_i \approx 
			\frac{1}{2} \sum_{m+n=i}^{N} \hat{\psi}_m\hat{v}_n +
			\frac{1}{2} \sum_{|m-n|=i}^{N} \hat{\psi}_m\hat{v}_n  \Longleftrightarrow  \widehat{\bm{(v\psi)}} \approx \mathbf{C}_v \bm{\hat{\Psi}}\\
			\label{eq:32c}
			\int_{-1}^{1}\psi(x) \mathrm{d} x = -2\sum_{
				\substack{n=0,\\
					n=\text{even}}
			}^{\infty}\frac{\hat \psi_n}{n^2-1}\approx-2\sum_{
				\substack{n=0,\\
					n=\text{even}}
			}^{N}\frac{\hat \psi_n}{n^2-1}=\mathbf{I}_N\bm{\hat{\Psi}}
		\end{gather}
	\end{subequations}
Eq.~\eqref{eq:32a} denotes the relationship between the spectral coefficients of a function and those of its derivative function. Similarly, Eq.~\eqref{eq:32b} describes the relationship between the spectral coefficients of a product of two functions and the spectral coefficients of one of the functions. \eqref{eq:32c} shows the relationship between the integral of a function and its spectral coefficients. The right-hand side of Eq.~\eqref{eq:32} contains the matrix-vector representations of the relationships.

Solving differential equations using the Chebyshev--Tau method starts with the variational form of the differential equation, namely:
\begin{equation}
		\label{eq:33}
		\begin{gathered}
			\int_{-1}^{1}\left[\frac{4}{|\Delta h|^2}\rho(x)\frac{\mathrm{d}}{\mathrm{d}x}\left(\frac{1}{\rho(x)}\frac{\mathrm{d}\psi(x)}{\mathrm {d}x}\right) +k^2\psi(x)-k_r^2 \psi(x) \right]\frac{T_i(x)}{\sqrt{1-x^2}}\mathrm{d}x=0\\
			x\in (-1,1), \quad i=0,1,\dots,N-2
		\end{gathered}
	\end{equation}
By substituting Eq.~\eqref{eq:31} into Eq.~\eqref{eq:33} and considering Eq.~\eqref{eq:32}, the modal equation can be directly discretized into the following matrix-vector form:
\begin{equation}
		\label{eq:34}
		\left(\frac{4}{|\Delta h|^2}\mathbf{C}_{\rho}\mathbf{D}_{N}\mathbf{C}_{1/\rho}\mathbf{D}_{N}+\mathbf{C}_{k^2}\right)
		\bm{\hat{\Psi}}
		= k_r^2 \bm{\hat{\Psi}}
	\end{equation}
where $\bm{\hat{\Psi}}$ is the column vector consisting of $\{\hat{\psi}_i\}_{i=0}^N$. For details regarding the discretization process, please see Eq.~(29) in reference \cite{Tuhw2021a}.

From a formal viewpoint, this is an ordinary matrix eigenvalue problem, and boundary constraints must be added to the actual solution. For the ocean acoustic waveguide in Eqs.~\eqref{eq:9} through \eqref{eq:13}, the modal equation Eq. \eqref{eq:6} must be established in $l$ layers. As shown in \autoref{Figure1}, in a range-independent segment, a single set of basis functions cannot span $l$ layers since the normal derivative of sound pressure is not continuously differentiable at the interfaces $\{h_\ell\}_{\ell=1}^{l-1}$. Thus, we use the domain decomposition strategy \cite{Min2005} in Eq.~\eqref{eq:6} and split the domain interval into $l$ subintervals. For every splitting event, the discontinuous point is the endpoint of one subinterval:
\begin{equation}
		\label{eq:35}
		\psi_\ell(z) = \psi_\ell(x) \approx\sum_{i=0}^{N_\ell}\hat{\psi}_{\ell,i}T_i(x_\ell),\quad
			x_\ell=-\frac{2}{h_\ell-h_{\ell-1}}z_\ell+\frac{h_\ell+h_{\ell-1}}{h_\ell-h_{\ell-1}},\quad h_{\ell-1}\leq z\leq h_\ell
	\end{equation}
where $N_\ell$ and $\{\hat{\psi}_{\ell,i}\}_{i=0}^{N_\ell}$ are the spectral truncated order and modal spectral coefficients in the $\ell$-th layer, respectively. Similar to Eq.~\eqref{eq:34}, the modal equation in the $\ell$-th layer can be directly discretized into the matrix-vector form:
\begin{equation}
			\label{eq:36}
			\mathbf{A}_\ell\bm{\hat{\Psi}}_\ell
			= k_r^2\bm{\hat{\Psi}}_\ell,\quad
			\mathbf{A}_\ell=\frac{4}{(h_\ell-h_{\ell-1})^2}\mathbf{C}_{\rho_\ell}\mathbf{D}_{N_\ell}\mathbf{C}_{1/\rho_\ell}\mathbf{D}_{N_\ell}+\mathbf{C}_{k_\ell^2}
	\end{equation}
where $\mathbf{A}_\ell$ is a square matrix of order $(N_\ell+1)$ and $\bm{\hat{\Psi}}_\ell$ is a column vector composed of $\{\hat{\psi}_{\ell,i}\}_{i=0}^{N_\ell}$.

Since the interface conditions are related to both the $(\ell-1)$-th and $\ell$-th layers, Eq.~\eqref{eq:36} of the $l$ layers should be simultaneously solved as follows:
\begin{equation}
		\label{eq:37}
		\left[\begin{array}{cccc}
			\mathbf{A}_1&\mathbf{0}&\mathbf{0}&\mathbf{0}\\
			\mathbf{0}&\mathbf{A}_2&\mathbf{0}&\mathbf{0}\\
			\mathbf{0}&\mathbf{0}&\ddots&\mathbf{0}\\
			\mathbf{0}&\mathbf{0}&\mathbf{0}&\mathbf{A}_l\\
		\end{array}\right]
		\left[\begin{array}{c}
			\bm{\hat{\Psi}}_1\\
			\bm{\hat{\Psi}}_2\\
			\vdots\\
			\bm{\hat{\Psi}}_l\\
		\end{array}
		\right]=k_r^2\left[\begin{array}{c}
			\bm{\hat{\Psi}}_1\\
			\bm{\hat{\Psi}}_2\\
			\vdots\\
			\bm{\hat{\Psi}}_l\\
		\end{array}
		\right]
	\end{equation}

The boundary conditions and interface conditions in Eqs.~\eqref{eq:10}--\eqref{eq:13} must also be expanded into the Chebyshev spectral space and expressed as row vectors. Let the $N=\sum_{\ell=1}^l(N_\ell+1)$-order square matrix on the left side of Eq.~\eqref{eq:37} be $\mathbf{L}$, and replace the last two rows of the first $(l-1)$ subblocks in the $\mathbf{L}$ matrix with the two interface conditions between the upper and lower layers; the last two rows of the last subblock are replaced with boundary conditions at the sea surface and floor, and the right-hand side of Eq.~\eqref{eq:37} is replaced accordingly. By rearranging the modified rows together by elementary row transformation, Eq.~\eqref{eq:37} can be rewritten into the form of the following block matrix:
\begin{equation}
		\label{eq:38}
		\left[
		\begin{array}{cc}
			\mathbf{L}_{11}&\mathbf{L}_{12}\\
			\mathbf{L}_{21}&\mathbf{L}_{22}\\
		\end{array}
		\right]\left[
		\begin{array}{c}
			\bm{\hat{\Psi}}^+\\
			\bm{\hat{\Psi}}^-
		\end{array}
		\right]=k_r^2\left[
		\begin{array}{c}
			\bm{\hat{\Psi}}^+\\
			\mathbf{0}
		\end{array}\right]
	\end{equation}
where $\mathbf{L}_{11}$ is a square matrix of order $\sum_{\ell=1}^l(N_\ell-1)$, $\mathbf{L}_{22}$ is a square matrix of order $2l$, $\mathbf{\hat{\Psi}}^+=[\hat{\psi}_{1,0},\hat{\psi}_{1,1},\cdots,\hat{\psi}_{1,N_1-2},\hat{\psi}_{2,0},\hat{\psi}_{2,1},\cdots,\hat{\psi}_{2,N_2-2},\cdots,\hat{\psi}_{l,0},\hat{\psi}_{l,1},\cdots,\hat{\psi}_{l,N_l-2}]^\mathrm{T}$ and $\mathbf{\hat{\Psi}}^-=[\hat{\psi}_{1,N_1-1},\hat{\psi}_{1,N_1},\hat{\psi}_{2,N_2-1},\hat{\psi}_{2,N_2},\cdots,\psi_{l,N_l-1},\hat{\psi}_{l,N_l}]^\mathrm{T}$. Solving this mixed linear eigensystem can yield the horizontal wavenumbers and spectral coefficients of the eigenmodes $(k_r,\bm{\hat{\Psi}}^+,\bm{\hat{\Psi}}^-)$. According to Eq.~\eqref{eq:37}, the subeigenmodes $\psi_\ell(x)$ of the $l$ layers are synthesized separately from the spectral coefficients $\{\bm{\hat{\Psi}}_\ell\}_{\ell=1}^l$, and then the complete modes $\psi(z)$ are obtained by splicing the submodes in the $l$ layers. Note that $\psi(z)$ obtained at this time is a discrete function value whose resolution depends on the physical spatial resolution of the Chebyshev inverse transform. In addition, for details on the treatment of the boundary conditions in Eqs.~\eqref{eq:10}, \eqref{eq:11} and \eqref{eq:13}, please see Eq.~(38) in reference \cite{Tuhw2021a}.

We emphasize that for the acoustic half-space boundary condition in Eq.~\eqref{eq:12}, since $\gamma_\infty$ contains the eigenvalue $k_r$ to be determined, Eq.~\eqref{eq:38} is no longer a general matrix eigenvalue problem and can be solved iteratively only by a root-finding algorithm. The greatest shortcoming of root-finding algorithms is that they must make a reasonable initial guess about the eigenvalue $k_r$ being sought \cite{Sabatini2019}. Since the prior estimate of $k_r$ is usually not available, many of the existing numerical programs following similar principles fail to converge to a specific root in some cases. To avoid the same problem when using the Chebyshev--Tau spectral method to solve for waveguides with an acoustic half-space, we consider an alternative approach: using $k_{z,\infty}=\sqrt{k_\infty^{2}-k_{r}^{2}}$ to transform the modal equation and Eq.~\eqref{eq:12} as follows \cite{Sabatini2019}:
\begin{subequations}
		\label{eq:39}
		\begin{gather}
			\label{eq:39a}
			\rho(z) \frac{\mathrm{d}}{\mathrm{d} z}\left(\frac{1}{\rho(z)} \frac{\mathrm{d} \psi}{\mathrm{d}z}\right)+\left(k^{2}(z)-k_{\infty}^{2}+k_{z,\infty}^{2}\right) \psi=0 \\
			\label{eq:39b}		
			\frac{\mathrm{i}\rho_\infty}{\rho_b(H)}\left.\frac{\mathrm{d} \psi(z)}{\mathrm{d} z}\right|_{z=H}+k_{z,\infty} \psi(H)=0
		\end{gather}
	\end{subequations}
For the acoustic half-space boundary, modal normalization should add the integral of $z\in[H,+\infty]$:
\begin{equation}
		\label{eq:40}
		\int_{0}^{H} \frac{\psi_{m}^{2}(z)}{\rho(z)} \mathrm{d} z +\int_{H}^{\infty} \frac{\psi_{m}^{2}(z)}{\rho(z)} \mathrm{d} z=\int_{0}^{H} \frac{\psi_{m}^{2}(z)}{\rho(z)} \mathrm{d} z +\frac{\psi_{m}^{2}(H)}{2 \rho_\infty\gamma_{\infty}} =1,\quad m=1,2,\dots
	\end{equation}

Eq.~\eqref{eq:39a} can naturally be discretized into the following form:
\begin{equation}
		\label{eq:41}
		\left[\mathbf{U}+k_{z,\infty}^{2} \mathbf{E}\right] \bm{\hat{\Psi}}=\mathbf{0},\quad \mathbf{U}=\mathbf{L}-k_{\infty}^{2}\mathbf{E},\quad \bm{\hat{\Psi}}=\left[\bm{\hat{\Psi}}_1,\bm{\hat{\Psi}}_2,\cdots,\bm{\hat{\Psi}}_l\right]^\mathrm{T}
	\end{equation}
Due to the addition of Eq.~\eqref{eq:39b} including $k_{z,\infty}$, Eq.~\eqref{eq:41} finally takes the following form:
\begin{equation}
		\label{eq:42}
		\left[\mathbf{U}+k_{z,\infty} \mathbf{V}+k_{z,\infty}^{2} \mathbf{W}\right] \bm{\hat{\Psi}}=\mathbf{0}
	\end{equation}
$\mathbf{U}$ in Eq.~\eqref{eq:42} is not exactly identical to that in Eq.~\eqref{eq:41}, as it has been modified by boundary conditions and interface conditions; nevertheless, we maintain the parameter name. $\mathbf{V}$ is a zero matrix of order $N$ with only the last row corresponding to the boundary condition in Eq.~\eqref{eq:39b}, and $\mathbf{W}$ is simply the identity matrix that has been changed by modifying the boundary conditions. This polynomial eigenvalue problem can be efficiently solved by the $\mathcal{QZ}$ algorithm; it can be transformed into a general matrix eigenvalue problem using the following formula:
\begin{subequations}
		\label{eq:43}
		\begin{gather}
			\tilde{\mathbf{U}} \tilde{\bm{\Psi}}=k_{z,\infty} \tilde{\mathbf{V}} \tilde{\bm{\Psi}}, \\
			\tilde{\mathbf{U}}=\left[\begin{array}{cc}
				-\mathbf{V} & -\mathbf{U} \\
				\mathbf{E} & 0
			\end{array}\right], \quad \tilde{\mathbf{V}}=\left[\begin{array}{cc}
				\mathbf{W} & 0 \\
				0 & \mathbf{E}
			\end{array}\right], \quad \tilde{\bm{\Psi}}=\left[\begin{array}{c}
				k_{z,\infty} \bm{\hat{\Psi}} \\
				\bm{\hat{\Psi}}
			\end{array}\right]
		\end{gather}
	\end{subequations}

It is necessary to take the inverse transform of the eigenvectors $\bm{\hat{\Psi}}_\ell$ to $[h_{\ell-1},h_\ell]$. The vectors $\{\bm{\Psi}_\ell\}_{\ell=1}^l$ are stacked into a single-column vector to form discrete $\psi(z)$; then, Eq.~\eqref{eq:40} is used to normalize $\bm{\Psi}$. After computing $k_{z,\infty}$, those elements with arguments in the interval $(-\pi/2,\pi/2]$ are selected, and the corresponding horizontal wavenumbers $k_{r,m}$ can be obtained by $k_r =\sqrt{ k_\infty^2-k_{z,\infty}^2}$. Finally, a set of eigenmodes $(k_r,\psi(z))$ is obtained.

The new formulation Eq.~\eqref{eq:39} of the modal equation Eq.~\eqref{eq:6} circumvents root-finding algorithms and does not require an initial guess for $k_{z,\infty}$, which is the most important advantage of this approach. Compared with the absorbing layer technique in COUPLE and the multilayer Legendre collocation method (MultiLC) \cite{Tuhw2021c,MultiLC}, this eigenvalue transformation can obtain more accurate horizontal wavenumbers and eigenmodes, but the increase in computational cost is also significant. As shown in Eq.~\eqref{eq:43}, the sizes of the matrices are doubled. Note that since the above algorithm can calculate waveguides in multilayer media, users can of course add a layer of medium as an absorbing layer to simulate an acoustic half-space, analogous to the COUPLE. We emphasize that the spectral coefficients of the eigenmodes obtained from the $J$ range-independent segments must be transformed using the same resolution in the vertical direction. Otherwise, the numerical quadrature of $\tilde{c}_{n m}$ in Eq.~\eqref{eq:19b} and $\hat{c}_{n m}$ in Eq.~\eqref{eq:23b} cannot be calculated.

\subsection{Numerical algorithm}
Summarizing the above derivation, we provide a complete description of the algorithm below:
\begin{enumerate}
\item
The environmental data are set up.

The data include the frequency $f$ and depth $z_\mathrm{s}$ of the sound source, total depth of the ocean $H$, topography of the seabed, number of acoustic profiles, and specific information of each group of acoustic profiles. In addition, the data should include the spectral truncated order ($\{N_\ell\}_{\ell=1}^l$), horizontal and vertical resolutions, number of coupled modes $M$, and type of bottom boundary condition. If the bottom is an acoustic half-space, the speed $c_\infty$, density $\rho_\infty$ and attenuation $\alpha_\infty$ in the half-space should also be specified.

\item
The ocean environment is segmented based on the seabed topography and sound speed profiles.

Jensen \cite{Jensen1998} established stair-step discretization criteria to accurately represent smoothly varying bathymetry in numerical models. A strict segmentation criterion is $\Delta r \le \lambda/4$, where $\lambda=\min(\{c_\ell(r,z)\}_{\ell=1}^l)/f$. Thus, we suppose that the entire waveguide is divided into $J$ segments.

\item
The Chebyshev--Tau spectral method is applied to form the mixed linear systems and solve for the horizontal wavenumbers and eigenmodes $\{k_{r,m}^j,\psi_m^j(z)\}_{m=1}^M$ of the $J$ flat segments.

The modal spectral coefficients $\bm{\hat{\Psi}}^j$ obtained for the $J$ segments should be transformed to a uniform vertical resolution. This process can be computed in parallel because the range-independent segments are irrelevant.

\item
The coupling submatrices $\{\mathbf{R}_{1}^{j}\}_{j=1}^{J-1}$, $\{\mathbf{R}_{2}^{j}\}_{j=1}^{J-1}$, $\{\mathbf{R}_{3}^{j}\}_{j=1}^{J-1}$, and $\{\mathbf{R}_{4}^{j}\}_{j=1}^{J-1}$ are calculated according to Eqs. \eqref{eq:15}, \eqref{eq:19b}, \eqref{eq:23b} and \eqref{eq:25}. This step is also naturally conducted in parallel.

\item
$\mathbf{D}$ and $\mathbf{s}$ are calculated using the boundary conditions, the global matrix is constructed according to Eq.~\eqref{eq:28}, and Eq.~\eqref{eq:28} is solved to obtain the coupling coefficients $\left(\{\mathbf{a}^j\}_{j=1}^J,\{\mathbf{b}^j\}_{j=1}^J\right)$ of $J$ segments. The global matrix is a band matrix of order $(2J-1)\times M$, and its bandwidth is $(3M-1)$. The inverse of a band matrix can be efficiently obtained using mature numerical algorithms and libraries.
\item
The sound field is synthesized.

The sound pressure field of each segment is calculated according to Eq.~\eqref{eq:14}, and the sound field of the first segment is corrected according to Eq.~\eqref{eq:29}. The sound fields of $J$ segments are individually embedded into the entire waveguide to obtain the final sound pressure field.
\end{enumerate}

\section{Numerical Simulation}
To validate the accuracy and performance of the numerical algorithm in solving range-dependent waveguide problems, the following tests and analyses are performed through six numerical experiments. In this article, the program developed based on the above numerical algorithm is named SPEC. We take the widely used KRAKEN program based on the finite difference method \cite{Kraken2001}, the COUPLE program based on the Galerkin method \cite{Couple}, and the RAM/RAMGeo programs \cite{RAM} based on the parabolic approximation as comparisons. The above codes are implemented in the FORTRAN language. In addition, the sound fields calculated by the commercial software COMSOL based on the finite element method are also used for comparison.

To present the acoustic field results, the transmission loss (TL) of the acoustic pressure is defined as $\text{TL}=-20\log_{10}(|p|/|p_0|)$ in units of decibels (dB), where $p_0=\exp(\mathrm{i}k_0)/(4 \pi)$ is the acoustic pressure at a range of 1 m from the point source. In actual displays, TL fields are often used to compare and analyze sound fields \cite{Jensen2011}.

\subsection{Slope terrain}
\begin{figure}[htbp]
		\centering
\subfigure[]{\begin{tikzpicture}[node distance=2cm,scale=0.8]
			\tikzstyle{every node}=[font=\footnotesize];
			\node at (1.5,0){0 $\text{m}$};
			\node at (1.4,-4.8){80 $\text{m}$};
			\node at (1.4,-6){100 $\text{m}$};
			\node at (2.2,0.4){0 $\text{m}$};
			\node at (14.3,0.4){2500 $\text{m}$};
			\node at (14.6,-1.8){30 $\text{m}$};
			\fill[cyan,opacity=0.7] (14,0)--(14,-1.8)--(11.2,-1.8)--(4.8,-4.8)--(2,-4.8)--(2,0)--cycle;
			\fill[brown,opacity=0.5] (2,-4.8)--(4.8,-4.8)--(11.2,-1.8)--(14,-1.8)--(14,-6)--(2,-6)--cycle;
			\draw[very thick, ->](2,0)--(14.5,0) node[right]{$r$};
			\draw[very thick, ->](2.02,0.1)--(2.02,-6.5) node[below]{$z$};
			\filldraw [red] (2.02,-1.5) circle [radius=2.5pt];
			\node[text=white] at (4,-1.5){$f=50 \mathrm{Hz}, z_\mathrm{s}=26 \mathrm{m}$};
			\draw[very thick](2,-4.8)--(4.8,-4.8);
			\draw[very thick](4.8,-4.8)--(11.2,-1.8);
			\draw[very thick](11.2,-1.8)--(14,-1.8);
			\draw[very thick](14.02,0.1)--(14.02,0);
			\draw[very thick](2.02,-6)--(14.02,-6);
			\draw[very thick](4.82,-5.9)--(4.82,-6);
			\draw[very thick](11.22,-5.9)--(11.22,-6);
			\node at (4.8,-6.4){500 $\text{m}$};
			\node at (11.2,-6.4){2000 $\text{m}$};
			\node at (8,-0.7){$c_1 = 1500 \mathrm{m/s}$};
			\node at (8,-1.2){$\rho_1 = 1.0 \mathrm{g/cm^3}$};
			\node at (8,-1.7){$\alpha_1 = 0.0 \mathrm{dB/\lambda}$};
			\node at (8,-4.1){$c_2 = 1800 \mathrm{m/s}$};
			\node at (8,-4.6){$\rho_2 = 1.5 \mathrm{g/cm^3}$};
			\node at (8,-5.1){$\alpha_2 = 2.0 \mathrm{dB/\lambda}$};
			\node at (8,-5.7){perfectly free boundary $p=0$};
		\end{tikzpicture}}
\subfigure[]{\begin{tikzpicture}[node distance=2cm,scale=0.8]
				\tikzstyle{every node}=[font=\footnotesize];
				\node at (1.5,0){0 $\text{m}$};
				\node at (1.4,-1.8){30 $\text{m}$};
				\node at (1.4,-6){100 $\text{m}$};
				\node at (2.2,0.4){0 $\text{m}$};
				\node at (14.3,0.4){2500 $\text{m}$};
				\node at (14.6,-4.8){80 $\text{m}$};
				\fill[cyan,opacity=0.7] (14,0)--(14,-4.8)--(11.2,-4.8)--(4.8,-1.8)--(2,-1.8)--(2,0)--cycle;
				\fill[brown,opacity=0.5] (2,-1.8)--(4.8,-1.8)--(11.2,-4.8)--(14,-4.8)--(14,-6)--(2,-6)--cycle;
				\draw[very thick, ->](2,0)--(14.5,0) node[right]{$r$};
				\draw[very thick, ->](2.02,0.1)--(2.02,-6.5) node[below]{$z$};
				\filldraw [red] (2.02,-1.5) circle [radius=2.5pt];
				\node[text=white] at (4,-1.5){$f=50 \mathrm{Hz}, z_\mathrm{s}=26 \mathrm{m}$};
				\draw[very thick](2,-1.8)--(4.8,-1.8);
				\draw[very thick](4.8,-1.8)--(11.2,-4.8);
				\draw[very thick](11.2,-4.8)--(14,-4.8);
				\draw[very thick](14.02,0.1)--(14.02,0);
				\draw[very thick](2.02,-6)--(14.02,-6);
				\draw[very thick](4.82,-5.9)--(4.82,-6);
				\draw[very thick](11.22,-5.9)--(11.22,-6);
				\node at (4.8,-6.4){500 $\text{m}$};
				\node at (11.2,-6.4){2000 $\text{m}$};
				\node at (7.8,-0.7){$c_1 = 1500 \mathrm{m/s}$};
				\node at (7.8,-1.2){$\rho_1 = 1.0 \mathrm{g/cm^3}$};
				\node at (7.8,-1.7){$\alpha_1 = 0.0 \mathrm{dB/\lambda}$};
				\node at (7.8,-4.1){$c_2 = 1800 \mathrm{m/s}$};
				\node at (7.8,-4.6){$\rho_2 = 1.5 \mathrm{g/cm^3}$};
				\node at (7.8,-5.1){$\alpha_2 = 2.0 \mathrm{dB/\lambda}$};
				\node at (8,-5.7){perfectly free boundary $p=0$};
			\end{tikzpicture}}
\caption{Schematic diagram of the upslope (a) and downslope (b) waveguides.}
		\label{Figure2}
\end{figure}
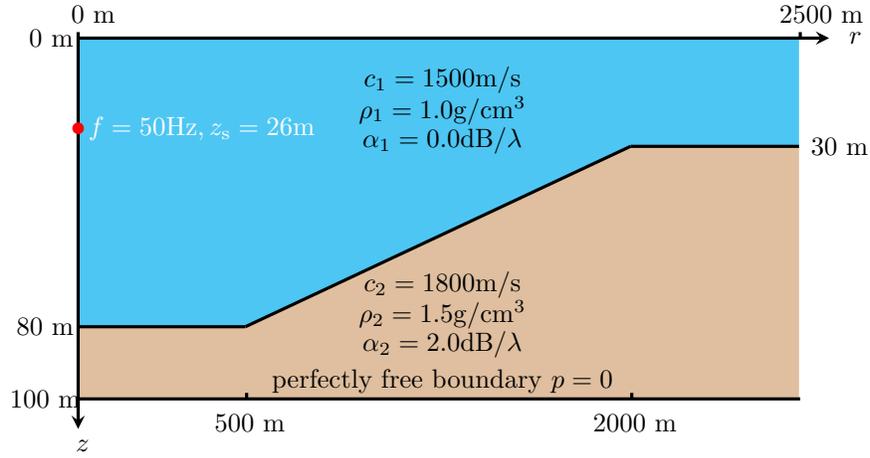
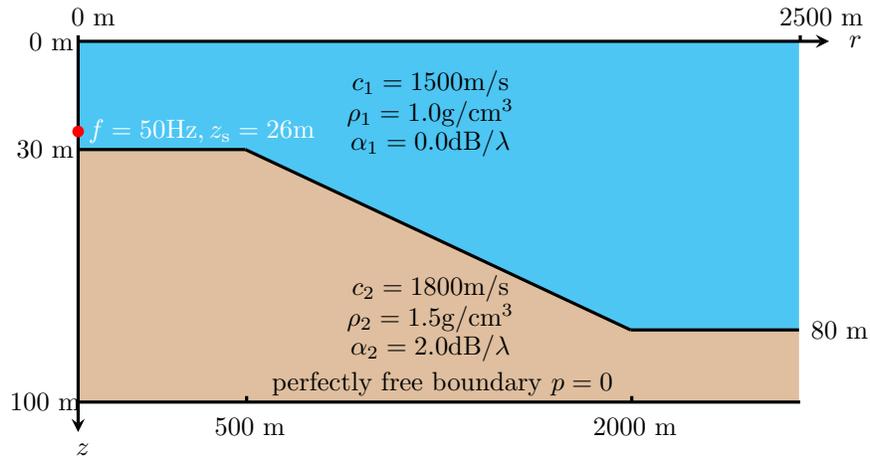

\begin{figure*}
\figline{
\fig{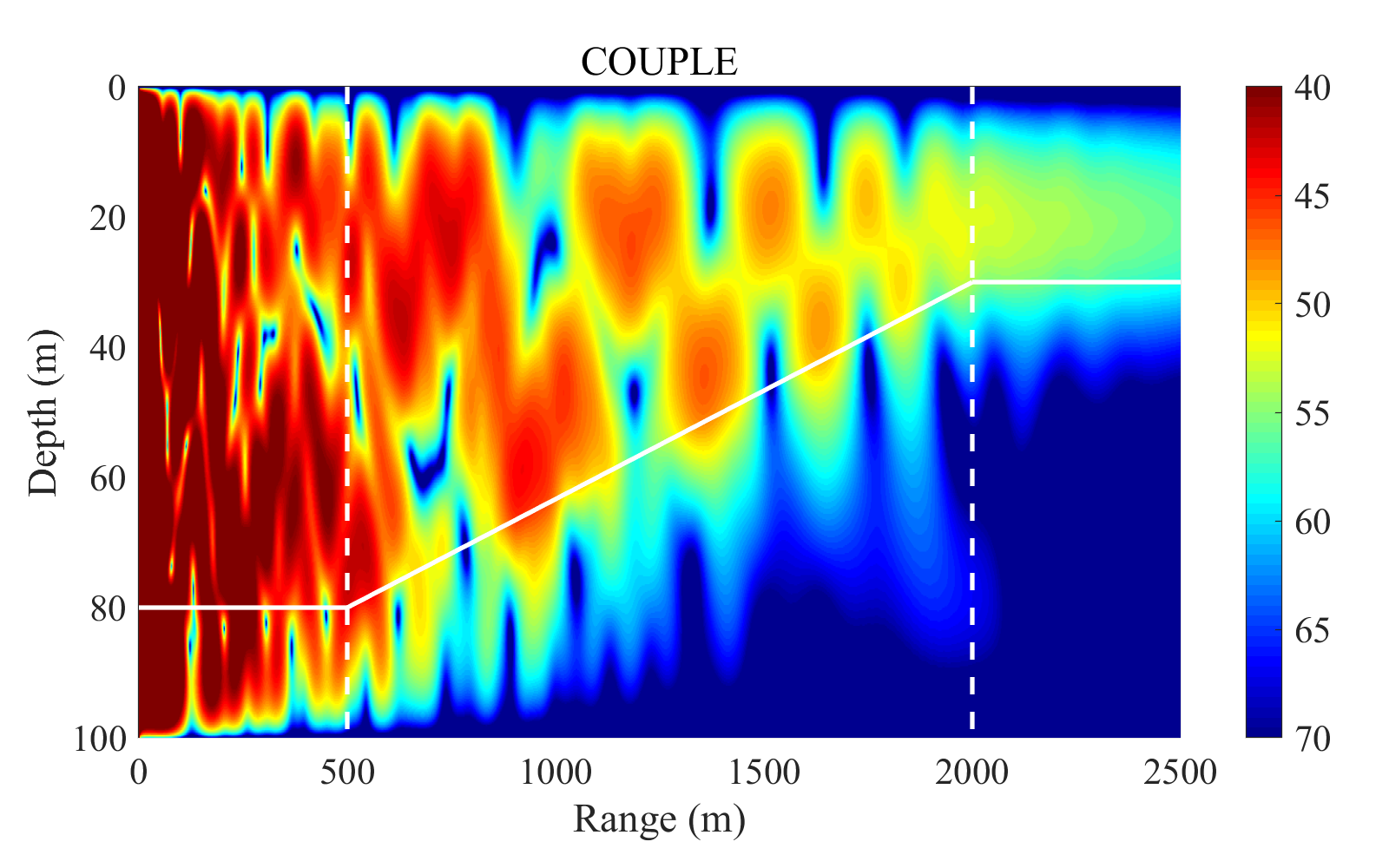}{8cm}{(a)}
\fig{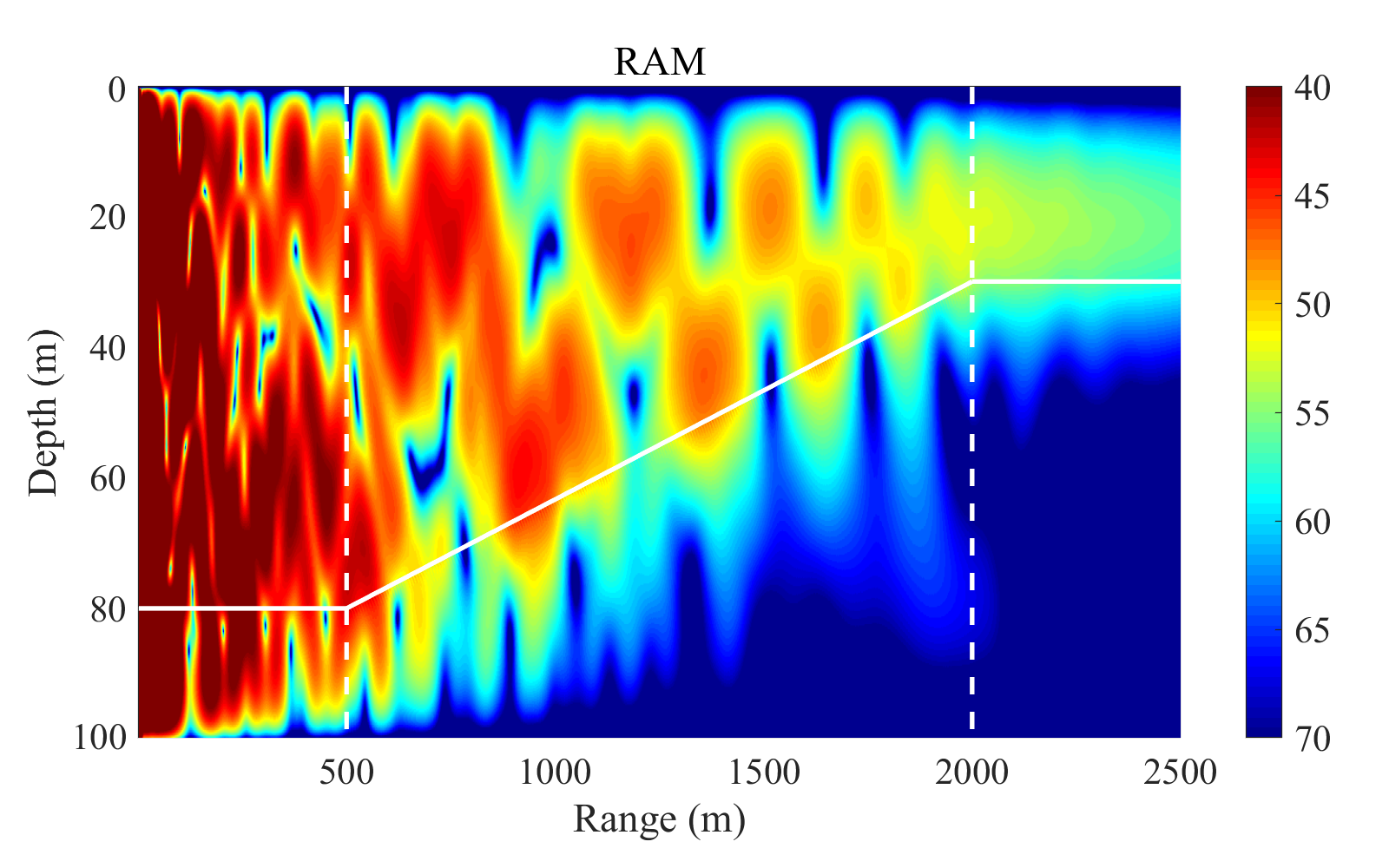}{8cm}{(b)}}
\figline{
\fig{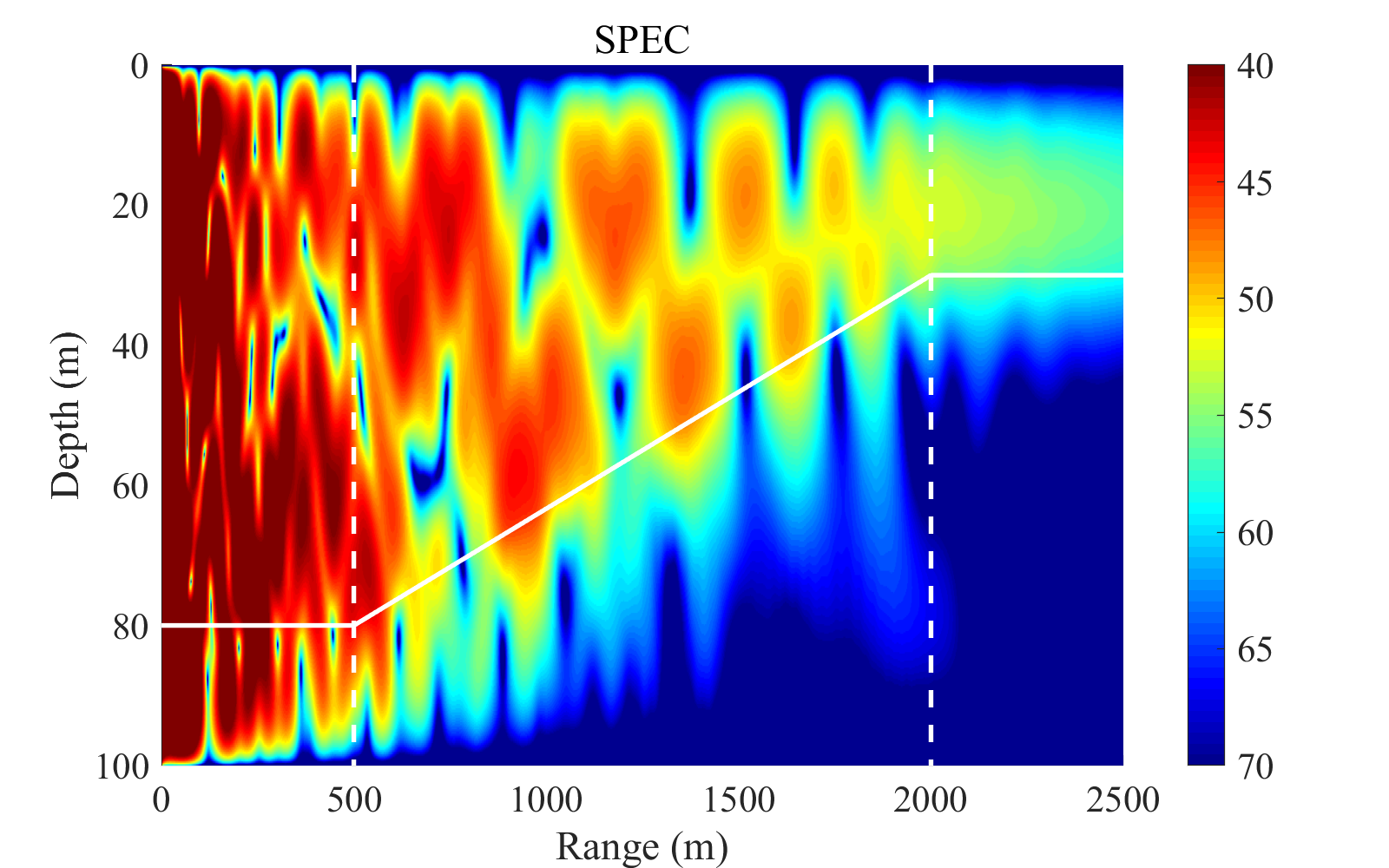}{8cm}{(c)}
\fig{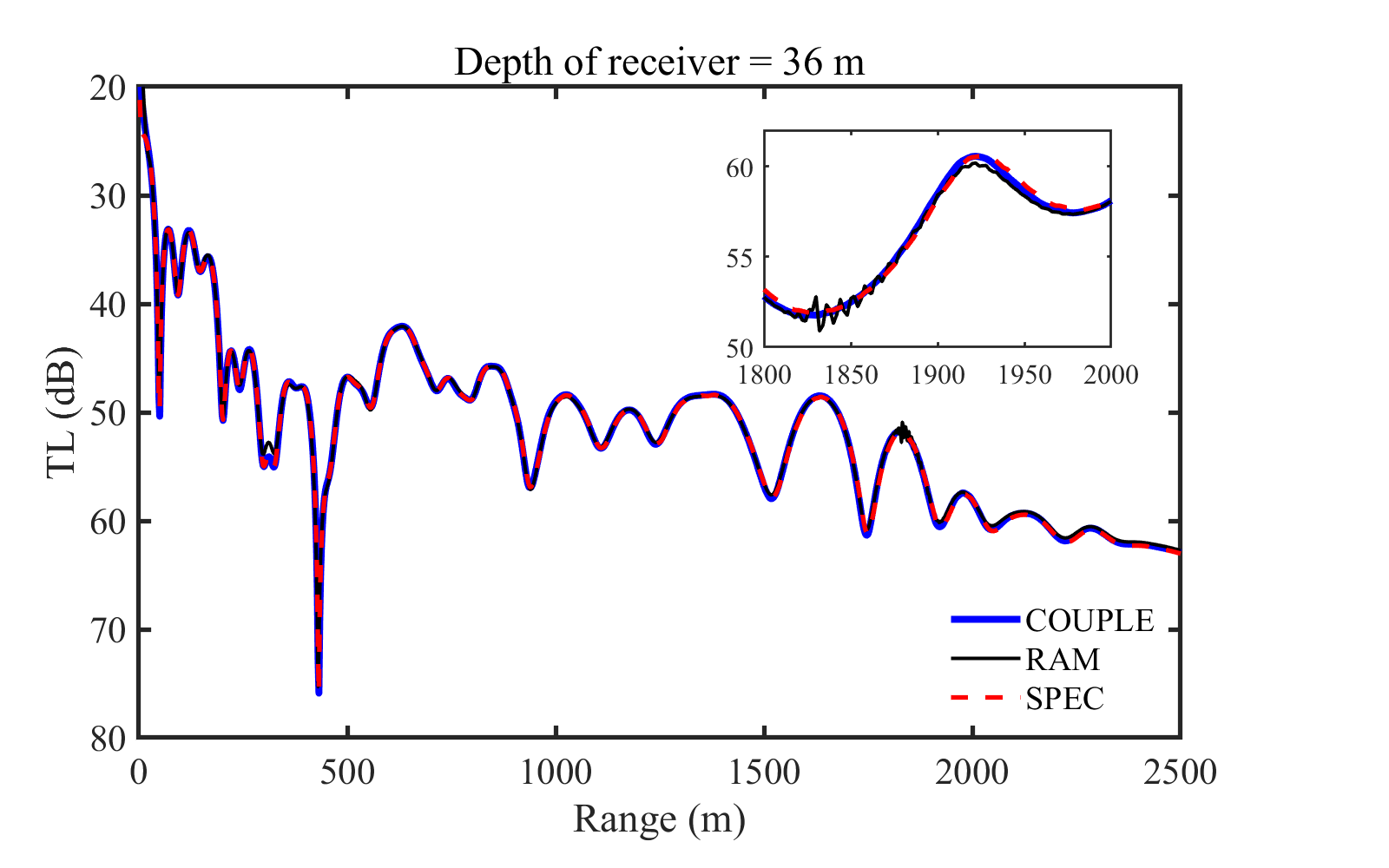}{8cm}{(d)}\label{Figure3d}}
\caption{Sound fields of the upslope waveguide ($f=50$ Hz, $z_\mathrm{s}=26$ m) calculated by COUPLE (a), RAM (b) and SPEC (c); TL curves at a depth of 36 m (d).}
		\label{Figure3}
\end{figure*}

Sloping terrain is one of the most common and classic range dependencies in underwater acoustic propagation. The specific configuration of this example is displayed in \autoref{Figure2}(a). We applied COUPLE, RAM and SPEC to calculate the sound fields of this example and set the receiver at a depth of 36 m. The COUPLE and SPEC programs use 6 modes, and both take the truncated order of 10 and 225 segments. The horizontal and vertical resolutions used by RAM are 2 m and 0.2 m, respectively. Overall, \autoref{Figure3} shows that the sound fields calculated by the three programs are highly consistent. The small window in \autoref{Figure3}(d) shows that SPEC is closer to COUPLE in a more detailed comparison, which may be because RAM ignores the backscattered waves.

\begin{figure*}
\figline{
\fig{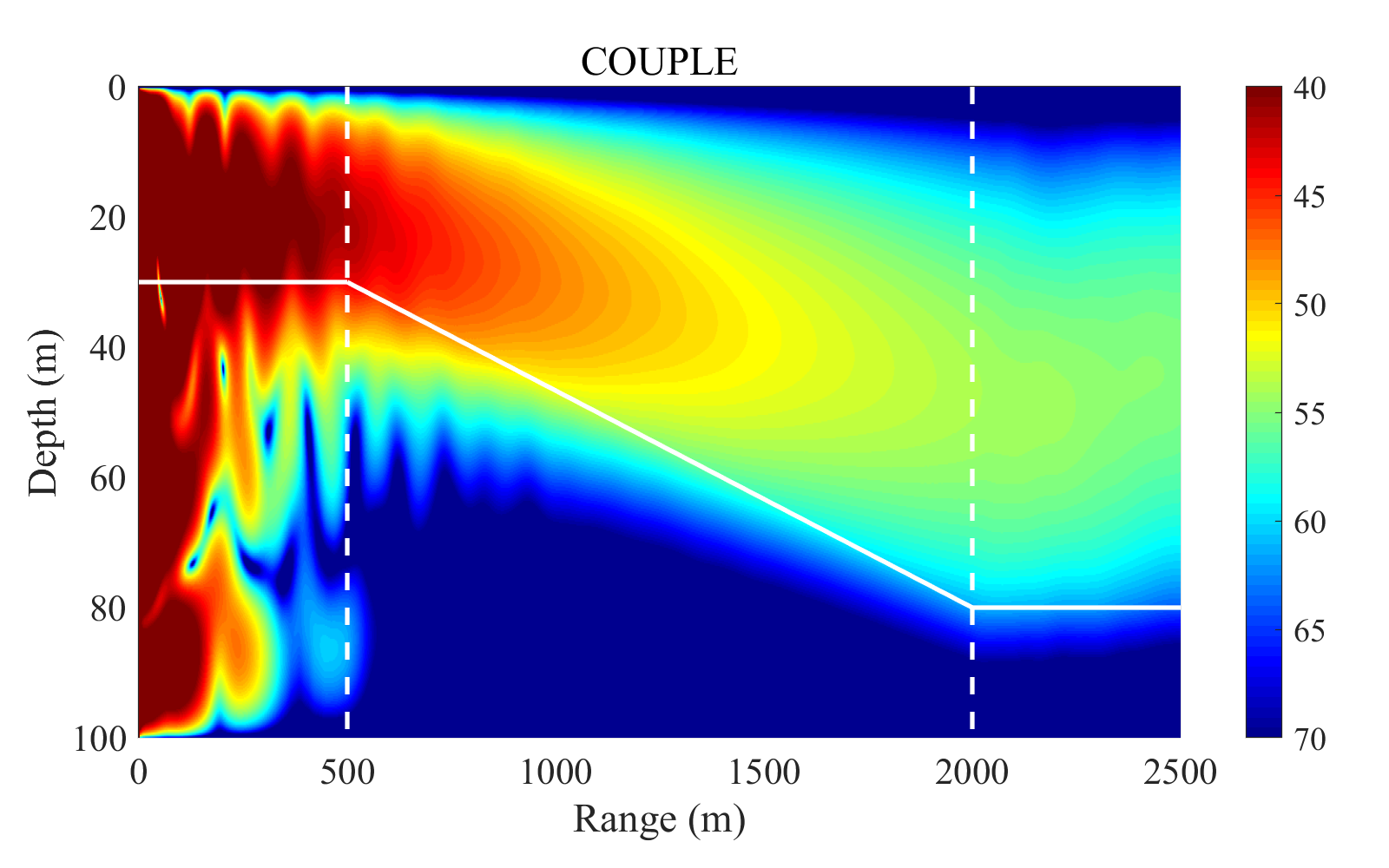}{8cm}{(a)}
\fig{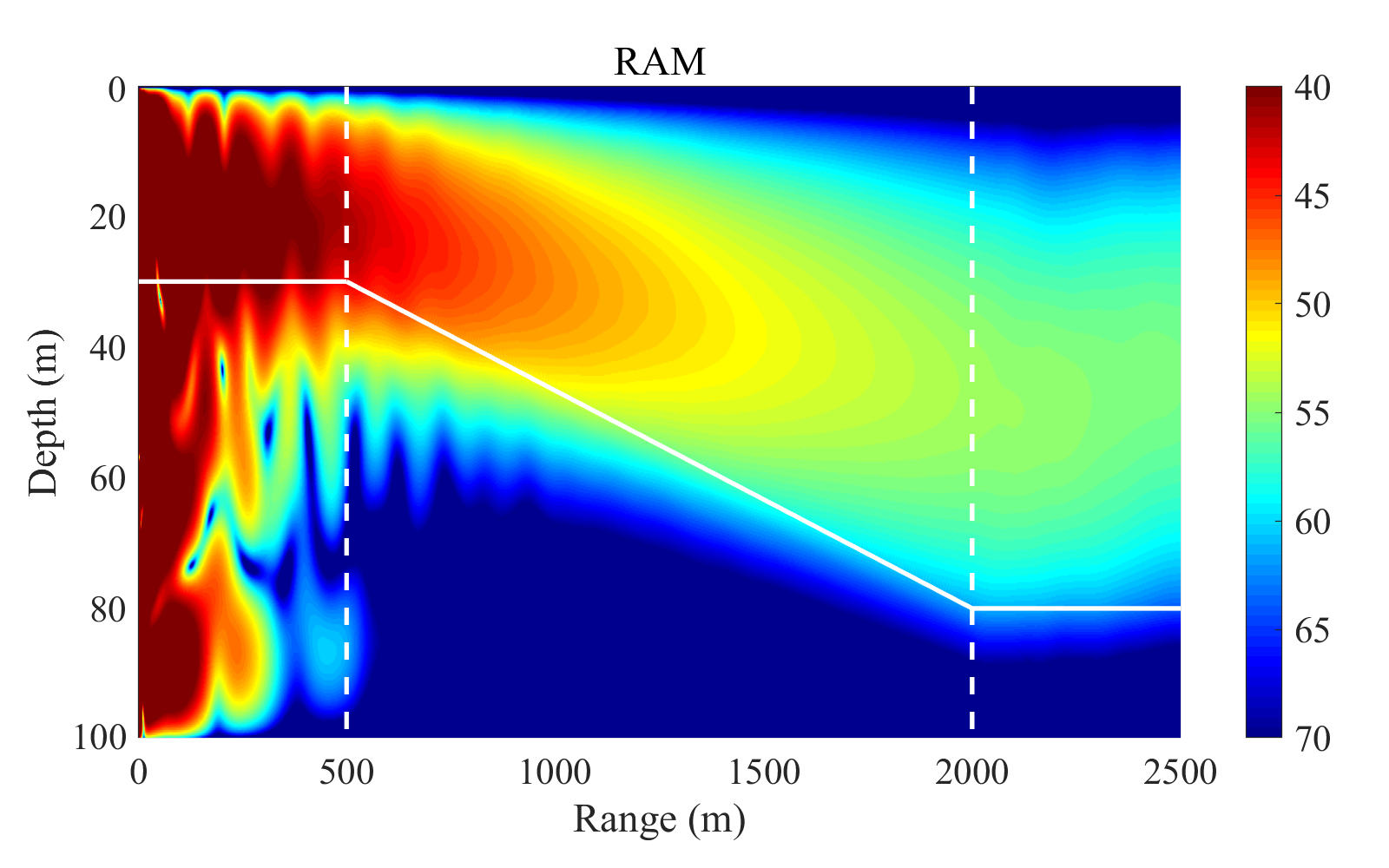}{8cm}{(b)}}
\figline{
\fig{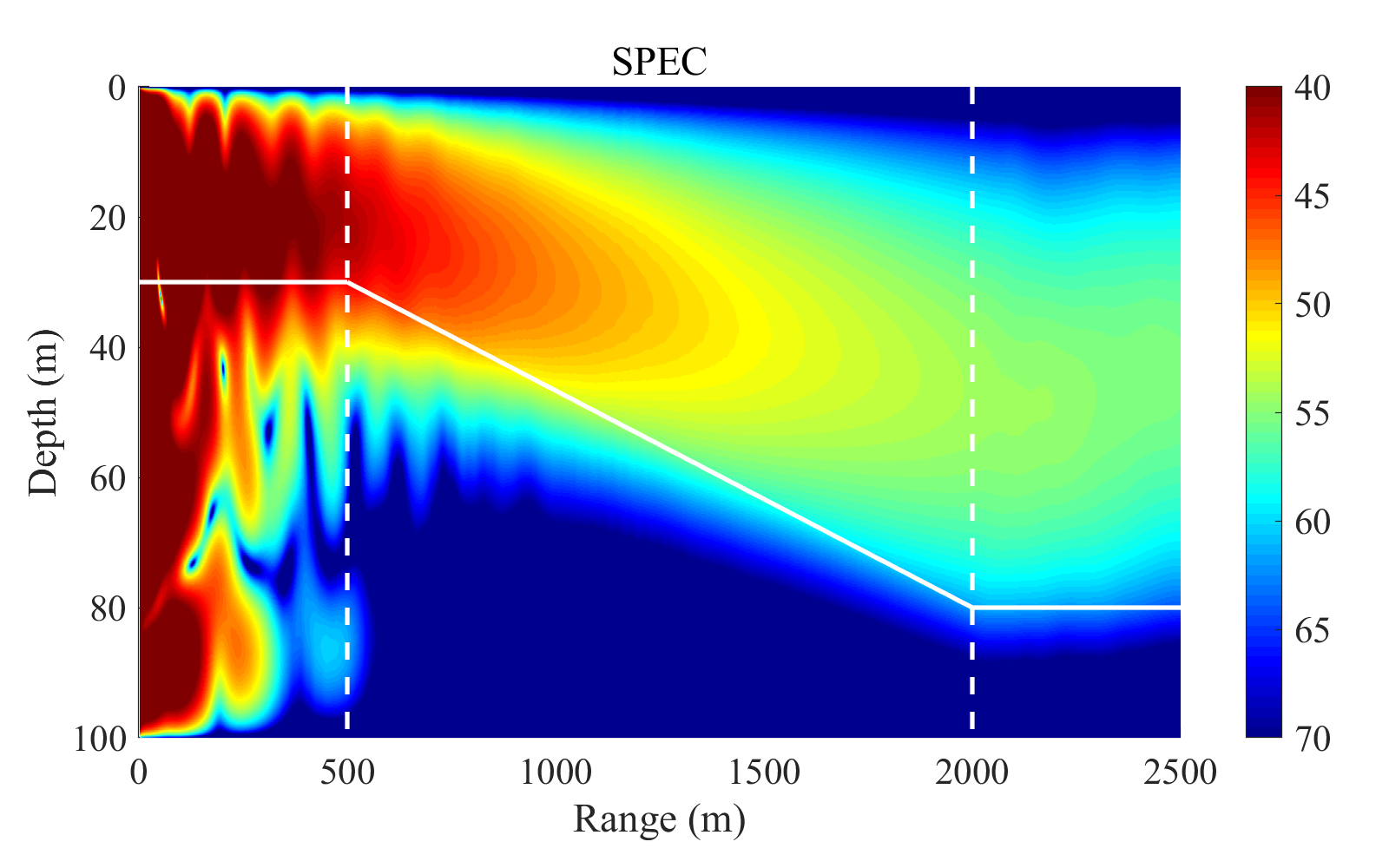}{8cm}{(c)}
\fig{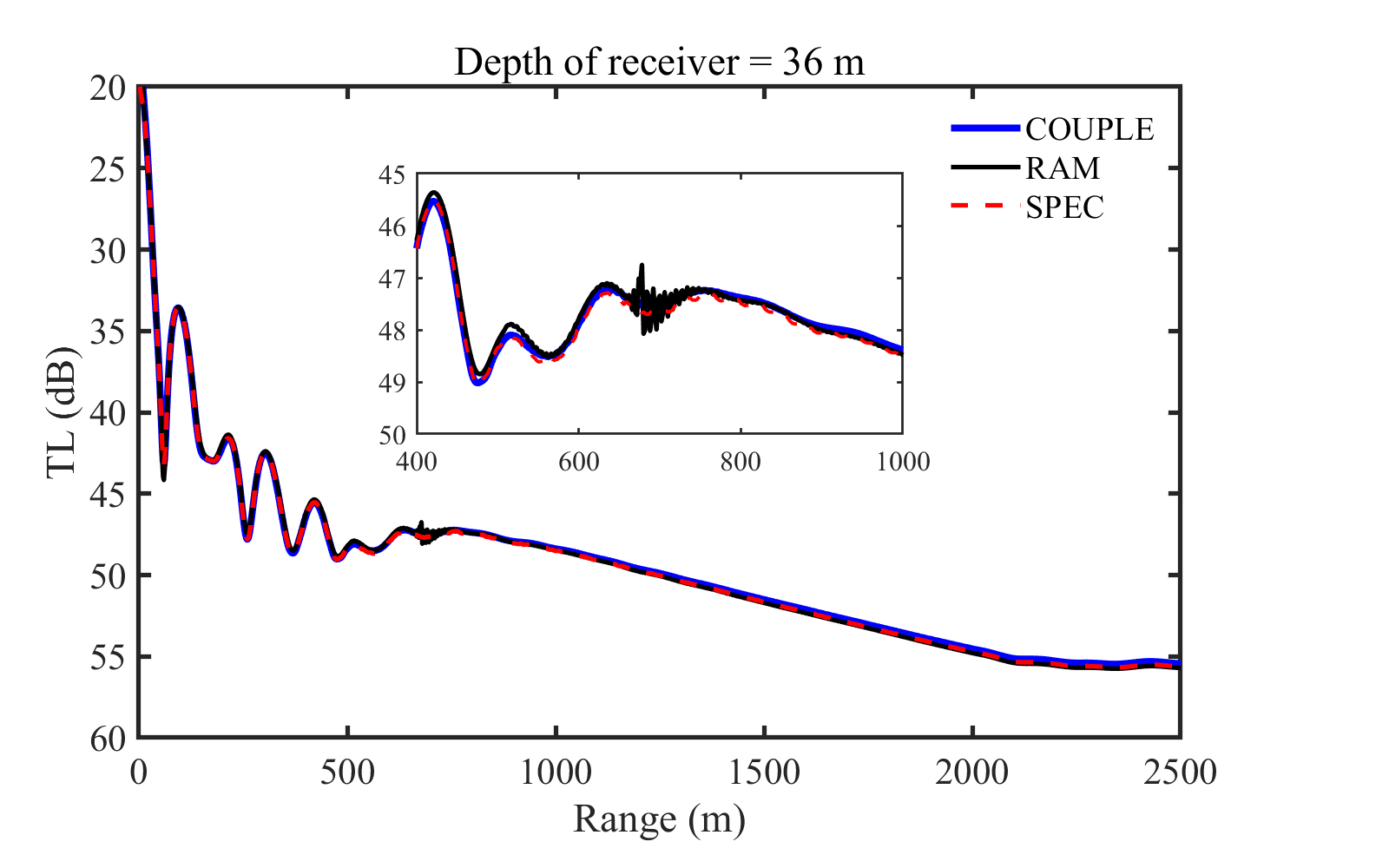}{8cm}{(d)}\label{Figure4d}}
\caption{Sound fields of the downslope waveguide ($f=50$ Hz, $z_\mathrm{s}=26$ m) calculated by COUPLE (a), RAM (b) and SPEC (c); TL curves at a depth of 36 m (d).}
		\label{Figure4}
\end{figure*}
As a comparison, we study the  downslope waveguide shown in \autoref{Figure2}(b), with the same configuration as the upslope example except for the terrain, and \autoref{Figure4} shows that the agreement between COUPLE, RAM and SPEC is once again very good.

\subsection{Seamount waveguide}
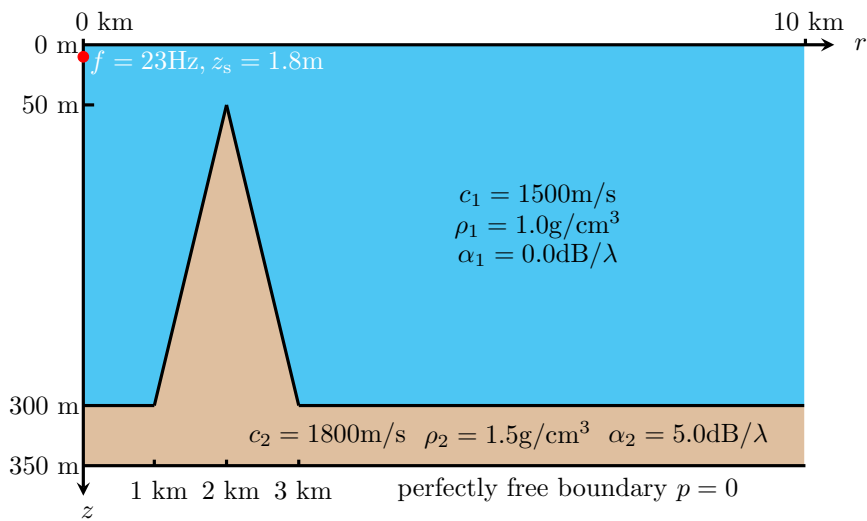
\begin{figure}[htbp]
		\centering
\begin{tikzpicture}[node distance=2cm,scale=0.8]
			\tikzstyle{every node}=[font=\footnotesize];
			\node at (1.5,0){0 $\text{m}$};
			\node at (1.4,-1){50 $\text{m}$};
			\node at (1.3,-6){300 $\text{m}$};
			\node at (1.3,-7){350 $\text{m}$};
			\node at (2.3,0.4){0 $\text{km}$};
			\node at (14,0.4){10 $\text{km}$};
			\fill[cyan,opacity=0.7] (2,0)--(2,-6)--(3.2,-6)--(4.4,-1)--(5.6,-6)--(14,-6)--(14,0)--cycle;
			\fill[brown,opacity=0.5] (2,-6)--(3.2,-6)--(4.4,-1)--(5.6,-6)--(14,-6)--(14,-7)--(2,-7)--cycle;
			\draw[very thick, ->](2,0)--(14.5,0) node[right]{$r$};
			\draw[very thick, ->](2.02,0.1)--(2.02,-7.5) node[below]{$z$};
			\filldraw [red] (2.02,-0.2) circle [radius=2.5pt];
			\node[text=white] at (4,-0.3){$f=23 \mathrm{Hz}, z_\mathrm{s}=1.8 \mathrm{m}$};
			\draw[very thick](2,-1)--(2.2,-1);
			\draw[very thick](2,-6)--(3.2,-6);
			\draw[very thick](3.2,-6)--(4.4,-1);
			\draw[very thick](4.4,-1)--(5.6,-6);
			\draw[very thick](5.6,-6)--(14,-6);
			\draw[very thick](14.02,0.1)--(14.02,0);
			\draw[very thick](2.02,-7)--(14.02,-7);
			\draw[very thick](3.2,-6.9)--(3.2,-7);
			\draw[very thick](4.4,-6.9)--(4.4,-7);
			\draw[very thick](5.6,-6.9)--(5.6,-7);
			\node at (3.2,-7.4){1 $\text{km}$};
			\node at (4.4,-7.4){2 $\text{km}$};
			\node at (5.6,-7.4){3 $\text{km}$};
			\node at (9.5,-2.5){$c_1 = 1500 \mathrm{m/s}$};
			\node at (9.5,-3.0){$\rho_1 = 1.0 \mathrm{g/cm^3}$};
			\node at (9.5,-3.5){$\alpha_1 = 0.0 \mathrm{dB/\lambda}$};
			\node at (6,-6.5){$c_2 = 1800 \mathrm{m/s}$};
			\node at (9,-6.5){$\rho_2 = 1.5 \mathrm{g/cm^3}$};
			\node at (12,-6.5){$\alpha_2 = 5.0 \mathrm{dB/\lambda}$};
			\node at (10,-7.4){perfectly free boundary $p=0$};
		\end{tikzpicture}
\caption{Schematic diagram of the seamount waveguide.}
		\label{Figure5}
\end{figure}

\begin{figure*}
\figline{
\fig{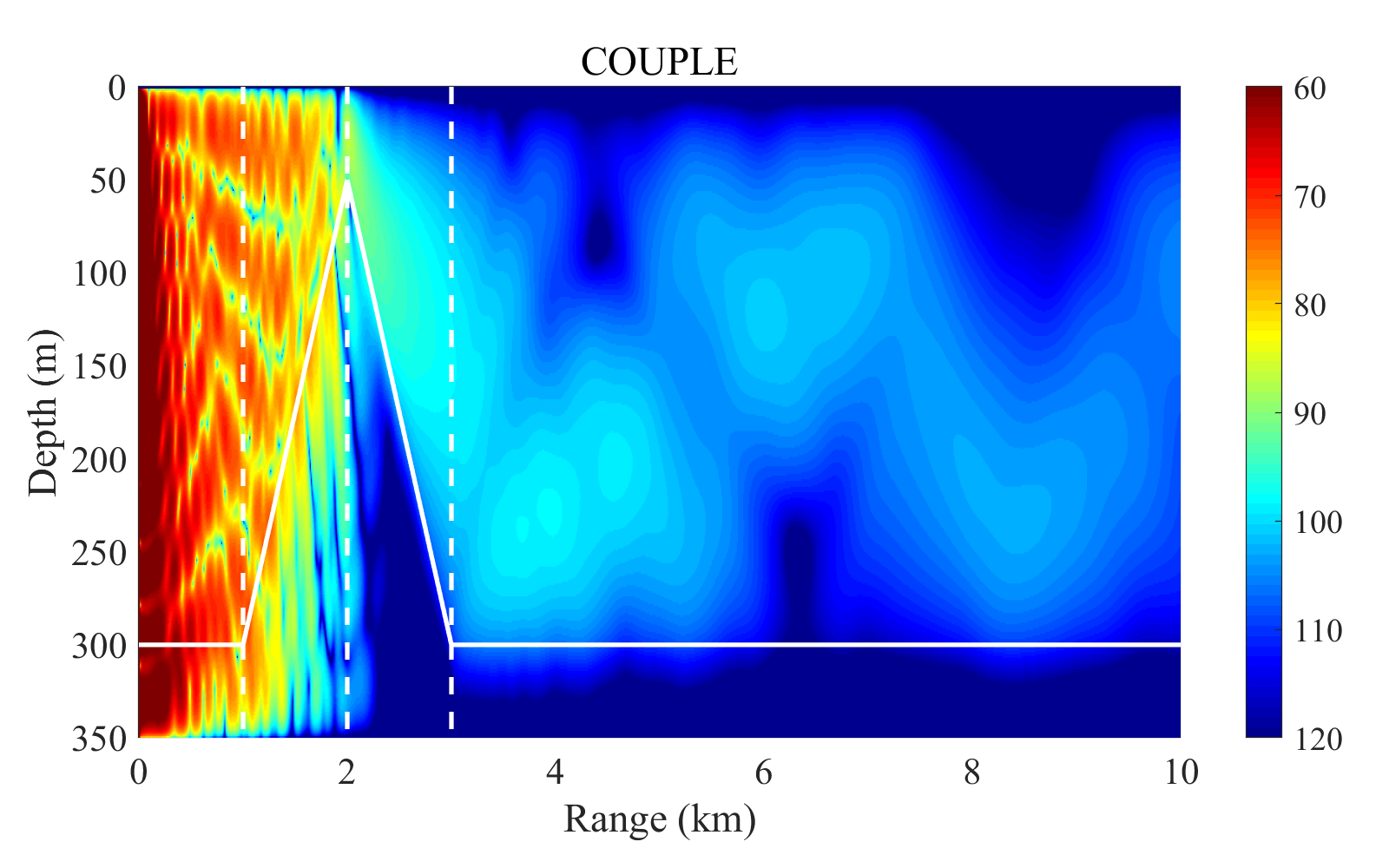}{8cm}{(a)}
\fig{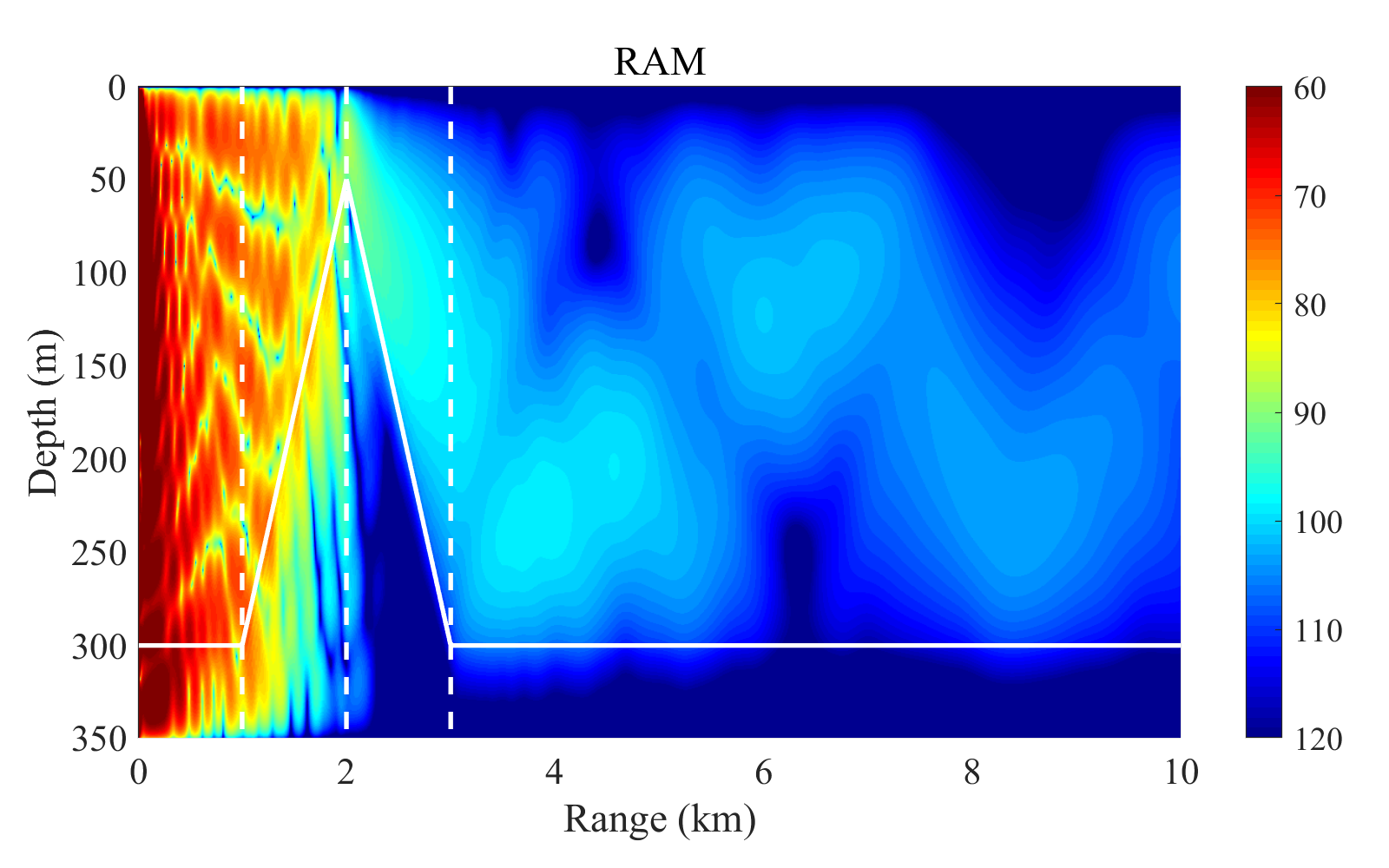}{8cm}{(b)}}
\figline{
\fig{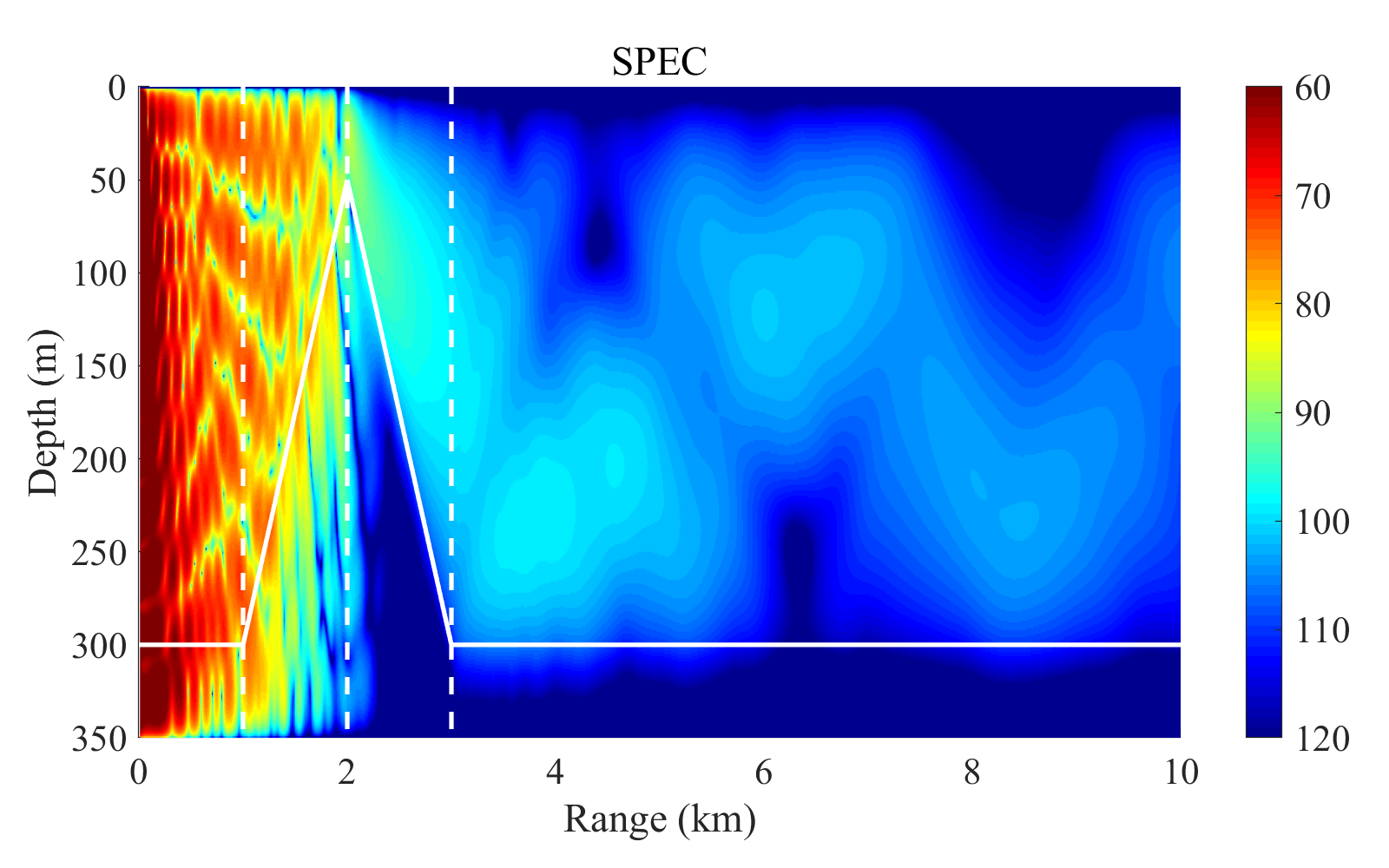}{8cm}{(c)}
\fig{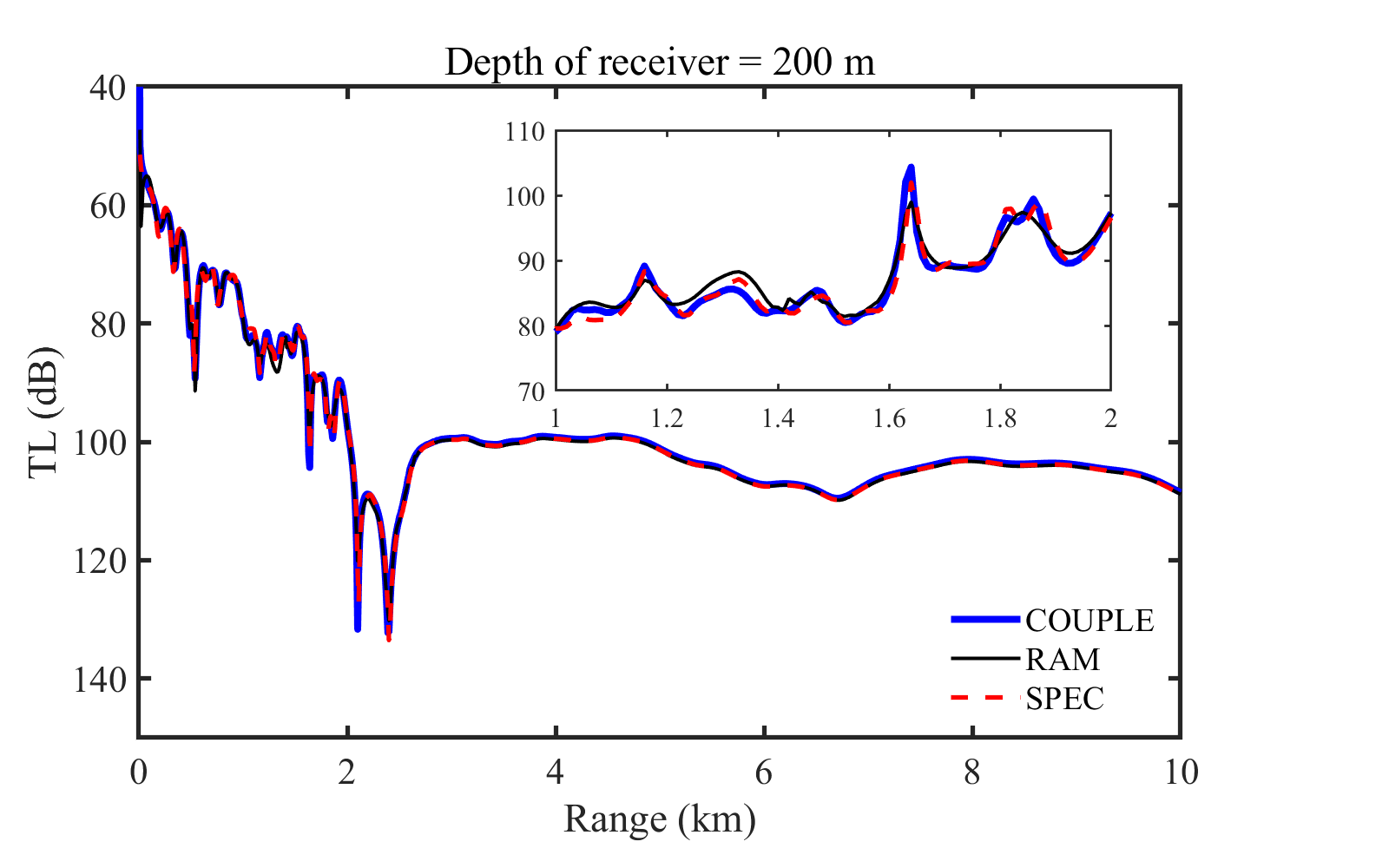}{8cm}{(d)}}
\caption{Sound fields of the seamount waveguide ($f=23$ Hz, $z_\mathrm{s}=1.8$ m) calculated by COUPLE (a), RAM (b) and SPEC (c); TL curves at a depth of 200 m (d).}
		\label{Figure6}
\end{figure*}

The topography of a seamount represents a typical range-dependent ocean environment. This example considers a seamount configuration, as shown in \autoref{Figure5}. Instead of the gentle slope of 1.9$^\circ$ in Example A, Example B involves a steep slope of 14$^\circ$. \autoref{Figure6} illustrates the sound fields calculated by COUPLE, RAM and SPEC and the TL curves at a depth of 200 m. The coupling of the COUPLE and SPEC programs includes 8 modes, the truncated order of the basis functions of both programs is 16, and the number of segments of both programs is 126. The horizontal and vertical resolutions used by RAM are 2 m and 0.2 m, respectively. Observation of the whole sound field shows that the results of the three programs are very similar, with only slight differences before crossing the seamount, as also indicated by the TL curve at a depth of 200 m. Good agreement of the three programs indicates that SPEC offers excellent accuracy.

\subsection{Warm-core eddy}
\begin{figure}[htbp]
		\centering
\subfigure[]{\begin{tikzpicture}[node distance=2cm,scale=0.8]
			\node at (1.6,0){0 $\text{m}$};
			\node at (1.2,-5){5000 $\text{m}$};		
			\node at (2,0.4){0};
			\node at (2.75,0.4){12.5};
			\node at (3.5,0.4){25};
			\node at (4.25,0.4){37.5};
			\node at (5,0.4){50};
			\node at (6.5,0.4){75};	
			\node at (8,0.4){100};
			\node at (9.5,0.4){125};												
			\node at (14.3,0.4){200 $\text{km}$};
			\fill[brown,opacity=0.7] (14,-6.2) rectangle (2,-5);
			\fill[cyan,opacity=0.7] (14,0)--(14,-5)--(2,-5)--(2,0)--cycle;
			\draw[very thick, ->](2,0)--(14.5,0) node[right]{$r$};
			\draw[very thick, ->](2.02,0.1)--(2.02,-6.5) node[below]{$z$};
			\filldraw [red](2.02,-0.3) circle [radius=2.5pt];
			\draw[very thick](2,-5)--(14,-5);
			\draw[dash dot, very thick](14.02,0.1)--(14.02,0);
			\draw[dashed, very thick](2.75,0)--(2.75,-5);
			\draw[dashed, very thick](3.5,0)--(3.5,-5);
			\draw[dashed, very thick](4.25,0)--(4.25,-5);
			\draw[dashed, very thick](5,0)--(5,-5);
			\draw[dashed, very thick](6.5,0)--(6.5,-5);
			\draw[dashed, very thick](8,0)--(8,-5);
			\draw[dashed, very thick](9.5,0)--(9.5,-5);	
			\node[text=white] at (3.4,-0.5){$f=50$ Hz};
			\node[text=white] at (3.5,-1){$z_\mathrm{s}=300$ m};	
			\node at (8,-5.6){$c_\infty=1800$ m/s, $\rho_\infty=2.0$ g/cm$^3$, $\alpha_\infty=0.1$ dB/$\lambda$};
		\end{tikzpicture}}
\subfigure[]{\includegraphics[width=0.65\linewidth]{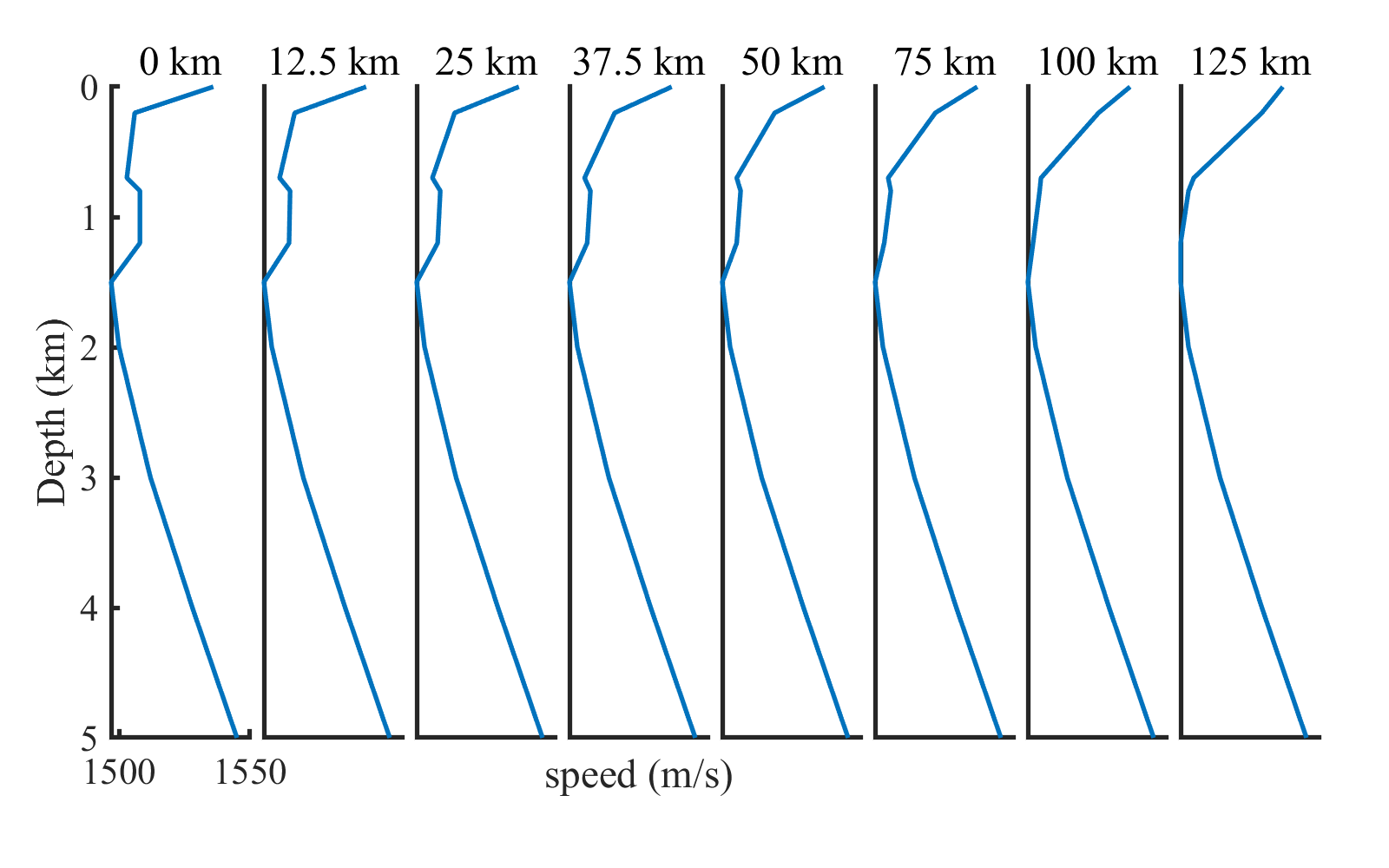}}
\caption{Schematic diagram of the warm-core eddy waveguide (a) and the sound speed profiles taken at the horizontal ranges (b).}
		\label{Figure7}
\end{figure}

Eddy currents are common hydrological phenomena in the ocean that alter the temperature and salinity of seawater, thereby altering the ocean's acoustic properties. Therefore, the propagation of sound through an eddy is different from that through seawater without an eddy. Here, we consider a warm-core eddy in \autoref{Figure7}(a), which is a classic example for range-dependent waveguides, as mentioned by Jensen et al. \cite{Jensen2011} and Porter \cite{Kraken2001}. The sound speed profiles of the warm-core eddy taken at the horizontal ranges are shown in \autoref{Figure7}(b).

\begin{figure*}
\figline{
\fig{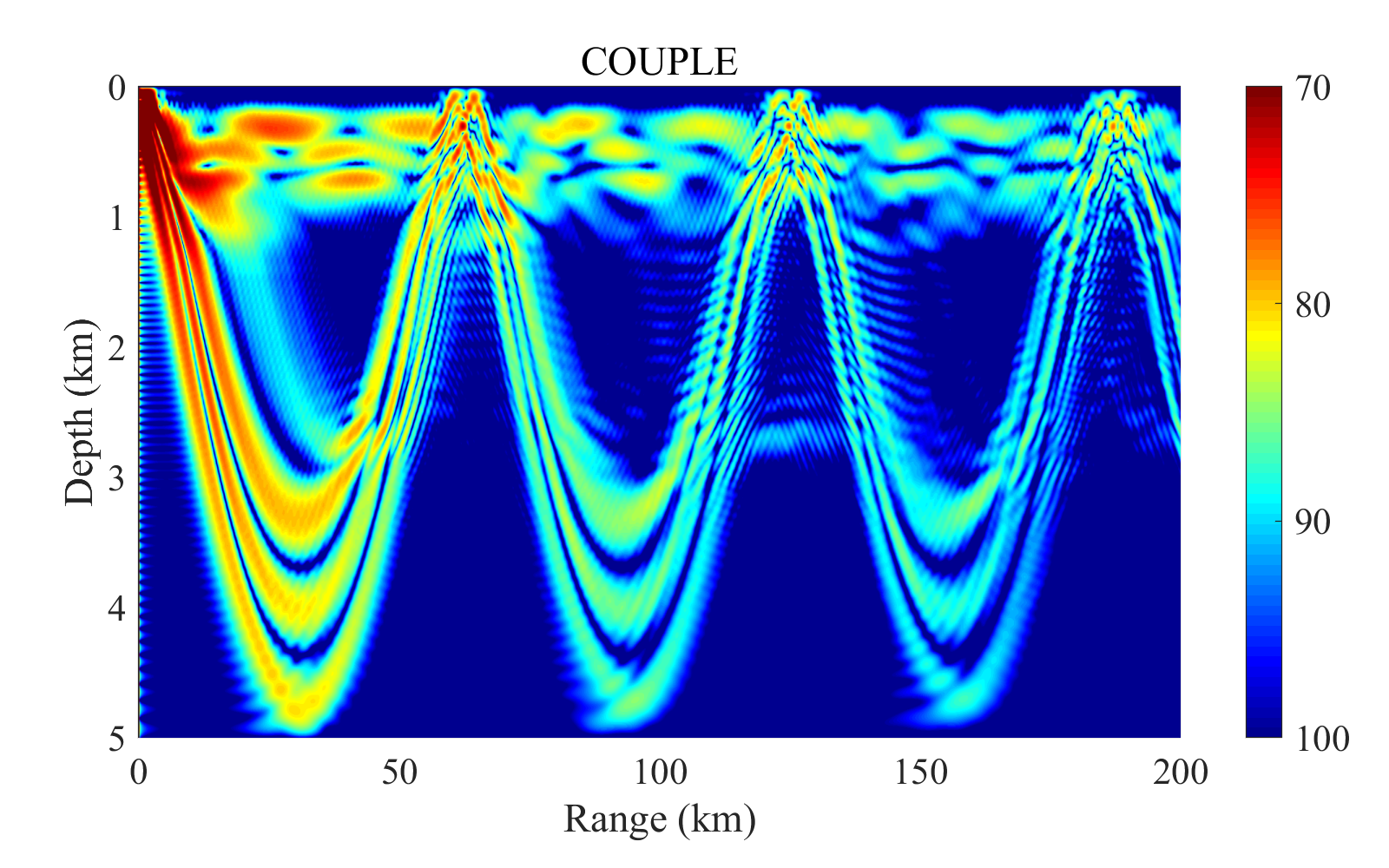}{8cm}{(a)}\label{Figure8a}
\fig{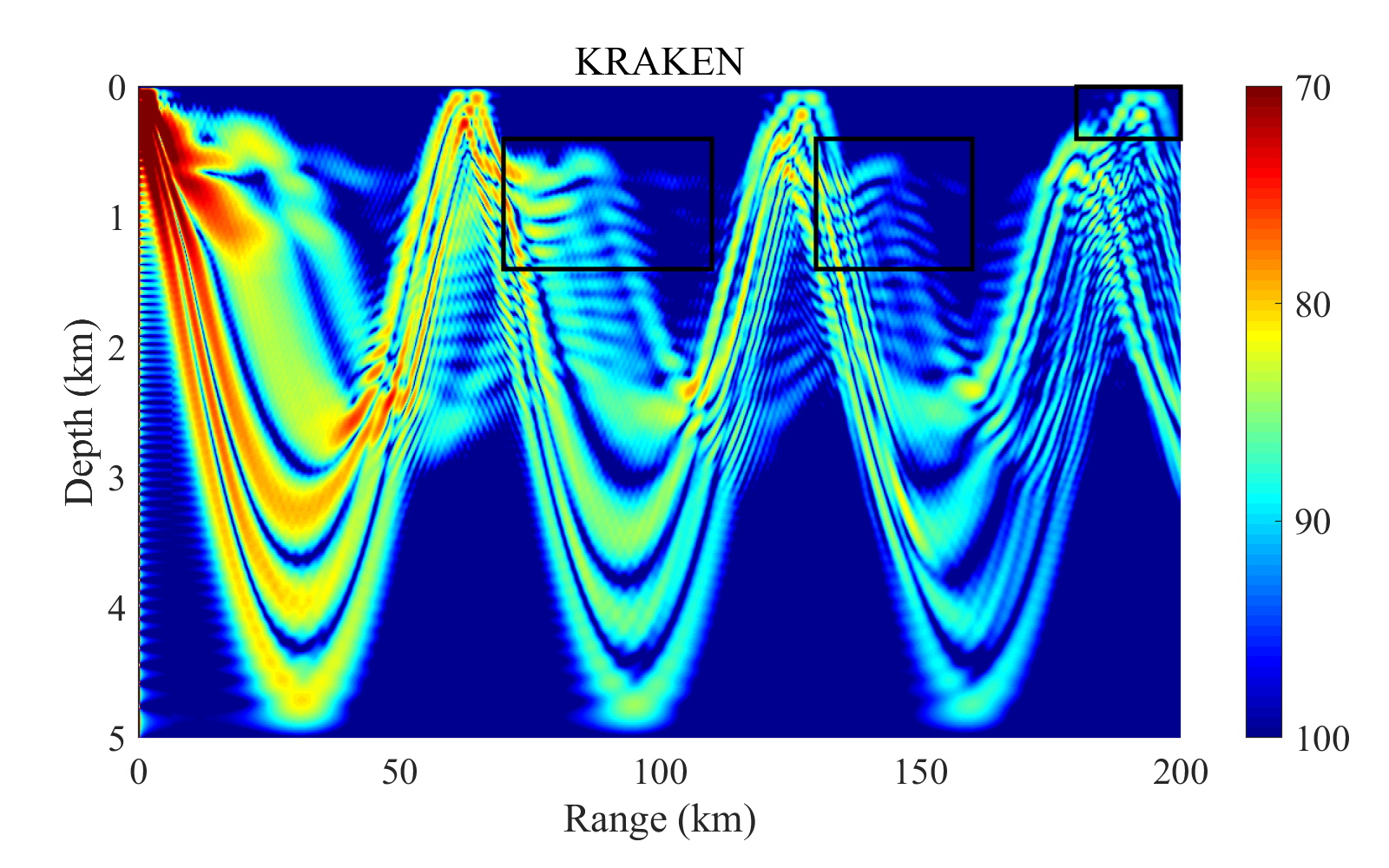}{8cm}{(b)}\label{Figure8b}}
\figline{
\fig{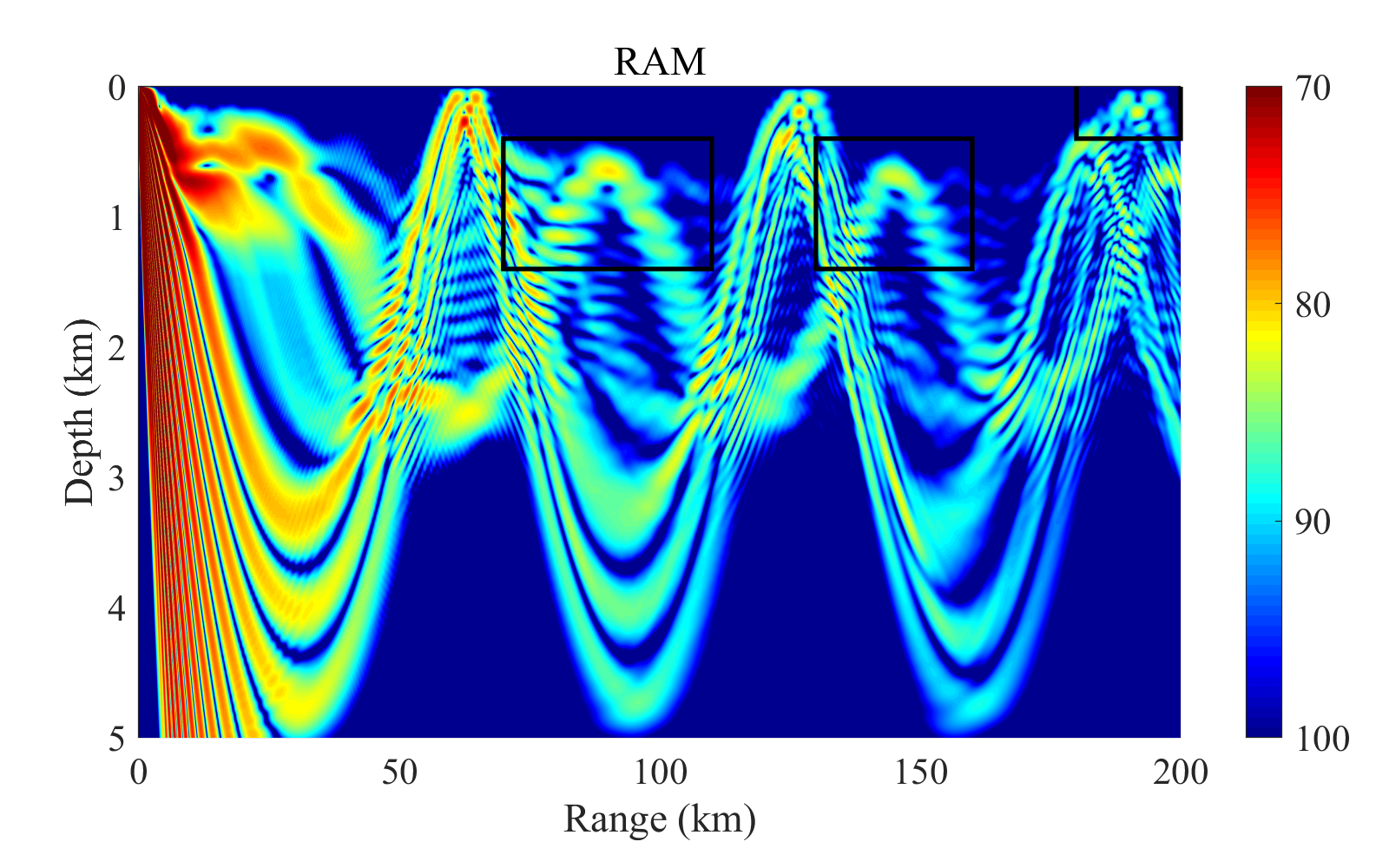}{8cm}{(c)}\label{Figure8c}
\fig{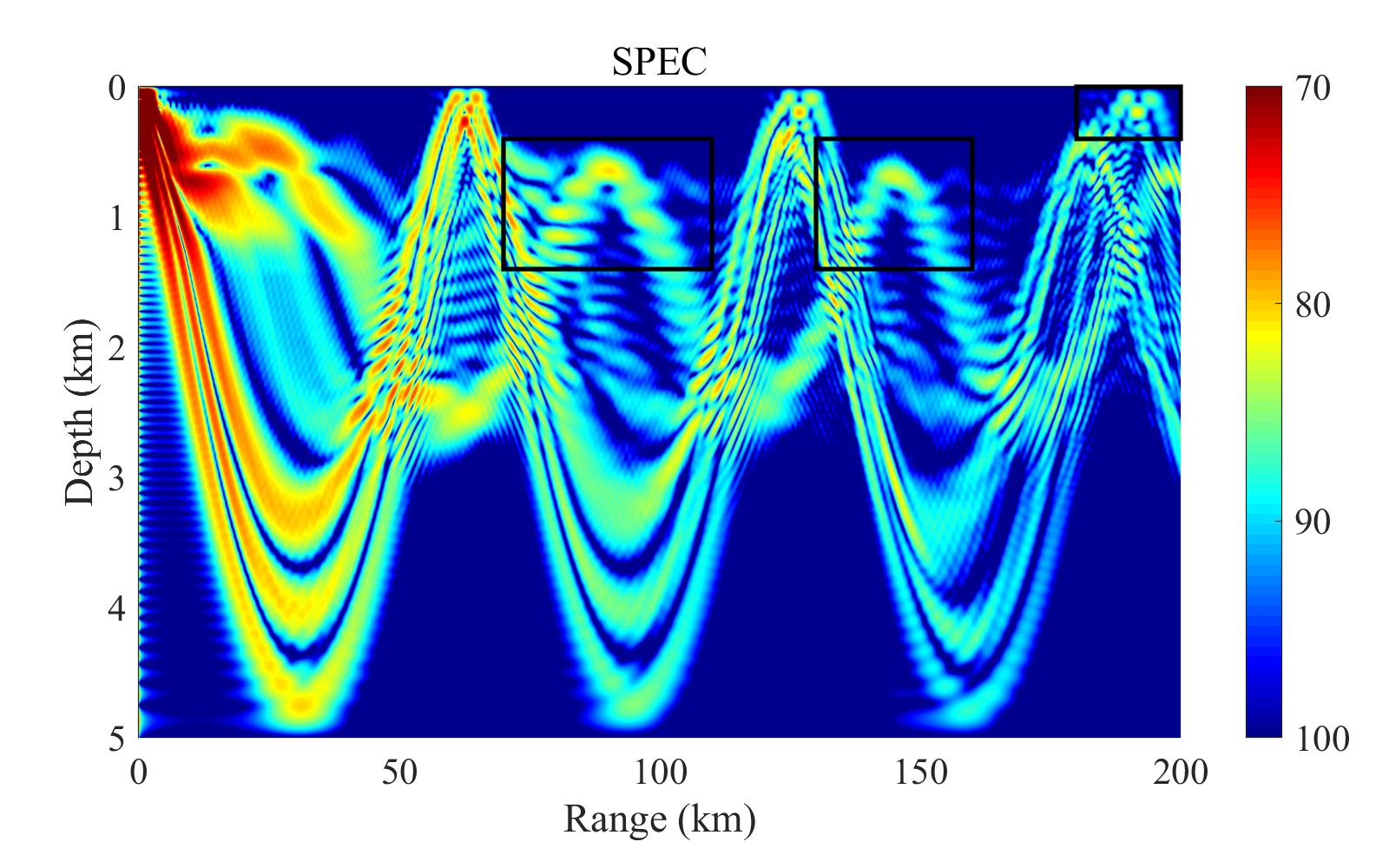}{8cm}{(d)}\label{Figure8d}}
\caption{Sound fields of the warm-core eddy waveguide ($f=50$ Hz, $z_\mathrm{s}=300$ m) calculated by COUPLE (a), KRAKEN (b), RAM (c) and SPEC (d); (a) shows the range-independent modes using the first sound speed profile throughout.}
		\label{Figure8}
\end{figure*}

\autoref{Figure8} plots the sound fields through the warm-core eddy calculated using the four numerical models. Since COUPLE is inconvenient for such a computationally expensive example, \autoref{Figure8}(a) shows the case where COUPLE uses only the range-independent simulation of the first sound speed profile. The number of discrete points used by KRAKEN is automatically selected by the program, while the spectral truncated order used by SPEC in the water column is 300. COUPLE, KRAKEN and SPEC all have a total of 63 modes involved in the simulation. A cursory observation shows that the results of COUPLE and the other three programs are quite different, which illustrates the effect of warm-core eddy currents on sound propagation. If the contribution of the `continuous spectrum' to the near field of the RAM is ignored, the sound field calculated by the RAM in \autoref{Figure8} is very similar to that of KRAKEN and SPEC. However, significant differences in the sound fields are still visible in the areas of the three black boxes in \autoref{Figure8}(b) to \autoref{Figure8}(d). In these regions, the results for SPEC and RAM are more consistent, possibly because the range dependence in KRAKEN is handled by the theory of one-way coupled modes.

\subsection{Multilayer parallel waveguide}
\begin{figure}[htbp]
		\centering
\subfigure[]{\begin{tikzpicture}[node distance=2cm,scale=0.8]
				\tikzstyle{every node}=[font=\footnotesize];
				\node at (1.5,0){0 $\text{m}$};
				\node at (1.3,-2){200 $\text{m}$};
				\node at (1.3,-3){300 $\text{m}$};
				\node at (1.2,-6){2000 $\text{m}$};				
				\node at (2.2,0.4){0 $\text{km}$};
				\node at (14.3,0.4){10 $\text{km}$};
				\node at (14.6,-1){100 $\text{m}$};
				\node at (14.6,-2){200 $\text{m}$};
				\fill[cyan,opacity=0.4] (14,0)--(14,-1)--(8,-2)--(2,-2)--(2,0)--cycle;
				\fill[cyan,opacity=0.7] (14,-1)--(14,-2)--(8,-3)--(2,-3)--(2,-2)--(8,-2)--cycle;
				\fill[brown,opacity=0.5] (14,-2)--(14,-6)--(2,-6)--(2,-3)--(8,-3)--cycle;
				\draw[very thick, ->](2,0)--(14.5,0) node[right]{$r$};
				\draw[very thick, ->](2.02,0.1)--(2.02,-6.5) node[below]{$z$};
				\filldraw [red] (2.02,-0.5) circle [radius=2.5pt];
				\node[text=white] at (4,-0.5){$f=25 \mathrm{Hz}, z_\mathrm{s}=50 \mathrm{m}$};
				\draw[very thick](2,-2)--(8,-2);
				\draw[very thick](8,-2)--(14,-1);
				\draw[very thick](2,-3)--(8,-3);
				\draw[very thick](8,-3)--(14,-2);
				\draw[very thick](8,-5.9)--(8,-6);				
				\draw[very thick](14.02,0.1)--(14.02,0);
				\node at (8,-6.4){5 $\text{km}$};
				\node at (4,-1){$c_1 = 1500 \mathrm{m/s}$};
				\node at (7,-1){$\rho_1 = 1.0 \mathrm{g/cm^3}$};
				\node at (10,-1){$\alpha_1 = 0.0 \mathrm{dB/\lambda}$};
				\node at (4,-2.5){$c_2 = 1700 \mathrm{m/s}$};
				\node at (7,-2.5){$\rho_2 = 1.1 \mathrm{g/cm^3}$};
				\node at (10,-2.2){$\alpha_2 = 0.1 \mathrm{dB/\lambda}$};
				\node at (4,-4.5){$c_3 = 2000 \mathrm{m/s}$};
				\node at (7,-4.5){$\rho_3 = 1.5 \mathrm{g/cm^3}$};
				\node at (10,-4.5){$\alpha_3 = 0.5 \mathrm{dB/\lambda}$};
				\node at (8,-5.7){perfectly free boundary $p=0$};				
				\draw[very thick](2.02,-6)--(14.02,-6);
		\end{tikzpicture}}
\subfigure[]{\begin{tikzpicture}[node distance=2cm,scale=0.8]
				\tikzstyle{every node}=[font=\footnotesize];
				\node at (1.5,0){0 $\text{m}$};
				\node at (14.6,-3.6){60 $\text{m}$};
				\node at (14.6,-5.4){90 $\text{m}$};				
				\node at (1.3,-6){100 $\text{m}$};
				\node at (2.2,0.4){0 $\text{m}$};
				\node at (14.3,0.4){2500 $\text{m}$};
				\node at (1.4,-4.8){80 $\text{m}$};
				\node at (1.4,-5.4){90 $\text{m}$};			
				\fill[cyan,opacity=0.4] (14,0)--(14,-3.6)--(11.6,-1.2)--(4.4,-4.2)--(2,-4.8)--(2,0)--cycle;
				\fill[cyan,opacity=0.7] (2,-4.8)--(4.4,-4.2)--(11.6,-1.2)--(14,-3.6)--(14,-5.4)--(11.6,-3)--(4.4,-5.1)--(2,-5.4)--cycle;
				\fill[brown,opacity=0.5] (2,-5.4)--(4.4,-5.1)--(11.6,-3)--(14,-5.4)--(14,-6)--(2,-6)--cycle;
				\draw[very thick, ->](2,0)--(14.5,0) node[right]{$r$};
				\draw[very thick, ->](2.02,0.1)--(2.02,-6.5) node[below]{$z$};
				\filldraw [red] (2.02,-1.98) circle [radius=2.5pt];
				\node[text=white] at (4,-1.98){$f=50 \mathrm{Hz}, z_\mathrm{s}=36 \mathrm{m}$};
				\draw[very thick](2,-4.8)--(4.4,-4.2);
				\draw[very thick](2,-5.4)--(4.4,-5.1);
				\draw[very thick](4.4,-4.2)--(11.6,-1.2);				
				\draw[very thick](4.4,-5.1)--(11.6,-3);
				\draw[very thick](4.4,-5.9)--(4.4,-6);
				\draw[very thick](11.6,-1.2)--(14,-3.6);
				\draw[very thick](11.6,-3)--(14,-5.4);
				\draw[very thick](2,-6)--(14,-6);				
				\draw[very thick](11.6,-5.9)--(11.6,-6);
				\draw[very thick](14.02,0.1)--(14.02,0);
				\node at (4.4,-6.4){500 $\text{m}$};
				\node at (11.6,-6.4){2000 $\text{m}$};				
				\node at (7.8,-0.5){$c_1 = 1500 \mathrm{m/s}$};
				\node at (7.8,-1.0){$\rho_1 = 1.0 \mathrm{g/cm^3}$};
				\node at (7.8,-1.5){$\alpha_1 = 0.0 \mathrm{dB/\lambda}$};
				\node at (6.5,-4){$c_2 = 1600 \mathrm{m/s}$};
				\node at (9,-3.2){$\rho_2 = 1.5 \mathrm{g/cm^3}$};
				\node at (11,-2.5){$\alpha_2 = 0.5 \mathrm{dB/\lambda}$};
				\node at (6.3,-5.2){$c_3 = 1800 \mathrm{m/s}$};
				\node at (9.2,-5.2){$\rho_3 = 2.0 \mathrm{g/cm^3}$};
				\node at (12.1,-5.2){$\alpha_3 = 1.0 \mathrm{dB/\lambda}$};				
				\node at (8,-5.7){perfectly rigid boundary $p=0$};
				\node[text=white] at (4.4,-5.4){85 $\text{m}$};	
				\node[text=white] at (4.4,-3.9){70 $\text{m}$};	
				\node[text=white] at (11.6,-1){20 $\text{m}$};	
				\node[text=white] at (11.6,-3.3){50 $\text{m}$};	
		\end{tikzpicture}}
\caption{Schematic diagram of the multilayer parallel (a) and multilayer undulating (b) waveguides.}
		\label{Figure9}
\end{figure}
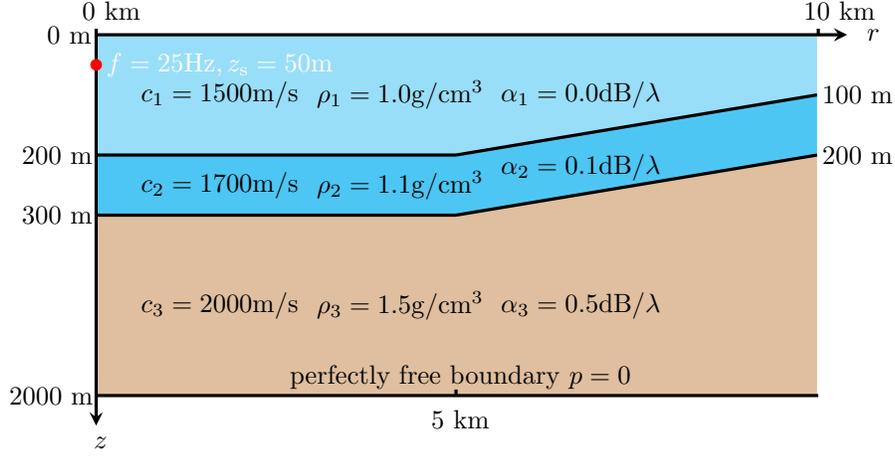
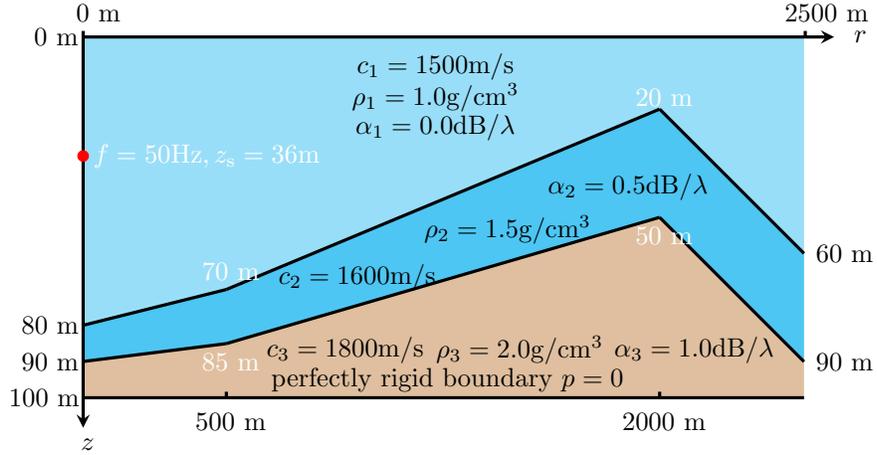

\begin{figure*}
\figline{
\fig{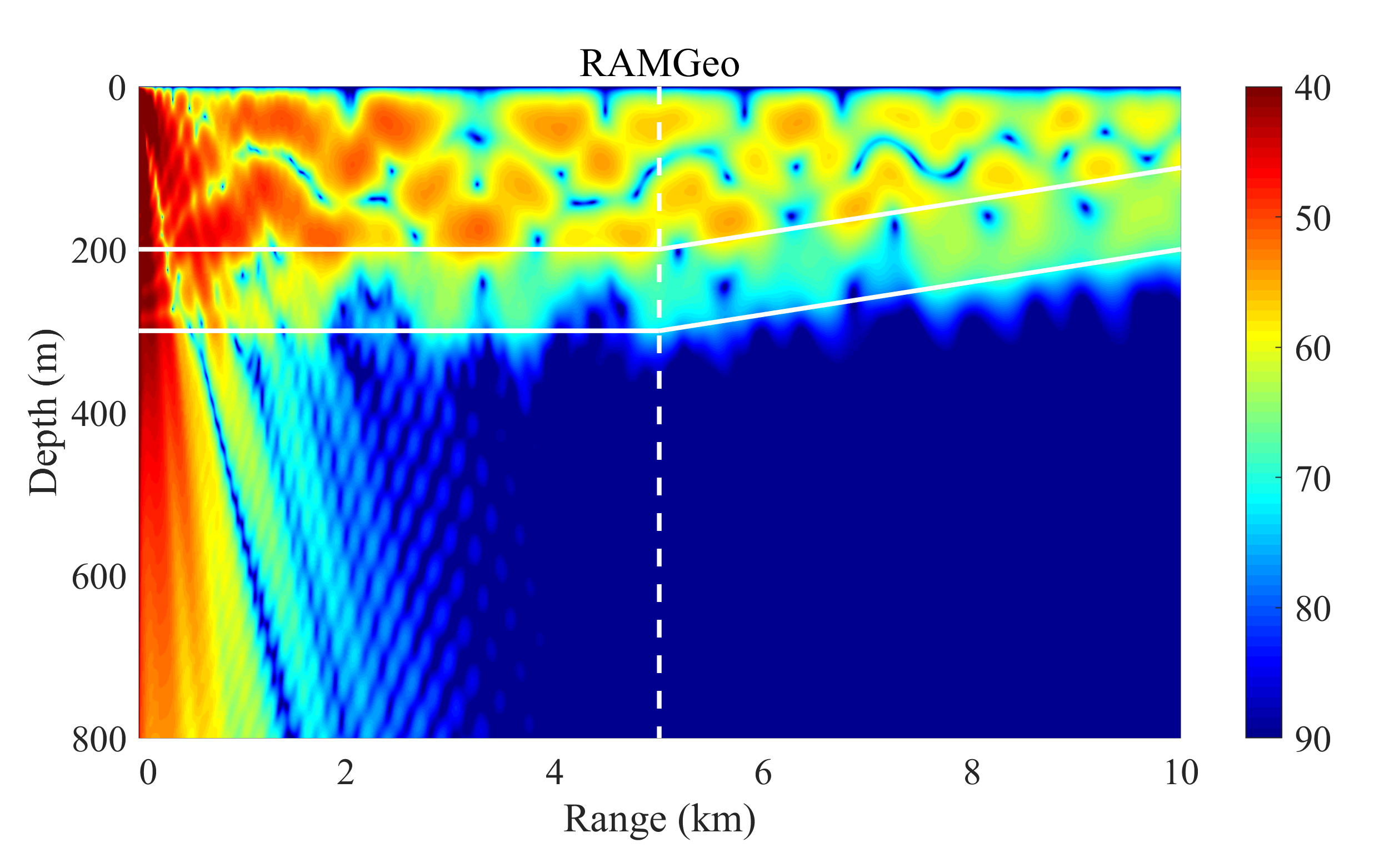}{8cm}{(a)}\label{Figure10a}
\fig{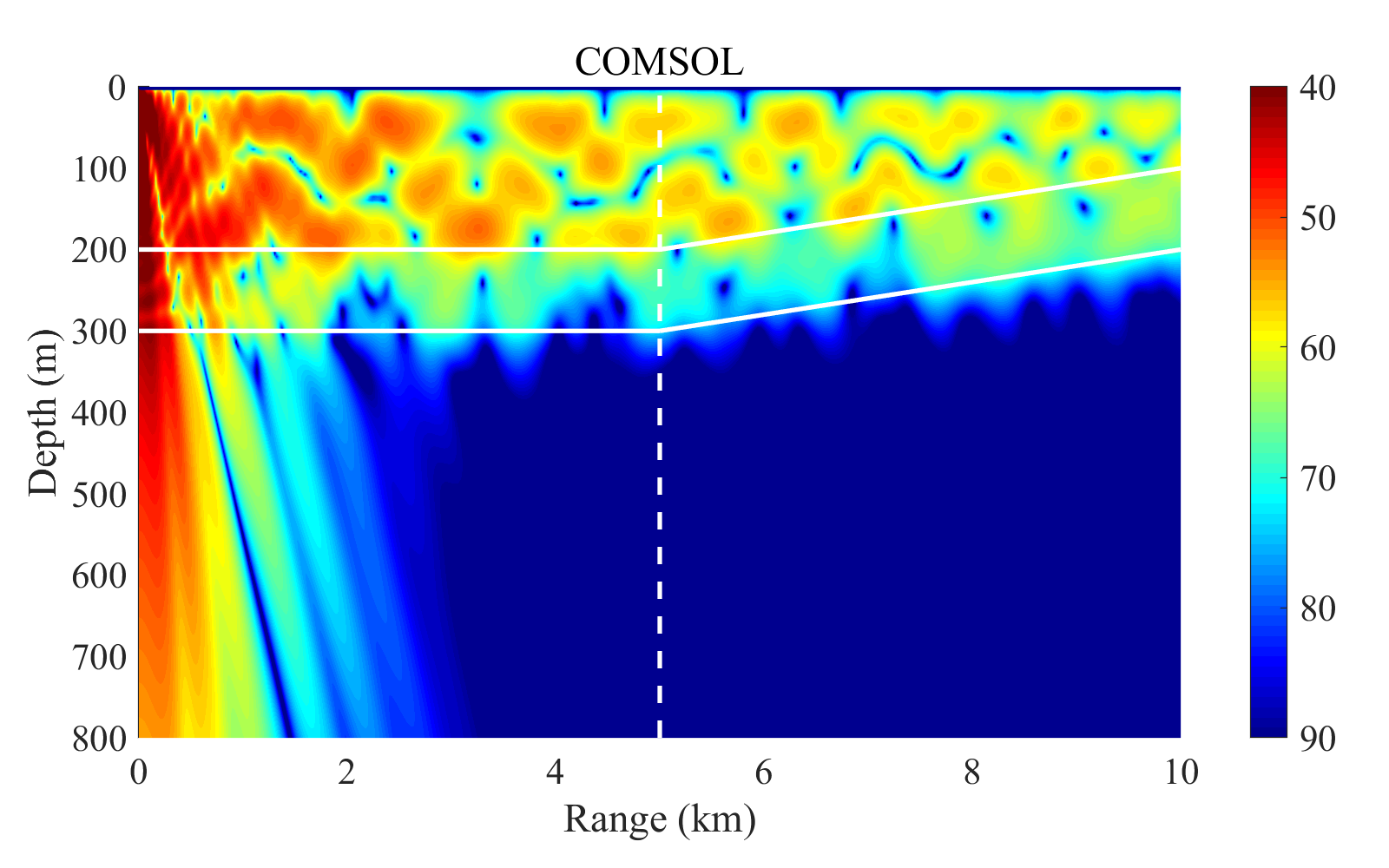}{8cm}{(b)}\label{Figure10b}}
\figline{
\fig{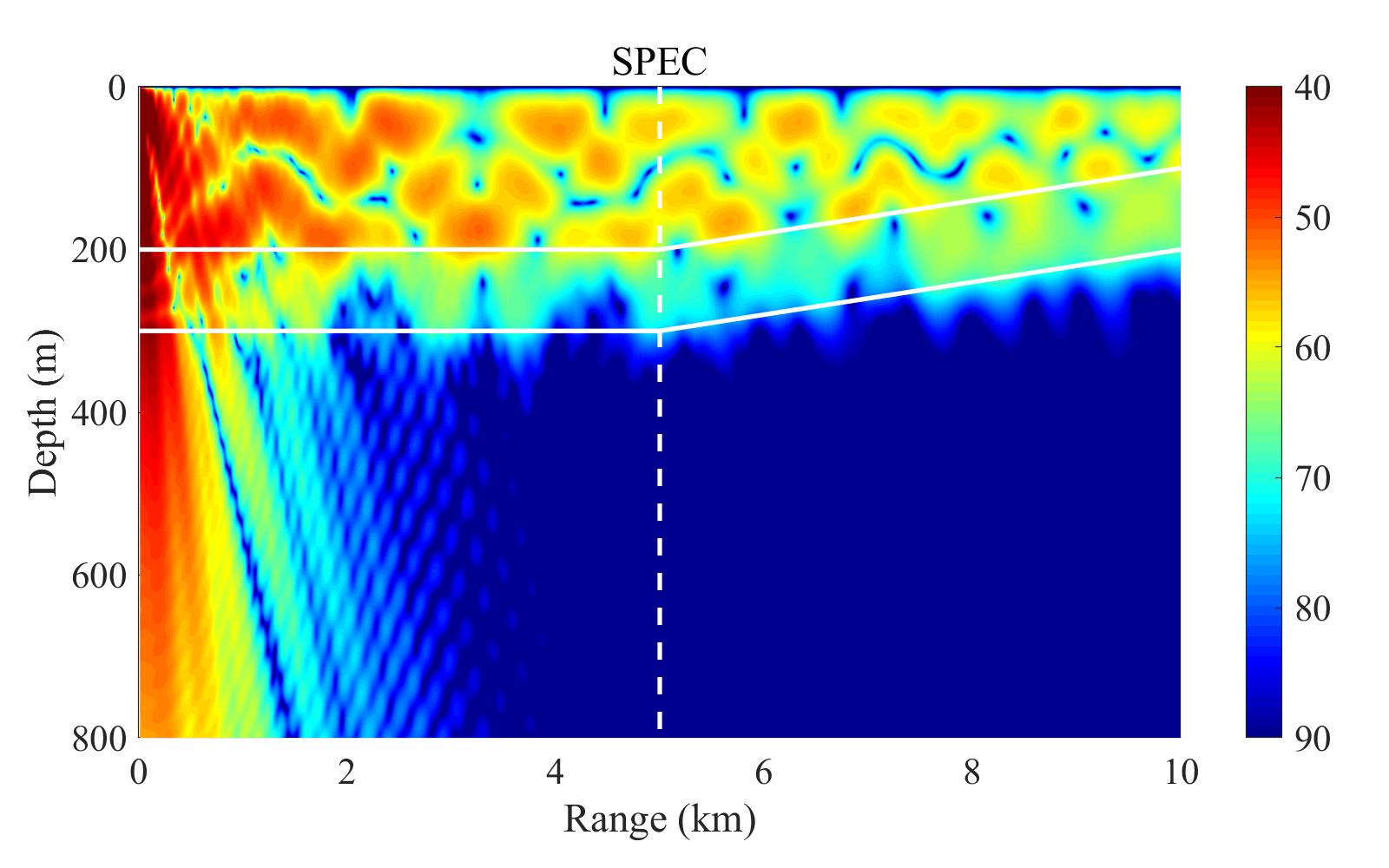}{8cm}{(c)}\label{Figure10c}
\fig{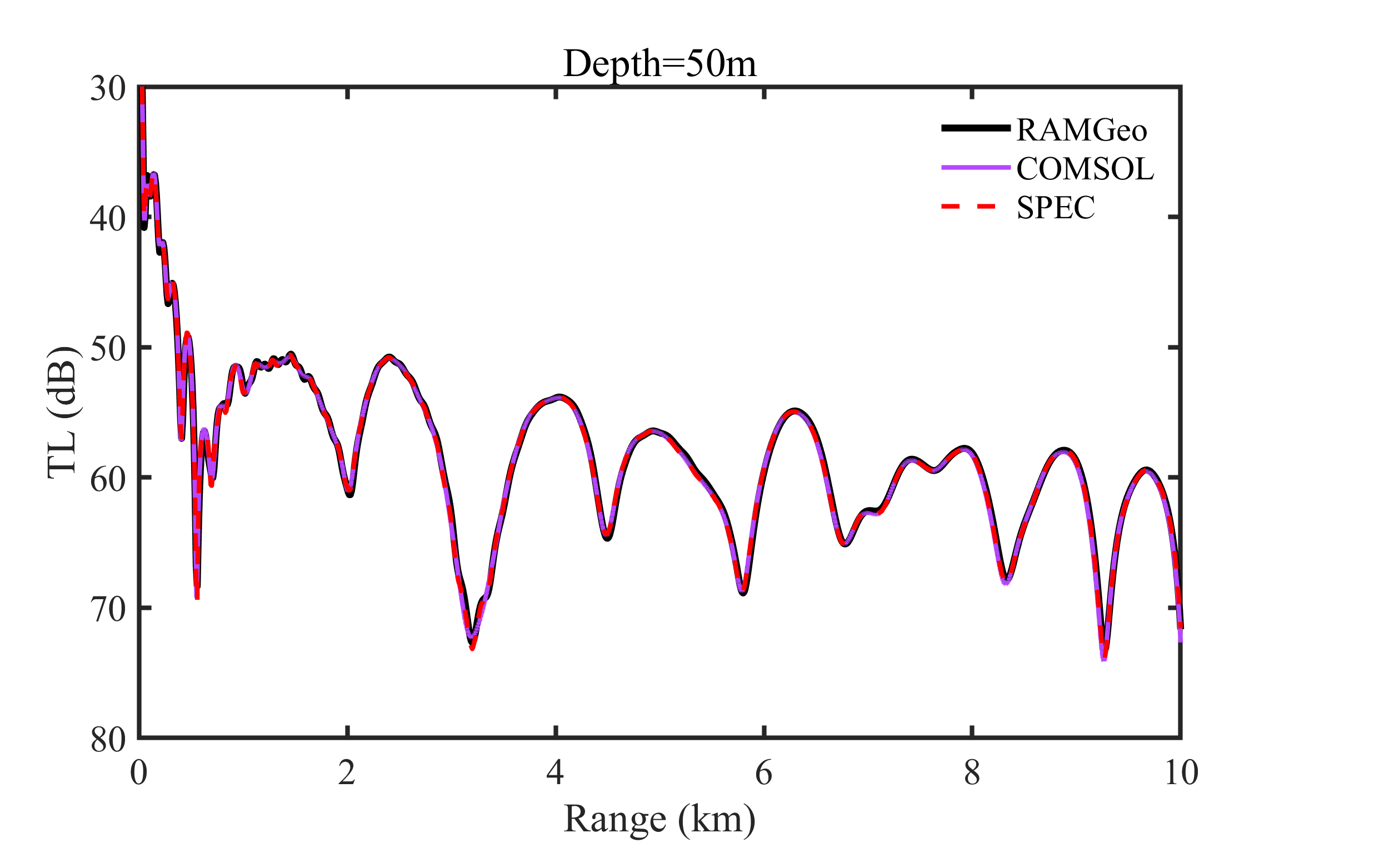}{8cm}{(d)}\label{Figure10d}}
\caption{Sound fields of the multilayer parallel waveguide ($f=25$ Hz, $z_\mathrm{s}=50$ m) calculated by RAMGeo (a), COMSOL (b) and SPEC (c); TL curves at a depth of 50 m (d).}
		\label{Figure10}
\end{figure*}

To demonstrate the capability of the SPEC to simulate range-dependent waveguides in multilayer media, two numerical experiments in \autoref{Figure9} are next considered. RAMGeo can handle multiple sediment layers that parallel the bathymetry. \autoref{Figure9}(a) shows the native example of RAMGeo. Since COUPLE cannot be used to calculate sound propagation in multilayer media, the results of COMSOL are presented here for verification. COMSOL, a commercial numerical simulation platform based on the finite element method, directly solves the Helmholtz equation of underwater acoustic propagation without errors caused by certain model assumptions. In \autoref{Figure10}, great agreement between these three programs is observed, except at certain peaks and troughs. Minor differences between the models can be confidently related to the use of completely different numerical methods and model assumptions.

\subsection{Multilayer undulating waveguide}

\begin{figure*}
\figline{
\fig{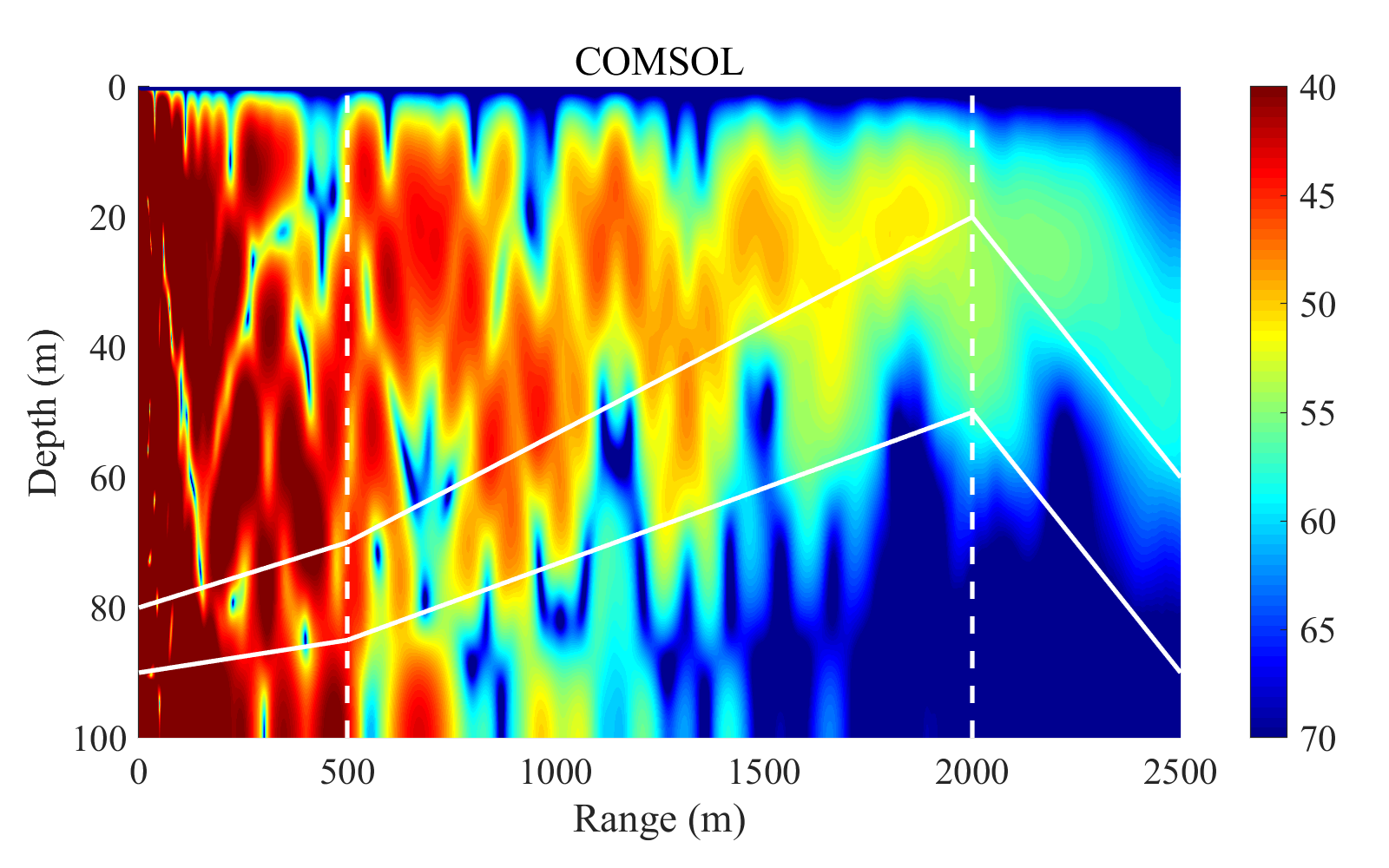}{8cm}{(a)}\label{Figure11a}
\fig{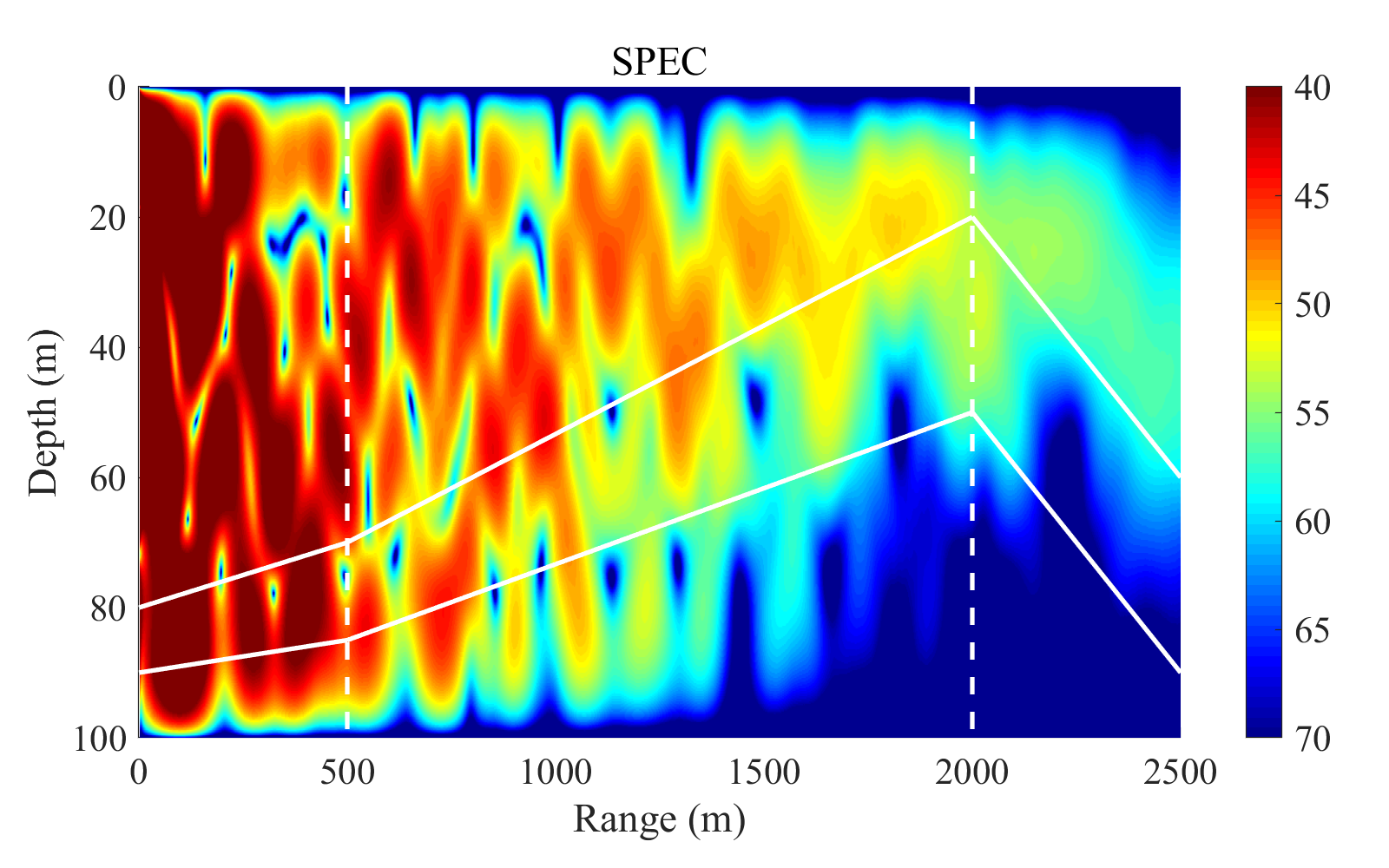}{8cm}{(b)}\label{Figure11b}}
\figline{
\fig{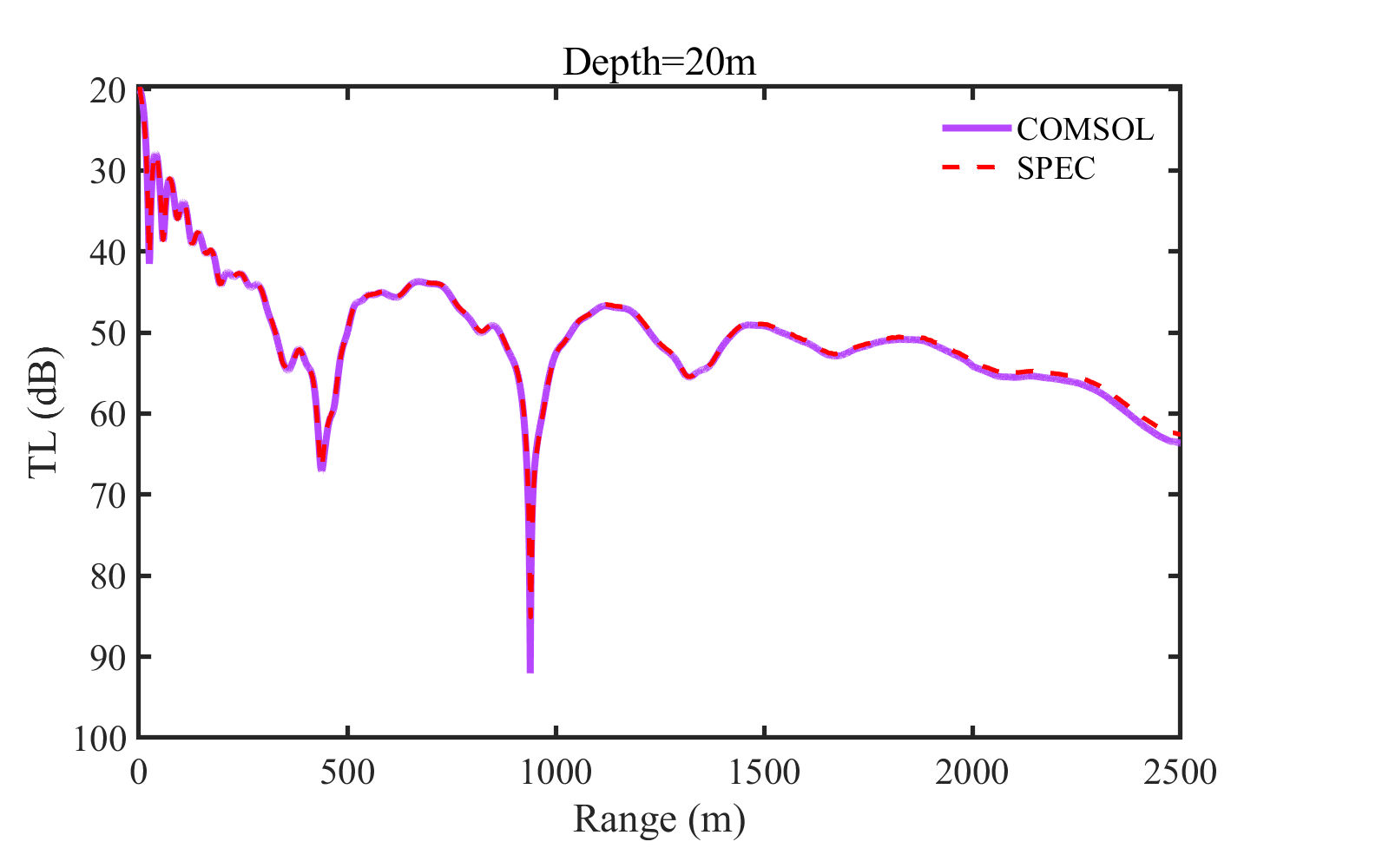}{8cm}{(c)}\label{Figure11c}
\fig{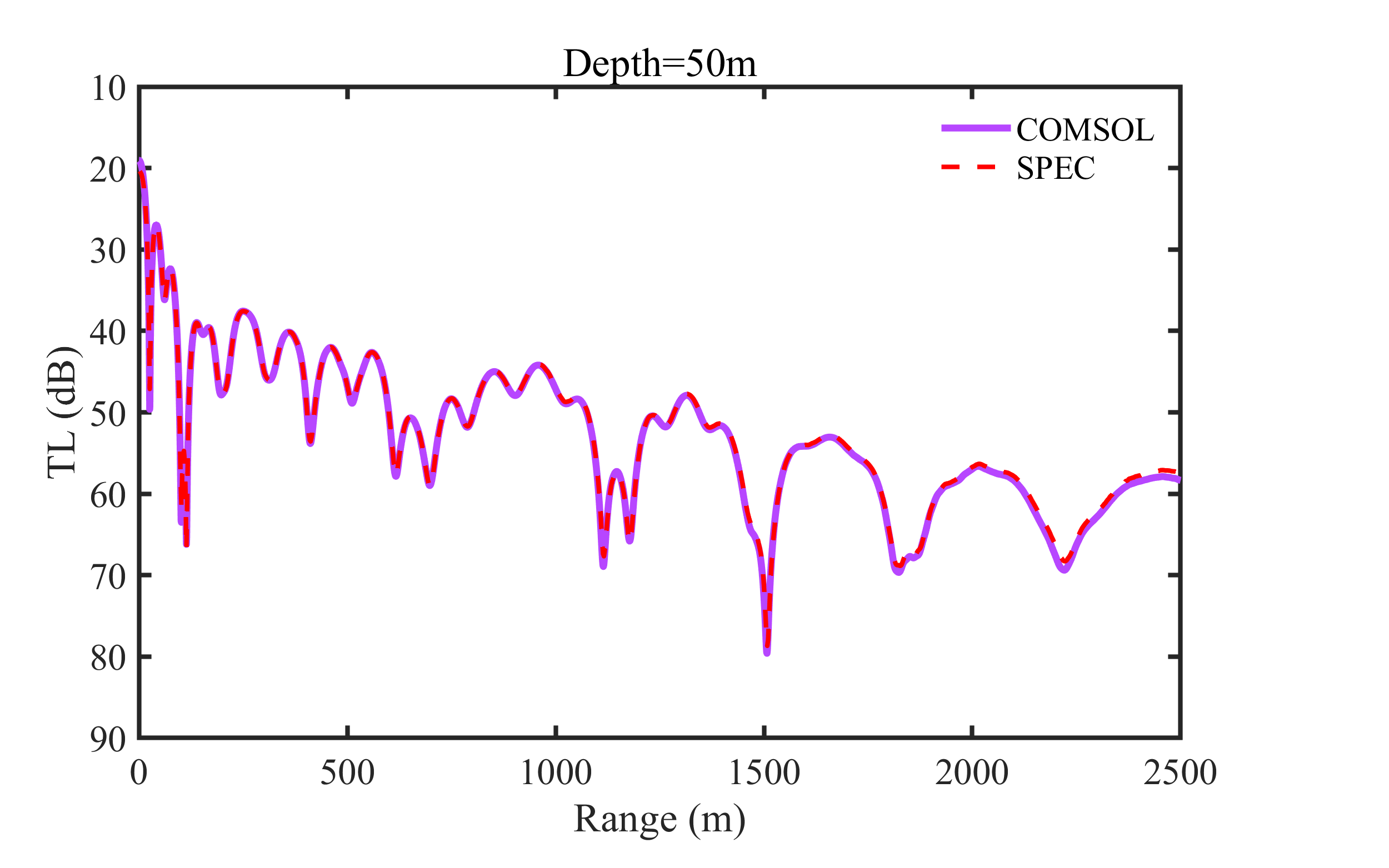}{8cm}{(d)}\label{Figure11d}}
\caption{Sound fields of the multilayer undulating waveguide ($f=50$ Hz, $z_\mathrm{s}=36$ m) calculated by COMSOL (a) and SPEC (b); TL curves at depths of 20 m (c) and 50 m (d).}
		\label{Figure11}
\end{figure*}

\autoref{Figure9}(b) depicts an example of random terrain relief, and the bathymetric nonparallel relief is a good test of the capabilities of the SPEC. \autoref{Figure11} illustrates the sound field and TL curves at different depths calculated by COMSOL and SPEC. The spectral truncated order in each layer adopted by SPEC is 20, and 7 modes are involved in the coupling. The similarities are striking despite small differences in the far field.

These numerical simulations strongly confirm the accuracy of the proposed algorithm and its implementation in this article and fully demonstrate that SPEC can handle these three types of seabed conditions with ease.

\section{Analysis and Parallelization}
	\subsection{Analysis}
To better analyze the computational cost of the algorithm proposed in this paper, Table \ref{tab1} shows the run times of the above examples. The tests were run on the Tianhe--2 supercomputer \cite{Top500}, and a single node of Tianhe--2 was equipped with two Xeon E5 12-core central processing units (CPUs) and 64 GB of shared memory. Each program was run ten times, and the running times listed in the table are the average results. The compiler used was gfortran 7.5.0, and all programs used for comparison were also compiled with this compiler. For the same experiments, under identical configurations, SPEC had a much shorter running time than COUPLE, which directly demonstrates the efficiency of the proposed algorithm.

\begin{table}[htbp]
		\caption{\label{tab1} Comparison among the running times of the numerical experiments (unit: seconds).}
		\begin{ruledtabular}
			\begin{tabular}{lccc}
				Example & SPEC& COUPLE& RAM/RAMGeo \\
				\hline
				Upslope  & 1.757  &13.485 &0.658\\
				Downslope& 2.130  &13.359 &0.687\\
				Seamount & 1.299  &7.722  &0.814\\
				Warm-core eddy& 108.683 &/      &12.157\\
				Multilayer parallel &643.068 &/ &2.224\\
				Multilayer undulating &4.958 &/ &/\\
			\end{tabular}
		\end{ruledtabular}
	\end{table}

From a computational cost perspective, the bulk of the calculations performed by this algorithm is divided into two parts: one part solves $J$ matrix eigenvalue problems for the horizontal wavenumbers and eigenmodes in the range-independent segments (see Eqs.~\eqref{eq:38} and \eqref{eq:43}), and the other part solves the global matrix of linear equations (see Eq.~\eqref{eq:28}). The number of calculations in the first part is high because the eigenvalues and eigenvectors of $J$ square matrices of order $N$ or $2N$ must be determined. The computational effort in the second part is spent on solving a banded sparse linear system of order $(2J-1)\times M$, the size of which depends on the number of segments $J$ and the number of modes to be coupled $M$. In other words, the main computational load of the algorithm is concentrated in the third and fifth steps. The test results in Table \ref{tab2} also support this analysis.
\begin{table}[htbp]
		\caption{\label{tab2} Running times of the two parts in SPEC (unit: seconds).}
		\begin{ruledtabular}
			\begin{tabular}{lccc}
				Example & Step 3 &Step 5 & Total \\
				\hline
				Upslope   & 1.183  &0.504 &1.757 \\
				Downslope & 1.199  &0.465 &2.130 \\
				Seamount  & 0.745  &0.291 &1.299 \\
				Warm-core eddy & 84.432 &15.244 &108.683\\
				Multilayer parallel &410.120 &181.023 &643.068\\
				Multilayer undulating &3.030 &0.996 &4.958\\
			\end{tabular}
		\end{ruledtabular}
	\end{table}

Similarly, the computational load of the COUPLE program is concentrated in these two steps. In terms of solving for the coupling coefficients, COUPLE uses the propagator matrix in Eq.~\eqref{eq:25} to recursively obtain the solution. This method requires solving $(J-1)$ $(2M\times 2M)$-order dense matrix linear equations, and there are many matrix transformation and matrix multiplication operations; this is another aspect of COUPLE that makes it more time-consuming than SPEC. In addition, due to the use of the normalization method in Eq.~\eqref{eq:16}, COUPLE must be segmented in a long-range range-independent region to prevent numerical overflow (such as $r$=0--500 m and $r$=2000--2500 m in \autoref{Figure2}). Since the normal modes of the range-independent region are exactly the same, such segmentation increases the computational cost. In contrast, SPEC uses the normalization in Eq.~\eqref{eq:15b}. It does not overflow, so it does not need to be segmented for range-independent regions, which reduces calculation requirements to a certain extent.

\subsection{Parallelization}
The third and fourth steps of the algorithm are naturally parallel. Therefore, we adopt the idea of multithreaded parallel acceleration and use OpenMP to accelerate SPEC. Table \ref{tab3} shows the effect of the multithreaded acceleration of SPEC. Generally, when 4--8 threads are used, SPEC can achieve a speedup of 3--5. This considerable acceleration effect further reflects the advantages of SPEC in the simulation of large-scale underwater acoustic propagation problems, which is particularly salient because multicore processors have become immensely popular, and it is not expensive to purchase hardware with 4--8 threads for personal computers.
\begin{table}[htbp]
		\caption{\label{tab3} Acceleration effect of SPEC using OpenMP multithreaded parallel computing technology (unit: seconds; the number in brackets is the speedup based on the running time of a single thread).}
		\begin{ruledtabular}
			\begin{tabular}{lrrrrr}
			\multirow{2}{*}{Example}&\multicolumn{4}{c}{Number of Threads}\\
			\cline{2-6}
			&1 &2 &4 &8 &16\\
			\hline
			Upslope &2.044 (1) &1.346 (1.52) &0.783 (2.61)&0.570 (3.59) &0.425 (4.81)\\
			Downslope &2.413 (1)&1.498 (1.61)&0.930 (2.59)&0.615 (3.92) &0.461 (5.23) \\
			Seamount &1.543 (1)&1.008 (1.53)&0.673 (2.29) &0.467 (3.30) &0.376 (4.10)\\
			Warm-core eddy &122.112 (1)&65.819 (1.86) &35.166 (3.47)&21.504 (5.68)&21.373 (5.71)\\
			Multilayer parallel &762.856 (1)&399.807 (1.92)&218.651 (3.52) &128.275 (5.99)&78.159 (9.84)\\
			Multilayer undulating &5.572 (1)&3.067 (1.82)&1.811 (3.08) &1.180 (4.72)&0.818 (6.82)\\
		\end{tabular}
		\end{ruledtabular}
	\end{table}

\begin{figure*}
\figline{
\fig{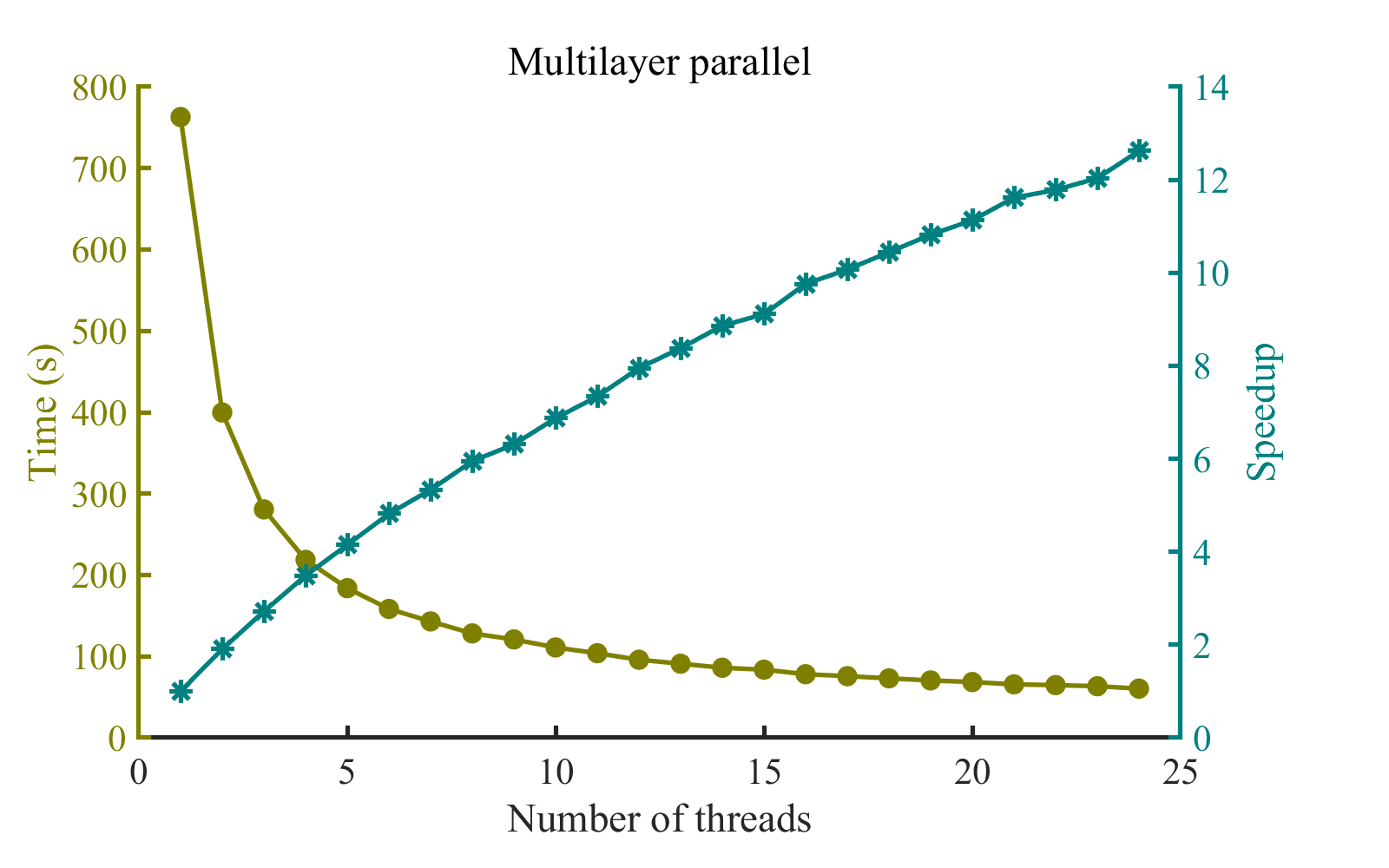}{8cm}{(a)}\label{Figure12a}
\fig{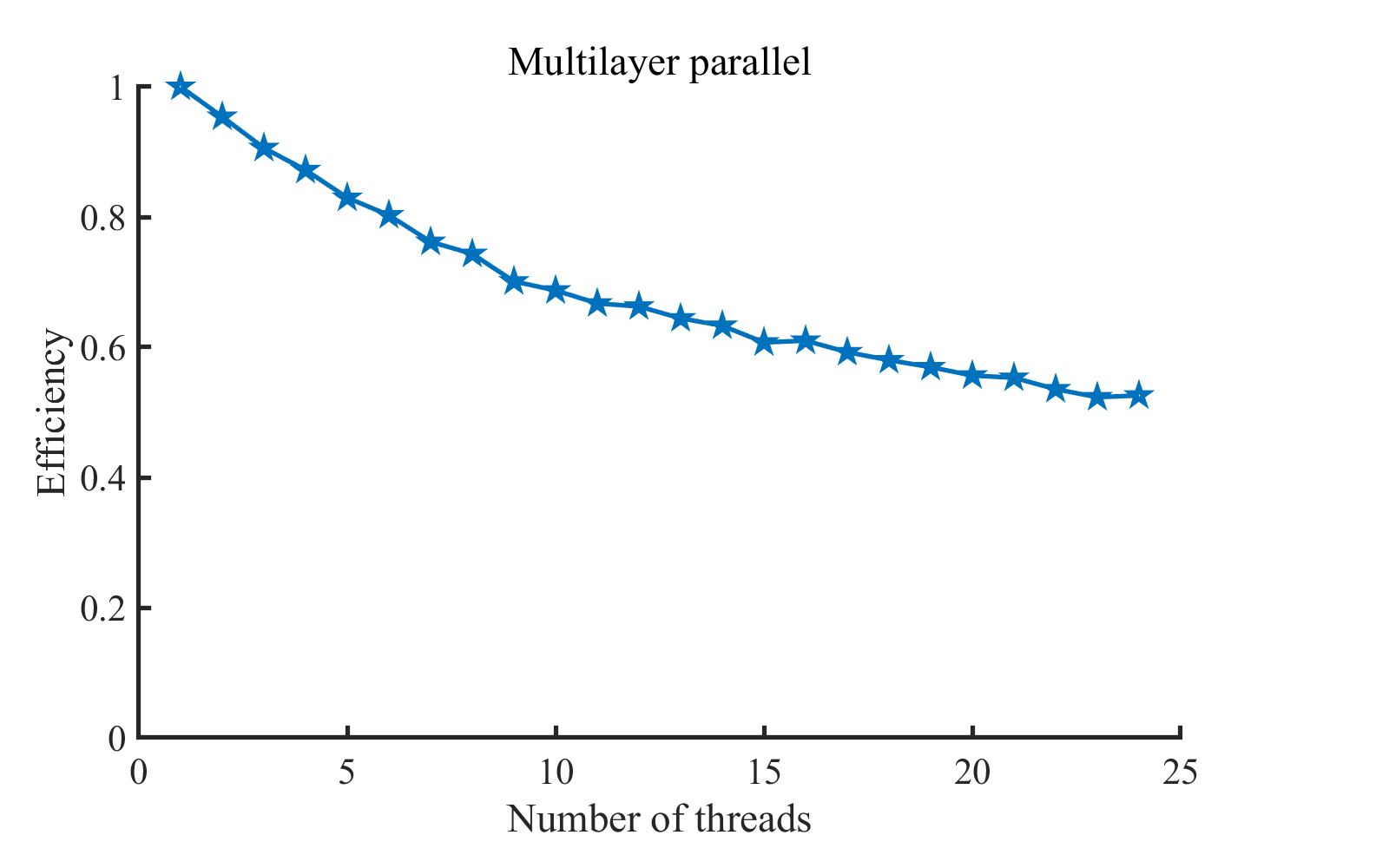}{8cm}{(b)}\label{Figure12b}}
\figline{
\fig{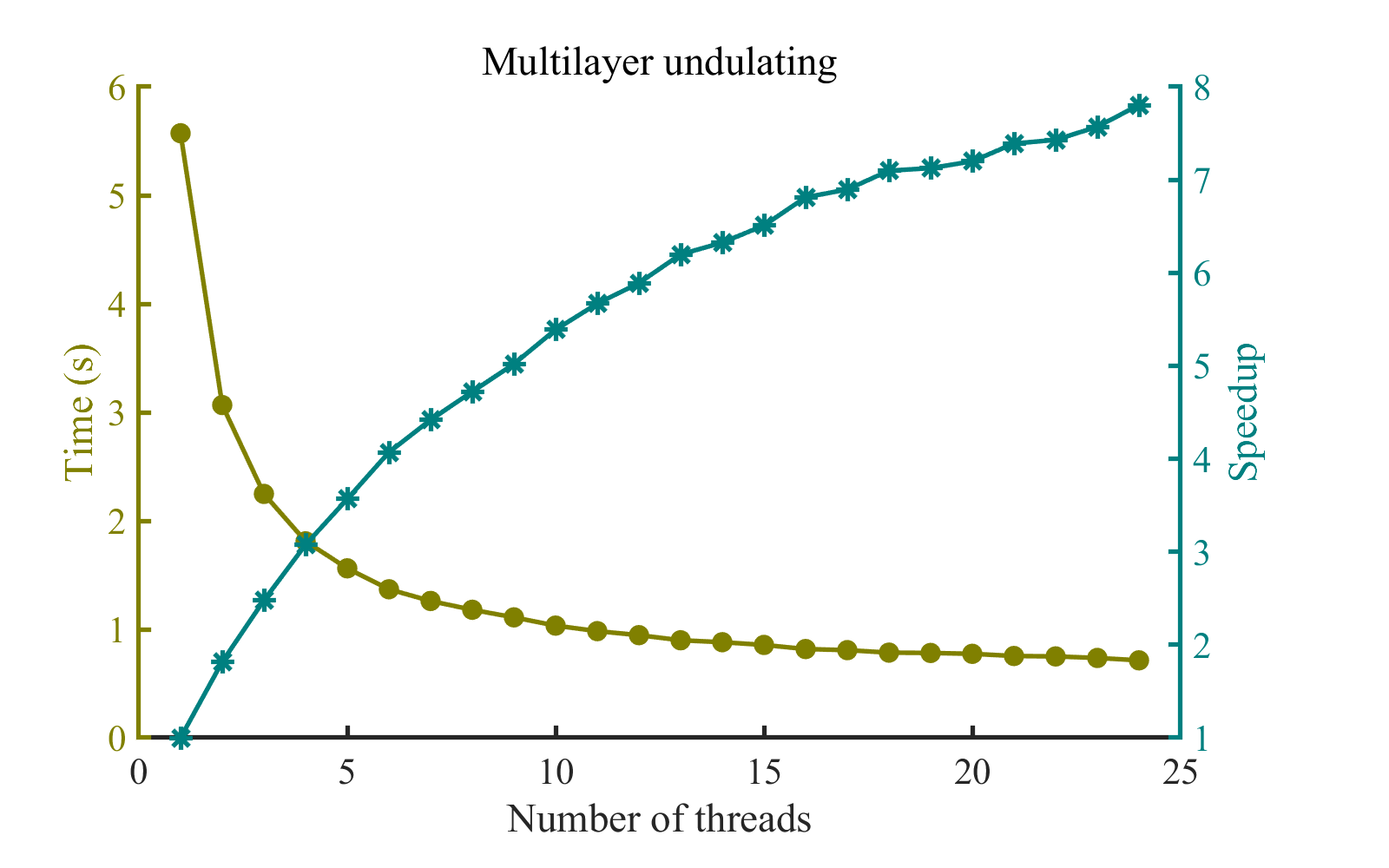}{8cm}{(c)}\label{Figure12c}
\fig{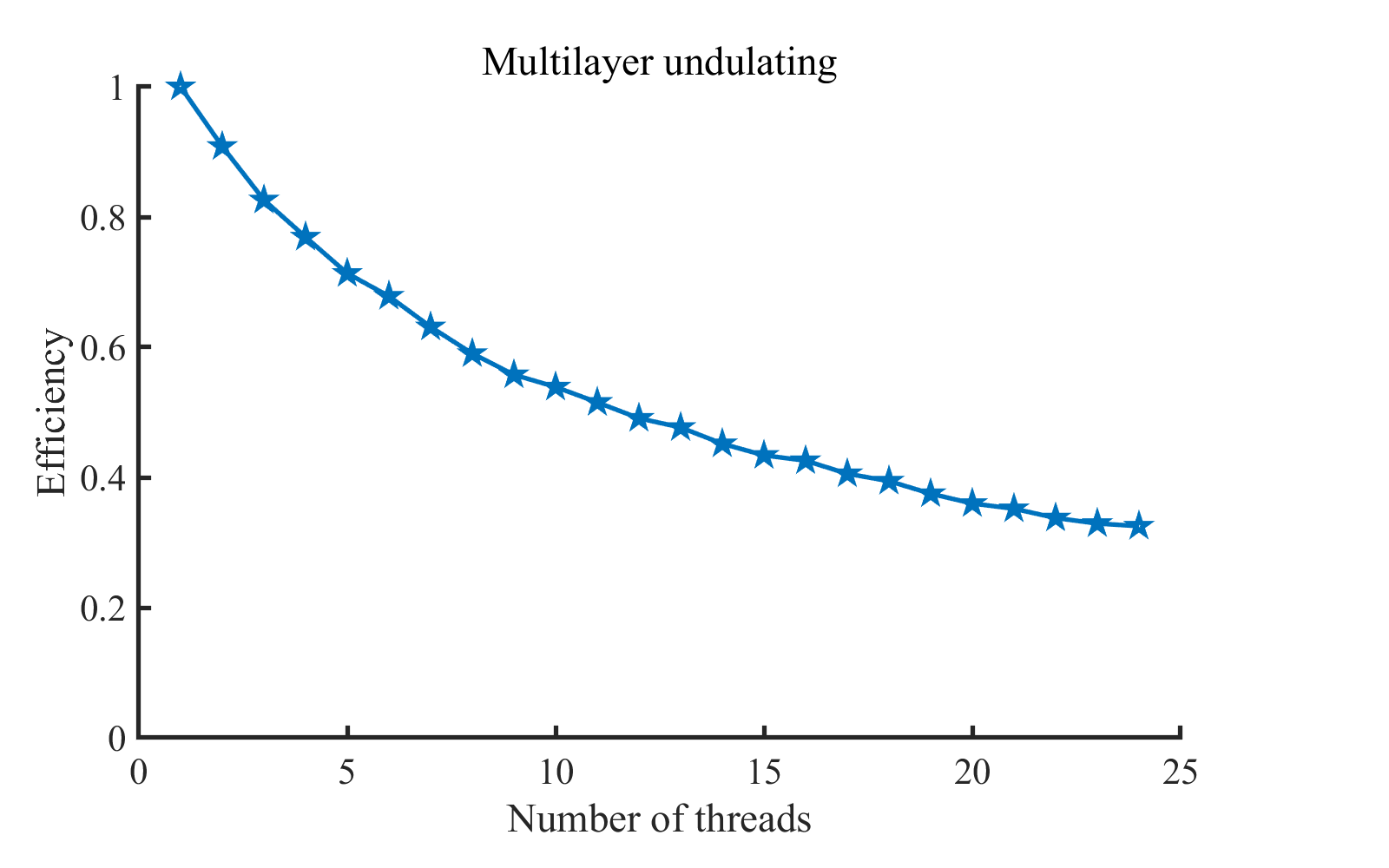}{8cm}{(d)}\label{Figure12d}}
\caption{Runtime/speedup (a) and parallel efficiency (b) for the multilayer parallel waveguide; runtime/speedup (c) and parallel efficiency (d) for the multilayer undulating waveguide.}
\label{Figure12}
\end{figure*}

In addition to runtime/speedup, a more common metric that can be used to measure parallel program performance is scalability. For parallel programs, scalability is well defined. If the fixed efficiency can be maintained when the number of threads is increased without increasing the size of the problem, then the program is strongly scalable. If the number of threads is increased, meaning that the efficiency value can be maintained only by increasing the problem size at the same rate, then the program is weakly scalable. For multithreaded parallel programs that use shared memory, strong scalability is more of a concern because the memory resources of the hardware are fixed, making it difficult to increase the resources to accommodate larger problems. Without loss of generality, we take the last two numerical experiments as examples to test and analyze the strong scalability of the SPEC program. The results of running time, speedup and speedup efficiency for the last two experiments for a fixed problem size are presented in \autoref{Figure12}. The acceleration effect of SPEC is the most significant when the number of threads is initially increased. As the number of threads increases, although the running time is still decreasing, the efficiency gradually decreases. According to Amdahl's law \cite{Amdahl1967}, ideally, the limit of the parallel speedup depends on the proportion of the parallelizable part of the program to the total program computation. In the SPEC program, the calculation of the coupling coefficients cannot be fully parallelized, which is why the multithread parallelism of SPEC has a ceiling. In addition, more threads necessitate more overhead to create threads.

The abovementioned scalability tests are performed under the condition that the running memory is compatible with the memory of Tianhe--2's single node. Furthermore, configurations with strong range dependencies that require a large number of segments undoubtedly result in very large systems of equations, which are almost impossible to process on consumer-grade CPUs, as large-scale matrix manipulations can make memory a bottleneck. However, for workstations with more memory and better single-core node performance, the conclusion of the above scalability test always holds true.

\section{Remarks and Summary}
\subsection{Remarks}
In general, the main contributions and highlights of the devised SPEC program developed based on the algorithm are as follows:
\begin{enumerate}
\item
The improved global matrix coupled-mode algorithm is implemented as a robust procedure. The improved range normalization is unconditionally stable; thus, SPEC does not experience the problem of numerical overflow.

\item
The global matrix formed during the calculation exhibits good sparsity and a banded shape, so SPEC can efficiently solve for the coupling coefficients. In addition, due to the existence of natural parallelism, SPEC can be easily run in parallel and achieve excellent acceleration effects.

\item
SPEC can be used to assess acoustic propagation in multilayer arbitrarily undulating media, so it can flexibly simulate more complicated and realistic ocean acoustic waveguides.

\item
The Chebyshev--Tau spectral method can accurately find the eigenmodes and eigenvalues for waveguides over a half-space bottom without using an iterative root-finding algorithm. Therefore, SPEC does not exhibit the iterative divergence problem caused by poor initial guesses in root-finding algorithms. At present, mature normal mode programs that have this advantage are virtually nonexistent.

\end{enumerate}

In terms of computational accuracy, SPEC is mainly controlled by the spectral truncated orders ($N_\ell$) and the number of horizontal segments ($J$). The former determines the accuracy of modal information in the segments, and the latter determines the accuracy of coupling coefficients. In terms of computational speed, the performance of SPEC varies with source frequency and ocean environment. For high-frequency/deep-sea waveguides, the number of modes $M$ increases, and SPEC requires more spectral truncated orders (meaning larger-scale matrix eigenvalue problems) to solve for local modes.

Note that compared with the KRAKEN program based on the finite difference method, the SPEC program based on the Chebyshev--Tau spectral method has no absolute advantage in solving for the local modes of high-frequency sound sources. KRAKEN is slow at low frequencies because it has considerable overhead. It is faster than the Chebyshev method at high frequencies because its matrices are sparse and easy to solve, whereas the Chebyshev matrices are full rank and take longer at higher frequencies. Therefore, for specific simulations, the two have a crossover point in this regard. High frequencies not only increase the number of modes $M$ but also increase the number of segments $J$, so it is necessary to obtain the modal information for more segments. The growth of $J$ and $M$ usually causes the global matrix size to skyrocket, resulting in a larger computational cost, which is also the most formidable limitation of coupled-mode theory.

\subsection{Summary}
In this article, we propose a new numerical algorithm for range-dependent waveguides in ocean acoustics. An improved global matrix of coupled modes is used to solve for the range dependence of the ocean environment, and the Chebyshev--Tau spectral method is used to solve for the normal modes in stepwise range-independent segments. Numerical simulations involving various range dependencies in deep and shallow ocean environments verified that our devised algorithm is reliable, practical, and efficient for range-dependent waveguides. Due to the natural parallelism of the main steps of the algorithm, we also leverage parallel computing technology to further accelerate the algorithm. At present, the algorithm is both comprehensive and efficient. SPEC compares well in accuracy to COUPLE, and its performance and capability are better than those of COUPLE. To a certain extent, it can be regarded as a modernized algorithm and replacement of COUPLE.

Due to the inherent limitations of coupled modes, this algorithm is still more computationally expensive than parabolic approximations (such as RAM) and ray models (such as Bellhop) in most cases. Therefore, it is valuable to further optimize the SPEC program for high-frequency deep-sea long-range waveguides.

\begin{acknowledgments}
We are very grateful to Michael B. Porter for providing the warm-core eddy sound speed profile data. We are also very grateful to Prof. William L. Siegmann for his critical comments on the first submission, which greatly broadened our academic horizons.

This work was supported by the National Natural Science Foundation of China [grant number 61972406] and the National Key Research and Development Program of China [grant number 2016YFC1401800].
\end{acknowledgments}

\appendix
\section{Group speeds of modes}
The Chebyshev--Tau spectral method can calculate the group speeds $c_g$ of the modes via Rayleigh's method (see Eq.~(5.189) in Ref.~\cite{Jensen2011}).
\begin{equation}
			\frac{1}{c_g} = \frac{\mathrm{d} k_{r}}{\mathrm{~d} \omega}=\frac{\omega}{k_{r}} \int_{0}^{H} \frac{\Psi^{2}(z)}{\rho(z) c^{2}(z)} \mathrm{d} z
		\end{equation}
Let the phase velocity $c_p=\omega/k_r$; thus:
\begin{equation}
			\frac{1}{c_pc_g} = \frac{\int_{0}^{H} \frac{\Psi^{2}(z)}{\rho(z) c^{2}(z)} \mathrm{d} z}{\int_{0}^{H} \Psi^{2}(z) \mathrm{d} z} = \frac{\int_{-1}^{1} \frac{\Psi^{2}(x)}{\rho(x) c^{2}(x)} \mathrm{d} x}{\int_{-1}^{1} \Psi^{2}(x) \mathrm{d} x}
		\end{equation}
On the Chebyshev spectral space, the integral operation satisfies the transformation relationship in Eq.~\eqref{eq:32c}. Therefore, the group speed can be calculated by the following equation:
\begin{equation}
			\frac{1}{c_pc_g} =\frac{\mathbf{I}_N 	\mathbf{C}_{1/\rho}\mathbf{C}_{1/c^2} \mathbf{C}_\Psi \mathbf{\Psi}}{\mathbf{I}_N\mathbf{C}_\Psi \mathbf{\Psi}}
		\end{equation}


\begin{thebibliography}{10}
\def\enquote#1,{``#1,''}
\def\enxquote#1{``#1''}
\expandafter\ifx\csname url\endcsname\relax
  \def\url#1{\texttt{#1}}\fi
\expandafter\ifx\csname urlprefix\endcsname\relax\def\urlprefix{URL }\fi
\providecommand{\bibinfo}[2]{#2}
\def\plainquote#1{``#1''}
\providecommand{\noopsort}[1]{}
\providecommand{\switchargs}[2]{#2#1}
\providecommand{\dourl}[1]{\href{http://#1}{\nolinkurl{#1}}}
  \def\eatspace #1{#1}

\bibitem{Jensen2011}
\bibinfo{author}{F.~B. Jensen}, \bibinfo{author}{W.~A. Kuperman},
  \bibinfo{author}{M.~B. Porter}, and \bibinfo{author}{H.~Schmidt},
  \emph{\bibinfo{title}{Computational Ocean Acoustics}}
  (\bibinfo{publisher}{Springer-Verlag}, \bibinfo{address}{New York},
  \bibinfo{year}{2011}).

\bibitem{Etter2018}
\bibinfo{author}{P.~C. Etter}, \emph{\bibinfo{title}{Underwater Acoustic
  Modeling and Simulation}}  (\bibinfo{publisher}{CRC Press},
  \bibinfo{address}{Boca Raton, USA}, \bibinfo{year}{2018}).

\bibitem{RAM}
\bibinfo{author}{M.~D. Collins}, \enxquote{\bibinfo{title}{User's guide for
  {RAM} versions 1.0 and 1.0p}}  (\bibinfo{year}{1999}),
  \dourl{https://oalib-acoustics.org/models-and-software/parabolic-equation/}.

\bibitem{Liuw2021}
\bibinfo{author}{W.~Liu}, \bibinfo{author}{L.~Zhang},
  \bibinfo{author}{W.~Wang}, \bibinfo{author}{Y.~Wang},
  \bibinfo{author}{S.~Ma}, \bibinfo{author}{X.~Cheng}, and
  \bibinfo{author}{W.~Xiao}, \enquote{\bibinfo{title}{A three-dimensional
  finite difference model for ocean acoustic propagation and benchmarking for
  topographic effects}},  \bibinfo{journal}{The Journal of the Acoustical
  Society of America} \textbf{150}(2), \bibinfo{pages}{1140--1156}
  (\bibinfo{year}{2021}) \dodoi{10.1121/10.0005853}.

\bibitem{Murphy1988a}
\bibinfo{author}{J.~E. Murphy} and \bibinfo{author}{S.~A. Chin-Bing},
  \enquote{\bibinfo{title}{A finite-element model for ocean acoustic
  propagation}},  \bibinfo{journal}{Mathematical and computer modelling}
  \textbf{11}(C), \bibinfo{pages}{70--74} (\bibinfo{year}{1988})
  \dodoi{10.1016/0895-7177(88)90457-8}.

\bibitem{Murphy1988b}
\bibinfo{author}{S.~A. Chin-Bing}, \enquote{\bibinfo{title}{Long-range,
  range-dependent, acoustic propagation simulation using a full-wave,
  finite-element model coupled with a one-way parabolic equation model}},
  \bibinfo{journal}{The Journal of the Acoustical Society of America}
  \textbf{84}(S1) (\bibinfo{year}{1988}) \dodoi{10.1121/1.2026549}.

\bibitem{Murphy1989}
\bibinfo{author}{J.~E. Murphy} and \bibinfo{author}{S.~A. Chin-Bing},
  \enquote{\bibinfo{title}{A finite-element model for ocean acoustic
  propagation and scattering}},  \bibinfo{journal}{The Journal of the
  Acoustical Society of America} \textbf{86}(4), \bibinfo{pages}{1478--1483}
  (\bibinfo{year}{1989}) \dodoi{10.1121/1.398708}.

\bibitem{Murphy1996}
\bibinfo{author}{J.~E. Murphy}, \bibinfo{author}{G.~Li}, \bibinfo{author}{S.~A.
  Chin-Bing}, and \bibinfo{author}{D.~B. King},
  \enquote{\bibinfo{title}{Multifilament source model for short-range
  underwater acoustic problems involving penetrable ocean bottoms}},
  \bibinfo{journal}{The Journal of the Acoustical Society of America}
  \textbf{99}(2), \bibinfo{pages}{845--850} (\bibinfo{year}{1996})
  \dodoi{10.1121/1.414660}.

\bibitem{Pekeris1948}
\bibinfo{author}{C.~L. Pekeris}, \enquote{\bibinfo{title}{Theory of propagation
  of explosive sound in shallow water}},  \bibinfo{journal}{Geological Society
  of America Memoirs} \textbf{27}(1), \bibinfo{pages}{1--117}
  (\bibinfo{year}{1948}) \dodoi{10.1130/mem27-2-p1}.

\bibitem{Pierce1954}
\bibinfo{author}{J.~R. Pierce}, \enquote{\bibinfo{title}{Coupling of modes of
  propagation}},  \bibinfo{journal}{Journal of Applied Physics} \textbf{25}(2),
  \bibinfo{pages}{179--183} (\bibinfo{year}{1954}) \dodoi{10.1063/1.1721599}.

\bibitem{Miller1954}
\bibinfo{author}{S.~E. Miller}, \enquote{\bibinfo{title}{Coupled wave theory
  and waveguide applications}},  \bibinfo{journal}{The Bell System Technical
  Journal} \textbf{33}(3), \bibinfo{pages}{661--719} (\bibinfo{year}{1954})
  \dodoi{10.1002/j.1538-7305.1954.tb02359.x}.

\bibitem{Rutherford1981}
\bibinfo{author}{S.~R. Rutherford} and \bibinfo{author}{K.~E. Hawker},
  \enquote{\bibinfo{title}{Consisted coupled mode theory of sound propagation
  for a class of nonseparable problems}},  \bibinfo{journal}{The Journal of the
  Acoustical Society of America} \textbf{70}(2), \bibinfo{pages}{554--564}
  (\bibinfo{year}{1981}) \dodoi{10.1121/1.386744}.

\bibitem{Fawcett1992}
\bibinfo{author}{J.~A. Fawcett}, \enquote{\bibinfo{title}{A derivation of the
  differential equations of coupled-mode propagation}},  \bibinfo{journal}{The
  Journal of the Acoustical Society of America} \textbf{92}(1),
  \bibinfo{pages}{290--295} (\bibinfo{year}{1992}) \dodoi{10.1121/1.404293}.

\bibitem{Evans1983}
\bibinfo{author}{R.~B. Evans}, \enquote{\bibinfo{title}{A coupled mode solution
  for acoustic propagation in a waveguide with stepwise depth variations of a
  penetrable bottom}},  \bibinfo{journal}{The Journal of the Acoustical Society
  of America} \textbf{74}, \bibinfo{pages}{188--195} (\bibinfo{year}{1983})
  \dodoi{10.1121/1.389707}.

\bibitem{Mattheij1985}
\bibinfo{author}{R.~M.~M. Mattheij}, \enquote{\bibinfo{title}{Decoupling and
  stability of algorithms for boundary value problems}},
  \bibinfo{journal}{SIAM Review} \textbf{27}(1), \bibinfo{pages}{1--44}
  (\bibinfo{year}{1985}) \dodoi{10.1137/1027001}.

\bibitem{Evans1986}
\bibinfo{author}{R.~B. Evans}, \enquote{\bibinfo{title}{The decoupling of
  stepwise coupled modes}},  \bibinfo{journal}{The Journal of the Acoustical
  Society of America} \textbf{80}, \bibinfo{pages}{1414--1418}
  (\bibinfo{year}{1986}) \dodoi{10.1121/1.394395}.

\bibitem{Couple}
\bibinfo{author}{R.~B. Evans},
  \enxquote{\bibinfo{title}{{\href{https://oalib-acoustics.org/Modes/index.html}{COUPLE}}:
  A coupled normal-mode code {(Fortran)}}}  (\bibinfo{year}{2007}),
  \dourl{https://oalib-acoustics.org/models-and-software/normal-modes/}.

\bibitem{Luowy2012a}
\bibinfo{author}{W.~Luo}, \bibinfo{author}{C.~Yang}, \bibinfo{author}{J.~Qin},
  and \bibinfo{author}{R.~Zhang}, \enquote{\bibinfo{title}{A numerically stable
  coupled-mode formulation for acoustic propagation in range-dependent
  waveguides}},  \bibinfo{journal}{Science China, Physics, Mechanics and
  Astronomy} \textbf{55}(4), \bibinfo{pages}{572--588} (\bibinfo{year}{2012})
  \dodoi{10.1007/s11433-012-4666-0}.

\bibitem{Luowy2012b}
\bibinfo{author}{W.~Luo}, \bibinfo{author}{C.~Yang}, \bibinfo{author}{J.~Qin},
  and \bibinfo{author}{R.~Zhang}, \enquote{\bibinfo{title}{A coupled-mode
  solution for sound propagation in range-dependent waveguides}}, in
  \emph{\bibinfo{booktitle}{AIP Conference Proceedings}},
  \bibinfo{publisher}{American Institute of Physics} (\bibinfo{year}{2012}),
  Vol. \bibinfo{volume}{1495}, pp. \bibinfo{pages}{313--320},
  \dodoi{10.1063/1.4765924}.

\bibitem{Luowy2012c}
\bibinfo{author}{W.~Luo}, \bibinfo{author}{C.~Yang}, \bibinfo{author}{J.~Qin},
  and \bibinfo{author}{R.~Zhang}, \enquote{\bibinfo{title}{Generalized
  coupled-mode formulation for sound propagation in range-dependent
  waveguides}},  \bibinfo{journal}{Chinese Physics Letters} \textbf{29}(1),
  \bibinfo{pages}{1--4} (\bibinfo{year}{2012})
  \dodoi{10.1088/0256-307X/29/1/014302}.

\bibitem{Luowy2012d}
\bibinfo{author}{W.~Luo}, \bibinfo{author}{C.~Yang}, \bibinfo{author}{J.~Qin},
  and \bibinfo{author}{R.~Zhang}, \enquote{\bibinfo{title}{Sound propagation in
  a wedge with a rigid bottom}},  \bibinfo{journal}{Chinese Physics Letters}
  \textbf{29}(10), \bibinfo{pages}{1--4} (\bibinfo{year}{2012})
  \dodoi{10.1088/0256-307X/29/10/104303}.

\bibitem{Yangcm2012}
\bibinfo{author}{C.~Yang}, \bibinfo{author}{W.~Luo}, and
  \bibinfo{author}{R.~Zhang}, \enquote{\bibinfo{title}{A coupled-mode method
  based on direct global matrix approach in range-dependent waveguides}},
  \bibinfo{journal}{Acta Acustica (in Chinese)} \textbf{37}(5),
  \bibinfo{pages}{465--474} (\bibinfo{year}{2012})
  \dodoi{10.15949/j.cnki.0371-0025.2012.05.001}.

\bibitem{Yangcm2015a}
\bibinfo{author}{C.~Yang}, \bibinfo{author}{W.~Luo},
  \bibinfo{author}{R.~Zhang}, \bibinfo{author}{L.~Lyu}, and
  \bibinfo{author}{F.~Qiao}, \enquote{\bibinfo{title}{An efficient coupled-mode
  formulation for acoustic propagation in inhomogeneous waveguides}},
  \bibinfo{journal}{Journal of Computational Acoustics} \textbf{23}(1550019),
  \bibinfo{pages}{1--18} (\bibinfo{year}{2015})
  \dodoi{10.1142/S0218396X15500198}.

\bibitem{Kraken2001}
\bibinfo{author}{M.~B. Porter}, \emph{\bibinfo{title}{The Kraken Normal Mode
  Program}}  (\bibinfo{publisher}{{SACLANT} Undersea Research Centre},
  \bibinfo{year}{2001}),
  \dourl{https://oalib-acoustics.org/models-and-software/normal-modes/}.

\bibitem{Tuhw2020a}
\bibinfo{author}{H.~Tu}, \bibinfo{author}{Y.~Wang}, \bibinfo{author}{W.~Liu},
  \bibinfo{author}{X.~Ma}, \bibinfo{author}{W.~Xiao}, and
  \bibinfo{author}{Q.~Lan}, \enquote{\bibinfo{title}{A {Chebyshev} spectral
  method for normal mode and parabolic equation models in underwater
  acoustics}},  \bibinfo{journal}{Mathematical Problems in Engineering}
  \bibinfo{pages}{7461314} (\bibinfo{year}{2020}) \dodoi{10.1155/2020/7461314}.

\bibitem{Tuhw2021a}
\bibinfo{author}{H.~Tu}, \bibinfo{author}{Y.~Wang}, \bibinfo{author}{Q.~Lan},
  \bibinfo{author}{W.~Liu}, \bibinfo{author}{W.~Xiao}, and
  \bibinfo{author}{S.~Ma}, \enquote{\bibinfo{title}{A {Chebyshev--Tau} spectral
  method for normal modes of underwater sound propagation with a layered marine
  environment}},  \bibinfo{journal}{Journal of Sound and Vibration}
  \textbf{492}, \bibinfo{pages}{115784} (\bibinfo{year}{2021})
  \dodoi{10.1016/j.jsv.2020.115784}.

\bibitem{Tuhw2021b}
\bibinfo{author}{H.~Tu}, \bibinfo{author}{Y.~Wang}, \bibinfo{author}{X.~Ma},
  and \bibinfo{author}{X.~Zhu}, \enquote{\bibinfo{title}{Applying the
  {Chebyshev--Tau} spectral method to solve the parabolic equation model of
  wide-angle rational approximation in ocean acoustics}},
  \bibinfo{journal}{Journal of Theoretical and Computational Acoustics}
  (\bibinfo{year}{2021}) \dodoi{10.1142/S2591728521500134}.

\bibitem{Wangyx2021a}
\bibinfo{author}{Y.~Wang}, \bibinfo{author}{H.~Tu}, \bibinfo{author}{W.~Liu},
  \bibinfo{author}{W.~Xiao}, and \bibinfo{author}{Q.~Lan},
  \enquote{\bibinfo{title}{Application of a {Chebyshev} collocation method to
  solve a parabolic equation model of underwater acoustic propagation}},
  \bibinfo{journal}{Acoustics Australia} \bibinfo{pages}{1--12}
  (\bibinfo{year}{2021}) \dodoi{10.1007/s40857-021-00218-5}.

\bibitem{SMPE}
\bibinfo{author}{H.~Tu},
  \enxquote{\bibinfo{title}{{\href{https://oalib-acoustics.org/Modes/index.html}{SMPE}}:
  Two spectral methods for solving the range-independent parabolic equation
  model in ocean acoustics}}  (\bibinfo{year}{2021}),
  \dourl{https://oalib-acoustics.org/models-and-software/parabolic-equation/}.

\bibitem{Tuhw2021c}
\bibinfo{author}{H.~Tu}, \bibinfo{author}{Y.~Wang}, \bibinfo{author}{Q.~Lan},
  \bibinfo{author}{W.~Liu}, \bibinfo{author}{W.~Xiao}, and
  \bibinfo{author}{S.~Ma}, \enquote{\bibinfo{title}{Applying a {Legendre}
  collocation method based on domain decomposition to calculate underwater
  sound propagation in a horizontally stratified environment}},
  \bibinfo{journal}{Journal of Sound and Vibration} \textbf{511},
  \bibinfo{pages}{116364} (\bibinfo{year}{2021})
  \dodoi{10.1016/j.jsv.2021.116364}.

\bibitem{Wangyx2021b}
\bibinfo{author}{Y.~Wang}, \bibinfo{author}{H.~Tu}, \bibinfo{author}{W.~Liu},
  \bibinfo{author}{W.~Xiao}, and \bibinfo{author}{Q.~Lan},
  \enquote{\bibinfo{title}{Two {Chebyshev} spectral methods for solving normal
  modes in atmospheric acoustics}},  \bibinfo{journal}{Entropy} \textbf{23},
  \bibinfo{pages}{705} (\bibinfo{year}{2021}) \dodoi{10.3390/e23060705}.

\bibitem{Dzieciuch1993}
\bibinfo{author}{M.~A. Dzieciuch}, \enquote{\bibinfo{title}{Numerical solution
  of the acoustic wave equation using {Chebyshev} polynomials with application
  to global acoustics}}, in \emph{\bibinfo{booktitle}{Proceedings of Oceans}},
  \bibinfo{publisher}{IEEE}, \bibinfo{address}{Victoria, BC, Canada}
  (\bibinfo{year}{1993}), pp. \bibinfo{pages}{267--271},
  \dodoi{10.1109/OCEANS.1993.326000}.

\bibitem{aw}
\bibinfo{author}{M.~A. Dzieciuch},
  \enxquote{\bibinfo{title}{{\href{https://oalib-acoustics.org/Modes/index.html}{aw}}:
  A {Matlab} code for computing normal modes based on {Chebyshev}
  approximations}}  (\bibinfo{year}{1993}),
  \dourl{https://oalib-acoustics.org/models-and-software/normal-modes/}.

\bibitem{rimLG}
\bibinfo{author}{R.~B. Evans},
  \enxquote{\bibinfo{title}{{\href{https://oalib-acoustics.org/Modes/index.html}{rimLG}}:
  A {Legendre--Galerkin} technique for differential eigenvalue problems with
  complex and discontinuous coefficients, arising in underwater acoustics}}
  (\bibinfo{year}{2020}),
  \dourl{https://oalib-acoustics.org/models-and-software/normal-modes/}.

\bibitem{NM-CT}
\bibinfo{author}{H.~Tu},
  \enxquote{\bibinfo{title}{{\href{https://oalib-acoustics.org/Modes/index.html}{NM-CT}}:
  A {Chebyshev--Tau} spectral method for normal modes of underwater sound
  propagation with a layered marine environment in {Matlab} and {Fortran}}}
  (\bibinfo{year}{2020}),
  \dourl{https://oalib-acoustics.org/models-and-software/normal-modes/}.

\bibitem{Sabatini2019}
\bibinfo{author}{R.~Sabatini} and \bibinfo{author}{P.~Cristini},
  \enquote{\bibinfo{title}{A multi-domain collocation method for the accurate
  computation of normal modes in open oceanic and atmospheric waveguides}},
  \bibinfo{journal}{Acta Acustica United with Acustica} \textbf{105},
  \bibinfo{pages}{464--474} (\bibinfo{year}{2019}) \dodoi{10.3813/AAA.919328}.

\bibitem{MultiLC}
\bibinfo{author}{H.~Tu},
  \enxquote{\bibinfo{title}{{\href{https://oalib-acoustics.org/Modes/index.html}{MultiLC}}:
  A {Legendre} collocation method based on domain decomposition to calculate
  underwater sound propagation in a horizontally stratified environment in
  {Matlab} and {Fortran}}}  (\bibinfo{year}{2021}),
  \dourl{https://oalib-acoustics.org/models-and-software/normal-modes/}.

\bibitem{Lanczos1938}
\bibinfo{author}{C.~Lanczos}, \enquote{\bibinfo{title}{Trigonometric
  interpolation of empirical and analytical functions}},
  \bibinfo{journal}{Journal of Mathematical Physics} \textbf{17},
  \bibinfo{pages}{123--199} (\bibinfo{year}{1938}).

\bibitem{Boyd2001}
\bibinfo{author}{J.~P. Boyd}, \emph{\bibinfo{title}{{Chebyshev} and {Fourier}
  Spectral Methods}}  (\bibinfo{publisher}{Second Edition, Dover},
  \bibinfo{address}{New York, USA}, \bibinfo{year}{2001}).

\bibitem{Canuto2006}
\bibinfo{author}{C.~Canuto}, \bibinfo{author}{M.~Y. Hussaini},
  \bibinfo{author}{A.~Quarteroni}, and \bibinfo{author}{T.~A. Zang},
  \emph{\bibinfo{title}{Spectral Methods Fundamentals in Single Domains}}
  (\bibinfo{publisher}{Spring-Verlag}, \bibinfo{address}{Berlin, German},
  \bibinfo{year}{2006}).

\bibitem{Min2005}
\bibinfo{author}{M.~S. Min} and \bibinfo{author}{D.~Gottlieb},
  \enquote{\bibinfo{title}{Domain decomposition spectral approximations for an
  eigenvalue problem with a piecewise constant coefficient}},
  \bibinfo{journal}{SIAM Journal on Numerical Analysis} \textbf{43},
  \bibinfo{pages}{502--520} (\bibinfo{year}{2005})
  \dodoi{10.1137/s0036142903423836}.

\bibitem{Jensen1998}
\bibinfo{author}{F.~B. Jensen}, \enquote{\bibinfo{title}{On the use of stair
  steps to approximate bathymetry changes in ocean acoustic models}},
  \bibinfo{journal}{The Journal of the Acoustical Society of America}
  \textbf{104}(3), \bibinfo{pages}{1310--1315} (\bibinfo{year}{1998})
  \dodoi{10.1121/1.424340}.

\bibitem{Top500}
\enxquote{\bibinfo{title}{Top500}}  (\bibinfo{year}{2022}),
  \dourl{https://www.top500.org/lists/top500/2022/06/}.

\bibitem{Amdahl1967}
\bibinfo{author}{G.~M. Amdahl}, \enquote{\bibinfo{title}{Validity of the single
  processor approach to achieving large scale computing capabilities}}, in
  \emph{\bibinfo{booktitle}{AFIPS}}, \bibinfo{publisher}{Association for
  Computing Machinery}, \bibinfo{address}{New York, United States}
  (\bibinfo{year}{1967}), pp. \bibinfo{pages}{483--485},
  \dodoi{10.1145/1465482.1465560}.

\end{thebibliography}

\end{document}